\pdfoutput=0
\documentclass[11pt]{article}
\usepackage{axodraw}
\usepackage{epsfig}
\usepackage{amsfonts}
\usepackage{amsmath}
\usepackage{bm,bbm}
\usepackage{cite}
\allowdisplaybreaks[1]

\input pix.sty

\newcommand{\kallen}{\sqrt{\lambda}}
\newcommand{\kallensq}{\lambda}
\newcommand{\K}{\rmii{$K$}}
\newcommand{\T}{\rmii{$T$}}
\newcommand{\muY}{\mu_\rmii{$Y$}}

\newcommand{\muL}{\mu_\rmii{$L$}}

\renewcommand{\eq}{eq.~}
\renewcommand{\eqs}{eqs.~}
\renewcommand{\se}{sec.~}
\renewcommand{\ses}{secs.~}
\renewcommand{\fig}{fig.~}
\renewcommand{\figs}{figs.~}

\newcommand{\rmO}{{\mathcal{O}}}
\newcommand{\bmu}{\bar\mu}

\def\lsi{\raise0.3ex\hbox{$<$\kern-0.75em\raise-1.1ex\hbox{$\sim$}}}
\def\gsi{\raise0.3ex\hbox{$>$\kern-0.75em\raise-1.1ex\hbox{$\sim$}}}

\newcommand{\gsim}{\mathop{\gsi}}

\newcommand{\sign}{\mathop{\mbox{sign}}}
\newcommand{\nF}{n_\rmii{F}}
\newcommand{\nB}{n_\rmii{B}}
\newcommand{\rmii}[1]{{\mbox{\tiny\rm{#1}}}}
\newcommand{\rmiii}[1]{{\mbox{\tiny{$\scriptstyle{\rm#1}$}}}}
\newcommand{\re}{\mathop{\mbox{Re}}}
\newcommand{\im}{\mathop{\mbox{Im}}}
\newcommand{\Tint}[1]{{\hbox{$\sum$}\!\!\!\!\!\!\!\int\,}_{\!\!\!\!\raise-0.9ex\hbox{$\scriptstyle{#1}$}}}
\newcommand{\Tinti}[1]{{{\Sigma}\!\!\!\!\raise0.3ex\hbox{$\int$}_\rmii{${#1}$}}}

\newcommand{\bi}{\begin{itemize}}
\newcommand{\ei}{\end{itemize}}
\newcommand{\hide}[1]{ }

\newcommand{\msl}[1]{\,\slash\!\!\!{#1}\,}
\newcommand{\scat}[1]{\mbox{scat}^{ }_\rmi{$#1$}}
\newcommand{\Sq}[2]{#2^2-#1} %% {#2^2-#1^{ }}
\newcommand{\s}[1]{s_{#1}}

\newcommand{\E}{\mathcal{E}}
\newcommand{\vare}{\varepsilon}
\newcommand{\U}{\mathcal{U}}

\renewcommand{\P}{\mathcal{P}}
\renewcommand{\K}{\mathcal{K}}
\newcommand{\MM}{M^2}

\newcommand{\muS}{\mu_\phi} % {\mu_\rmiii{S}}
\newcommand{\mellT}{m_{\ell T}}

\newcommand{\mQ}{m_\gamma} % {m_g} % {m_\rmii{$Q$}} % {m_\rmiii{Q}}
\newcommand{\ala}{{\ell}} % {{\ell_a}}

\newcommand{\aS}{\phi} % {{S}}

\newcommand{\aQ}{\gamma} % {g} % {Q}

\newcommand{\bla}{{\tilde\ell}}
\newcommand{\bS}{\tilde\phi} % {\tilde S}
\newcommand{\cQ}{\gamma} % {g} % {\rmiii{Q}}
\newcommand{\cS}{\phi} % {\rmiii{S}}
\def\TAsc(#1,#2)(#3,#4,#5)%
{\SetWidth{2.0}\CArc(#1,#2)(#3,#4,#5)\SetWidth{1.0}}
\def\Lwidth{3}

\def\TAgl(#1,#2)(#3,#4,#5){\SetWidth{2.0}\PhotonArc(#1,#2)(#3,#4,#5){\Lwidth}%
{6.283 #3 mul 360 div #4 #5 sub #4 #5 sub mul sqrt mul Tdensity mul}%
\SetWidth{1.0}}
\def\TLgl(#1,#2)(#3,#4){\SetWidth{2.0}\Photon(#1,#2)(#3,#4){\Lwidth}
{#1 #3 sub #1 #3 sub mul #2 #4 sub #2 #4 sub mul add sqrt Tdensity mul}%
\SetWidth{1.0}}
\renewcommand{\picc}[1]{\;\parbox[c]{60pt}{\begin{picture}(60,30)(0,-10)
\SetWidth{1.0}\SetScale{1.3} #1 \end{picture}}\; }
\def\Lwidth{1.3}

\renewcommand{\pic}[1]{\;\parbox[c]{30pt}{\begin{picture}(30,30)(0,-3)
\SetWidth{1.0}\SetScale{0.8} #1 \end{picture}}\;}
\renewcommand{\picb}[1]{\;\parbox[c]{45pt}{\begin{picture}(45,30)(0,-3)
\SetWidth{1.0}\SetScale{1.0} #1 \end{picture}}\;}
\def\Cut{\picc{%
 \Asc(30,5)(22.3,27,90)%
 \Asc(30,5)(22.3,90,153)%
 \Asc(30,25)(22.3,207,270)%
 \Asc(30,25)(22.3,270,333)%
 \Lsc(30,2.7)(30,27.3)%
 \Line(0,14)(10,14)%
 \Line(0,16)(10,16)%
 \Line(50,14)(60,14)%
 \Line(50,16)(60,16)%
 \SetWidth{1.0}%
 \Line(10,5)(50,25)
 \Text(28,40)[c]{$\scriptstyle a_1$}
 \Text(52,40)[c]{$\scriptstyle a_2$}
 \Text(45,15)[c]{$\scriptstyle a_3$}
 \Text(28,0)[c]{$\scriptstyle a_4$}
 \Text(52,0)[c]{$\scriptstyle a_5$}
 \Text(5,26)[c]{$\scriptstyle a_6$}
 \Text(74,26)[c]{$\scriptstyle a_6$}
}}
\def\decD{\picb{%
 \Line(5,14)(16,14)%
 \Line(5,16)(16,16)%
 \Lqu(16,16)(35,25)%
 \Lsc(16,14)(35,5)%
}}
\def\decE{\picb{%
 \Line(5,14)(16,14)%
 \Line(5,16)(16,16)%
 \Lqu(16,16)(45,30)%
 \Lsc(16,14)(30,7.5)%
 \Lgl(30,7.5)(45,15)%
 \Lsc(30,7.5)(45,0)%
}}
\def\decF{\picb{%
 \Line(5,14)(16,14)%
 \Line(5,16)(16,16)%
 \Line(16,16)(30,22.5)%
 \Lqu(30,22.5)(45,30)%
 \Lgl(30,22.5)(45,15)%
 \Lsc(16,14)(45,0)%
}}
\makeatletter \@addtoreset{equation}{section} \makeatother
\renewcommand{\theequation}{\arabic{section}.\arabic{equation}}
\makeatletter
\renewcommand\section{\@startsection {section}{1}{\z@}%
                                   {-5.5ex \@plus -1ex \@minus -.2ex}% bfr-
                                   {2.3ex \@plus.2ex}%
                                   {\normalfont\large\bfseries}}
\renewcommand\subsection{\@startsection{subsection}{2}{\z@}%
                                     {-3.25ex\@plus -1ex \@minus -.2ex}%
                                     {1.5ex \@plus .2ex}%
                                     {\normalfont\normalsize\bfseries}}
\renewcommand\thesection {\@arabic\c@section}
\renewcommand\thesubsection   {\thesection.\@arabic\c@subsection}
\renewcommand{\@seccntformat}[1]{%
\csname the#1\endcsname.\hspace{1.0em}}
\makeatother
\begin{document}

\flushbottom

\begin{titlepage}

\begin{flushright}
% OUTLINE  \\ 
% DRAFT \\ 
% arXiv:2107.07132\\ 
September 2021
\end{flushright}
\begin{centering}
\vfill

{\Large{\bf
 Efficient numerical integration of thermal interaction rates 
}} 

\vspace{0.8cm}

G.~Jackson$^\rmi{a}$ and 
M.~Laine$^\rmi{b}$
 
\vspace{0.8cm}

$^\rmi{a}$%
{\em
Institute for Nuclear Theory, Box 351550, University of Washington, \\ 
Seattle, WA 98195-1550, United States \\}

\vspace{0.3cm}

$^\rmi{b}$%
{\em
AEC, 
Institute for Theoretical Physics, University of Bern, \\ 
Sidlerstrasse 5, CH-3012 Bern, Switzerland \\}

\vspace*{0.8cm}

\mbox{\bf Abstract}
 
\end{centering}

\vspace*{0.3cm}
 
\noindent
In many problems in particle cosmology, interaction rates are dominated by 
${2}\leftrightarrow{2}$ scatterings, or get a substantial contribution from 
them, given that ${1}\leftrightarrow{2}$ and ${1}\leftrightarrow{3}$ 
reactions are phase-space suppressed. We describe an algorithm 
to represent, regularize, and evaluate a class of  
thermal ${2}\leftrightarrow{2}$ and 
${1}\leftrightarrow{3}$ interaction rates for general momenta, 
masses, chemical potentials, and helicity projections. 
A key ingredient is  an automated inclusion of  
virtual corrections to ${1}\leftrightarrow{2}$ scatterings, 
which eliminate logarithmic and double-logarithmic IR divergences from 
the real ${2}\leftrightarrow{2}$ and 
${1}\leftrightarrow{3}$ processes. 
We also review 
thermal and chemical potential induced contributions 
that require resummation if plasma particles are ultrarelativistic. 

\vfill

%% %\noindent
%% %PACS numbers: 
%% %11.10.Wx, %        Finite temperature field theory
%% %11.15.Ha, %        Lattice gauge theory  
%% %12.38.Bx, %        Perturbative calculations in QCD
%% %12.38.Mh, %        Quark--gluon plasma
%% %14.40.Nd, %        Bottom mesons
%% %\\
%% %Keywords: Thermal Field Theory, CP violation, Neutrino Physics, Resummation
 
\end{titlepage}

\tableofcontents

%%%%%%%%%%%%%%%%%%%%%%%%%%% SECTION %%%%%%%%%%%%%%%%%%%%%%%%%%%%%%%%%%%%%%
%
\section{Introduction}

In an interacting system, the simplest processes leading to 
thermal reaction (or damping, or interaction) rates 
are $1\to 2$ decays as well as $2\to 1$ ``inverse decays''. 
However, the phase space for such processes is kinematically constrained, 
as exemplified by the impossibility of a massive particle 
emitting a photon. Therefore, processes which at first sight appear
to be of higher order but do not suffer from 
a similarly stringent constraint, notably $2\leftrightarrow 2$ scatterings, 
may play a surprisingly important role in thermal systems. 

There are a number of problems for which the importance of  
various scattering types has been analyzed in quantitative detail. 
In the context of massless thermal QCD, this was first
done for photon production, showing that 
all $2\leftrightarrow 2$ scatterings,
as well as a certain subclass of  
$1+n\leftrightarrow 2+n$ reactions, with $n \ge 0$, 
had to be included on 
equal footing in order to determine the complete
leading order rate~\cite{amy1,amy2}. The story is similar for GeV-scale
right-handed neutrino production from a Standard Model plasma 
in the early universe, where $2\leftrightarrow 2$ scatterings
involving weak gauge bosons 
are numerically the largest individual 
contribution~\cite{bb1,bb2}, or for the flavour 
equilibration rate of the most weakly interacting Standard Model 
particles, right-handed electrons~\cite{eR}. 
There are even cases in which 
{\em only} $2\leftrightarrow 2$ scatterings need to be 
included at leading order, 
notably the production rates of axions, gravitinos, 
or gravitons~\cite{axion,
gravitino1,gravitino2,gravitino3,gravitino4,gravity_lo}, 
or particles which interact via the Fermi model, 
like neutrinos at low temperatures. 

We may expect 
$2\leftrightarrow 2$ scatterings to prevail quite generally, 
including in many freeze-in dark matter computations
(cf.,\ e.g.,\ ref.~\cite{freezein}) 
or leptogenesis scenarios
(cf.,\ e.g.,\ ref.~\cite{klaric}). 
Yet, $2\leftrightarrow 2$ scatterings  
are often not properly accounted for. 
For instance, popular dark matter packages (cf.,\ e.g.,\ refs.~\cite{MO,DS})
typically treat $2\leftrightarrow 2$ scatterings through 
some empirical approximation,
or interpolation between tabulated points. However, as has been
pointed out e.g.\ in ref.~\cite{salvio}, the fact that virtual
corrections to $1\leftrightarrow 2$ scatterings are omitted, 
implies that such results
suffer from uncancelled IR divergences if some masses are small, 
and might therefore significantly 
{\em overestimate} the magnitude of 
$2\leftrightarrow 2$ reactions.\footnote{% 
  These packages differ from ours
  in other aspects as well, notably by considering
  momentum averages, whereas we compute momentum-dependent
  rates, or by focussing on  
  processes respecting an $R$-symmetry, whereas we consider
  the minimal case of a single particle species freezing in or out. 
  The latter setups coincide if the $R$-charged sector contains
  one species which interacts more weakly than the others. 
  }

There should be no technical obstacle
to fully including the $2\leftrightarrow 2$ scatterings.
In fact, 
if all plasma particles can be treated 
as massless whereas the particle of interest 
is relativistic, with a mass $M\sim \pi T$, then   
$2\leftrightarrow 2$, 
$1\leftrightarrow 3$, 
as well as the corresponding virtual corrections to 
$1\leftrightarrow 2$ 
interaction rates, 
can be reduced to a convergent two-dimensional integral
representation~\cite{master,relativistic,greg}. 
However, this has not been worked out
for a massive and/or charge-asymmetric plasma; furthermore, the handling
of any new functional form of a matrix element squared requires
substantial hand work. It would be welcome to develop 
a user-friendly approach, requiring
minimal input for a given problem, and applicable for
general masses, with the price that a three-dimensional
integration is performed numerically. 

The purpose of the present paper
is to show how this task can be met. 
We provide a prescription for how the 
interaction rates originating from $2\leftrightarrow 2$ as well as
the related $1\leftrightarrow 3$ processes 
can be evaluated. 
To achieve this, the poles that appear in the matrix elements squared
must be cancelled by the identification and inclusion of the pertinent
virtual corrections to $1\leftrightarrow 2$ processes.
For the subsequent numerical evaluation,  
we generalize the parametrizations 
that were introduced in ref.~\cite{bb2},  
to a situation with an arbitrary mass and 
chemical potential spectrum of the participating particles. 
The numerical and algebraic scripts
implementing these ingredients to a particular example 
are made publicly available, 
with the hope that they can be adapted to other
problems with modest effort. 

The presentation is organized as follows. 
We start with an exposition of how to concisely represent 
$2\leftrightarrow 2$ and $1\leftrightarrow 3$ scatterings, and how to 
cancel the poles that appear in them, in \se\ref{se:2to2}.
A regularization permitting to evaluate various contributions  
{\em before} the cancellation of poles is introduced in 
\se\ref{se:IR}.
Subsequently parametrizations for the 
remaining integrations are introduced, in \se\ref{se:phasespace}.
Even if all poles have been cancelled, the results still suffer
from divergences in certain limits, as reviewed in \se\ref{se:resum}.
Sec.~\ref{se:code} summarizes 
the algorithms implementing the ingredients
of \ses\ref{se:2to2}--\ref{se:phasespace}. 
Conclusions are offered in \se\ref{se:concl}, deferring formal 
proofs of thermal crossing relations to appendix~A,  
integration prescriptions for specific channels to appendix~B, 
and angular averages for virtual corrections to appendix~C. 

%%%%%%%%%%%%%%%%%%%%%%%%%%% SECTION %%%%%%%%%%%%%%%%%%%%%%%%%%%%%%%%%%%%%%
%
\section{Thermally averaged 
${2}\leftrightarrow{2}$ and ${1}\leftrightarrow{3}$ reactions}
\la{se:2to2}

%%%%%%%%%%%%%%%%%%%%%%%%%%% SUBSECTION %%%%%%%%%%%%%%%%%%%%%%%%%%%%%%%%%%%
%
\subsection{General structure and crossing relations}
\la{ss:master}

We are interested in the interaction rate of a non-equilibrium
particle, whose four-momentum is denoted by
$\K = (\omega,\vc{k})$ and mass squared by 
$M^2 \equiv \K^2 = \omega^2 - k^2$.
The interaction rate parametrizes a rate equation, describing
how this particle is approaching thermal equilibrium, or trying
to keep up with its changing temperature or density. 
It is not necessary for
us to specify the exact form of the rate equation, but it could
be of Boltzmann type, or have a more general appearance,  
involving density matrices. In quantum field theory, 
the interaction rate can be defined as the imaginary part (or cut)
of a retarded correlation function~\cite{dbtheory1}. 

An essential assumption we make is that 
the coupling between the non-equilibrium particle and the plasma
is much weaker than the generic couplings between the plasma
particles. We work to leading (2nd) order in the weak coupling 
associated with the non-equilibrium particle, whereas for plasma
couplings, higher orders can be considered as well.\footnote{% 
 Actually, our formalism applies somewhat more generally. 
 For instance, for the dynamics of multiple generations of 
 (degenerate or non-degenerate) right-handed neutrinos, 
 it can be shown~\cite{simultaneous,dbX}
 that the rate equations take a form
 in which many important 
 4th order effects factorize 
 into products of 2nd order rate coefficients, 
 for which our discussion applies.
 } 

Given that the plasma particles are assumed to be in equilibrium, their phase
space distributions take standard forms. 
For a simultaneous treatment of bosons and fermions, we denote
\be
 n^{ }_{\sigma}(\epsilon) \; \equiv \; \frac{\sigma}{e^{\epsilon/T} - \sigma}
 \;, \quad
 \bar{n}^{ }_{\sigma}(\epsilon) \; \equiv \; 1 + n^{ }_{\sigma}(\epsilon) 
 \;, \quad
 \sigma = \pm
 \;, \la{n_sigma}
\ee
where $T$ is the temperature. 
The distributions are related by 
$ 
 n^{ }_{\sigma}(-\epsilon) = -\, \bar{n}^{ }_{\sigma}(\epsilon)
$.
Thermal averages of reaction rates
involving the non-equilibrium particle, 
whose physical role becomes apparent around 
\eqs\nr{operator} and \nr{master},  
are defined as\hspace*{0.1mm}\footnote{%
 The overall factor $\frac{1}{2}$ can be viewed as a convention, 
 but it also guarantees a straightforward connection to Boltzmann
 equations on one hand (so that $\Theta$ in \eq\nr{master} corresponds
 to a ``matrix element squared''), and to the retarded correlators 
 considered in appendix~A on the other 
 (originating then as $\im\frac{1}{\scriptstyle\sum_i \epsilon_i - i 0^+}  
 = \pi\delta(\sum_i\epsilon_i) = \frac{1}{2} (2\pi) \delta(\sum_i\epsilon_i)$,
 with $2\pi$ included as 
 part of the usual energy-momentum conservation constraint).
 } 
\ba
 \scat{1\to3}(a,b,c) & \equiv & 
 \frac{1}{2} \int \! {\rm d}\Omega^{ }_{1\to3} 
          \, \mathcal{N}^{ }_{a,b,c} 
 \;,  \la{scatdef} \\ 
%%%%
 {\rm d}\Omega^{ }_{1\to3} & \equiv & 
 \frac{1}{(2\pi)^9}
 \frac{{\rm d}^3\vc{p}^{ }_a}{2 \epsilon^{ }_{a}}
 \frac{{\rm d}^3\vc{p}^{ }_b}{2 \epsilon^{ }_{b}}
 \frac{{\rm d}^3\vc{p}^{ }_c}{2 \epsilon^{ }_{c}}
 \,(2\pi)^4 \delta^{(4)}(\K - \P^{ }_a
                       - \P^{ }_b - \P^{ }_c )
 \;, \la{dO1to3} \\[2mm] 
%%%%
 \mathcal{N}^{ }_{a,b,c} 
 & \equiv & 
   \bar{n}^{ }_{\sigma_a}(\epsilon^{ }_{a} - \mu^{ }_a)
 \,\bar{n}^{ }_{\sigma_b}(\epsilon^{ }_{b} - \mu^{ }_b)
 \,\bar{n}^{ }_{\sigma_c}(\epsilon^{ }_{c} - \mu^{ }_c)
 \nn 
 & - & 
   n^{ }_{\sigma_a}(\epsilon^{ }_{a} - \mu^{ }_a)
 \,n^{ }_{\sigma_b}(\epsilon^{ }_{b} - \mu^{ }_b)
 \,n^{ }_{\sigma_c}(\epsilon^{ }_{c} - \mu^{ }_c)
 \;, \la{calN1to3} \\[2mm]
%%%%%%%%%%%%%%%%%%%%%%%%%%%%%%%%%%%%%%%%%%%
 \scat{2\to2}(-a;b,c) & \equiv & 
 \frac{1}{2} \int \! {\rm d}\Omega^{ }_{2\to2} 
          \, \mathcal{N}^{ }_{a;b,c} 
 \;, \\ 
%%%%
 {\rm d}\Omega^{ }_{2\to2} & \equiv & 
 \frac{1}{(2\pi)^9}
 \frac{{\rm d}^3\vc{p}^{ }_a}{2 \epsilon^{ }_{a}}
 \frac{{\rm d}^3\vc{p}^{ }_b}{2 \epsilon^{ }_{b}}
 \frac{{\rm d}^3\vc{p}^{ }_c}{2 \epsilon^{ }_{c}}
 \,(2\pi)^4 \delta^{(4)}(\K + \P^{ }_a 
                       - \P^{ }_b - \P^{ }_c )
 \;, \la{dO2to2} \\[2mm] 
%%%%
 \mathcal{N}^{ }_{a;b,c} 
 & \equiv & 
 n^{ }_{\sigma_a}(\epsilon^{ }_{a} + \mu^{ }_a)
 \,\bar{n}^{ }_{\sigma_b}(\epsilon^{ }_{b} - \mu^{ }_b)
 \,\bar{n}^{ }_{\sigma_c}(\epsilon^{ }_{c} - \mu^{ }_c)
 \nn 
 & - & 
 \bar{n}^{ }_{\sigma_a}(\epsilon^{ }_{a} + \mu^{ }_a)
 \,n^{ }_{\sigma_b}(\epsilon^{ }_{b} - \mu^{ }_b)
 \,n^{ }_{\sigma_c}(\epsilon^{ }_{c} - \mu^{ }_c)
 \;, \la{calN2to2} \\[2mm]
%%%%%%%%%%%%%%%%%%%%%%%%%%%%%%%%%%%%%%%%%%%%%%%
 \scat{3\to1}(-a,-b;c) & \equiv & 
 \frac{1}{2} \int \! {\rm d}\Omega^{ }_{3\to1} 
          \, \mathcal{N}^{ }_{a,b;c} 
 \;, \\ 
%%%%
 {\rm d}\Omega^{ }_{3\to1} & \equiv & 
 \frac{1}{(2\pi)^9}
 \frac{{\rm d}^3\vc{p}^{ }_a}{2 \epsilon^{ }_{a}}
 \frac{{\rm d}^3\vc{p}^{ }_b}{2 \epsilon^{ }_{b}}
 \frac{{\rm d}^3\vc{p}^{ }_c}{2 \epsilon^{ }_{c}}
 \, (2\pi)^4 \delta^{(4)}(\K + \P^{ }_a 
                       + \P^{ }_b - \P^{ }_c )
 \;, \la{dO3to1} \\[2mm] 
%%%%
 \mathcal{N}^{ }_{a,b;c} 
 & \equiv & 
 n^{ }_{\sigma_a}(\epsilon^{ }_{a} + \mu^{ }_a)
 \, n^{ }_{\sigma_b}(\epsilon^{ }_{b} + \mu^{ }_b)
 \, \bar{n}^{ }_{\sigma_c}(\epsilon^{ }_{c} - \mu^{ }_c)
 \nn
 & - & 
 \bar{n}^{ }_{\sigma_a}(\epsilon^{ }_{a} + \mu^{ }_a)
 \,\bar{n}^{ }_{\sigma_b}(\epsilon^{ }_{b} + \mu^{ }_b)
 \,n^{ }_{\sigma_c}(\epsilon^{ }_{c} - \mu^{ }_c)
 \;, \hspace*{6mm} \la{calN3to1}
%%%%%%%%%%%%%%%%%%%%%%%%%%%%%%%%%%%%%%%%%%%%%%%
\ea
where $\P^{ }_a = (\epsilon^{ }_a,\vc{p}^{ }_a)$ 
are four-momenta, with $\P_a^2 = m_a^2$.
All plasma particles are assumed to be in chemical equilibrium, 
so that their chemical potentials are constrained by linear relations. 

It is apparent from \eqs\nr{scatdef}--\nr{calN3to1} that a minus sign
in front of a label belonging to an initial state, e.g.\ $-a$, 
corresponds to an inversion of the sign of 
the corresponding four-momentum and chemical potential. 
This pattern continues to hold once
matrix elements squared are added. 
While the chemical potentials have been treated
within the functions $\mathcal{N}$, it is convenient to 
interpret $\scat{n\to m}$ as an ``operator'', acting on the momenta
that appear in the matrix element squared, with the rule that
\be
 \scat{n\to m}(...,-a,...;...) \, \Phi(...,\P^{ }_a,...)
 \; \equiv \; 
 \frac{1}{2} \int \! {\rm d}\Omega^{ }_{n\to m} 
 \, \mathcal{N}^{ }_{... a ... ; ...}
 \, \Phi(...,-\P^{ }_a,...)
 \;.
 \la{operator} 
\ee

We now posit that for considering $2\leftrightarrow 2$ and 
$1\leftrightarrow 3$ scatterings, it is sufficient 
to define the integrand needed for ${1\to3}$ decays, 
with the other processes following from it through crossing symmetries. 
The reason for selecting the ${1\to3}$ decays is that they display maximal 
symmetries, with all thermalized particles appearing in the final state
(of course, in phase space integrations, inverse processes are 
included as well, cf.\ \eq\nr{calN1to3}).
A formal proof of this statement can be found in appendix~A.1. 

Concretely, if we denote by $\Theta^{ }_{ }(\P^{ }_a,\P^{ }_b,\P^{ }_c)$
the matrix element squared for $1\to 3$ decays,
expressed in terms of physical variables 
(momenta, masses, energies, and helicities), then, 
in accordance with \eq\nr{cut},  
the full interaction rate\hspace*{0.1mm}\footnote{%
  We are somewhat lax with terminology here: 
  depending on how $\Theta$ is chosen, the ``rate proper'' may
  still require a division of $\Gamma$ by $\omega$, as appears for instance
  in the collision term of a Boltzmann equation. 
  }
reads (for $\omega > 0$)
\ba
 \Gamma^\rmi{Born}_{2\leftrightarrow 2,1\leftrightarrow 3}
 & =  & 
 \; \bigl[ \; 
  \scat{1\to3}(a^{ },b^{ },c^{ })
 \nn 
 &  & \;   
  +\, \scat{2\to2}(-a^{ };b^{ },c^{ })
  +\, \scat{2\to2}(-b^{ };c^{ },a^{ })
  +\, \scat{2\to2}(-c^{ };a^{ },b^{ })
 \nn 
 &  & \;  
  +\, \scat{3\to1}(-a^{ },-b^{ };c^{ })
  +\, \scat{3\to1}(-b^{ },-c^{ };a^{ })
  +\, \scat{3\to1}(-c^{ },-a^{ };b^{ })
  \, \bigr]
 \nn[1mm] 
 & \times & 
  \, \Theta^{ }_{ }(\P^{ }_a,\P^{ }_b,\P^{ }_c)
 \;. \la{master}
\ea
Mathematically, if $\omega < 0$ were viable, there would also be 
an 8th channel, of type $\scat{4\to0}$.

%%%%%%%%%%%%%%%%%%%%%%%%%%%%%%% FIGURE %%%%%%%%%%%%%%%%%%%%%%%%%%%%%%%%%%%%%%
%
\begin{figure}[t]
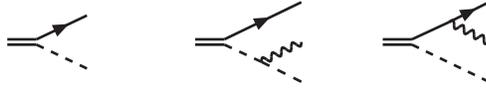


%\hspace*{1.5cm}%
%\begin{minipage}[c]{3cm}
\begin{eqnarray*}
%%%%%%%%%%
% \scat{1\to3}(\ala,\aQ,\aS): 
&& 
 \decD
 \hspace*{0.70cm}
 \decE
 \hspace*{0.70cm}
 \decF
 \hspace*{-0.00cm}
 \nn 
%%%%%%%%%%
\end{eqnarray*}
%\end{minipage}

\vspace*{-6mm}

\caption[a]{\small 
 As an example of amplitudes needed for determining the basic input 
 for our algorithm, we show the
 $1\rightarrow 2$ process leading to 
 $
   \Gamma^\rmi{Born}_{1\to 2} = 
   \scat{1\to2}(\ala,\aS) \, 4 \E\cdot\P^{ }_\ala 
 $, 
 and
 the $1\rightarrow 3$ processes leading to \eq\nr{example_1to3}.  
 Double lines correspond to right-handed neutrinos; 
 arrowed lined to leptons ($\ala$); 
 wiggly lines to gauge fields ($\aQ$);
 dashed lines to scalars ($\aS$). 
} 
\la{fig:1to3}
\end{figure}
%
%%%%%%%%%%%%%%%%%%%%%%%%%%%%%%%%%%%%%%%%%%%%%%%%%%%%%%%%%%%%%%%%%%%%%%%%%%%%%

As an example, consider right-handed neutrino interactions in the 
symmetric phase of a Standard Model plasma~\cite{bb1,bb2}. Then 
$1\to 3$ decays, illustrated in \fig\ref{fig:1to3},  
can go to a final state with 
a lepton, a gauge boson, and a scalar particle
($a^{ },b^{ },c^{ } = \ala,\aQ,\aS$).\footnote{%
 In the full physical computation, 
 there are also channels in which $\aQ,\aS$ are replaced  
 by a fermion-antifermion pair,
 originating from a scalar decay, however the algebraic 
 structure of the corresponding matrix element squared is rather simple, 
 and brings in nothing new to the case that we consider. 
 } 
The matrix element squared $|\mathcal{M}|^2$ for this process, 
summed over final-state helicities and with the substitution 
$\sum_\tau u^{ }_{\vc{k}\tau} \bar{u}^{ }_{\vc{k}\tau}
\to \msl{\E}$ for the initial-state helicity sum, 
but for convenience factoring out Yukawa couplings from the first vertex, 
leads to 
\ba
 \Gamma^\rmi{Born}_{1\to 3}
 & \equiv & 
 \scat{1\to3}(\ala,\aQ,\aS) \, 
 \Theta^{ }_{ }(\P^{ }_\ala,\P^{ }_\aQ,\P^{ }_\aS)
 \nn[2mm] 
  & \equiv &  
 \scat{1\to3}(\ala,\aQ,\aS) \, 
  2 (g_1^2 + 3 g_2^2)
  \la{example_1to3}
  \\[2mm] & \times & \!\!\! 
                \biggl\{ 
% \nn & & +
                     - \frac{
                        \E\cdot\P^{ }_\ala
                       }{ 
                         { s^{ }_{\ala\aQ} } }
                     + \frac{
                       (\MM - m_\cS^2 )\,
                        \E\cdot(\P^{ }_\aQ + 2 \P^{ }_\ala)
                       }{ 
                         { s^{ }_{\ala\aQ} } ({ s^{ }_{\aQ\aS} - m_\cS^2 }) }
%  \nn & &  
                 -      \frac{ 
                          \E\cdot(\P^{ }_\ala + \K )    
                             }{ { s^{ }_{\aQ\aS} - m_\cS^2 } }
                 -      \frac{ 
                         2 m_\cS^2 \, \E\cdot\P^{ }_\ala  
                             }{ ({ s^{ }_{\aQ\aS} - m_\cS^2 })^2 }
                       \biggr\} 
 \;, \nonumber  \hspace*{7mm}
\ea
where $s^{ }_{ab} \equiv (\P^{ }_a + \P^{ }_b)^2$,
% $a \in \{e,\mu,\tau\} $ denotes a leptonic flavour, 
$g^{ }_{1,2}$ are the U$^{ }_\rmii{Y}$(1) and 
SU$^{ }_\rmii{L}$(2) gauge couplings, respectively, 
and $m_\cS$ is the (thermally corrected) mass of the scalar particle.  
We note that the representation can be put in alternative forms
by making use of the identity
$
 s^{ }_{\ala\aQ} + s^{ }_{\aQ\aS} + s^{ }_{\aS\ala}
 = 
 \MM + m_\cS^2
$. 
The metric signature ($+$$-$$-$$-$) is assumed throughout. 

For the following, it is helpful to rewrite the physical cross section
in \eq\nr{example_1to3}, by denoting the locations of the propagator poles
with specific symbols, which may be called auxiliary masses. 
This is helpful not only as a labelling of pole locations, but also serves
as an intermediate IR regulator (cf.\ \se\ref{se:IR}), 
and in addition permits to take parametric
derivatives with respect to specific propagators, as is needed
later on. With this motivation, we replace
\ba
 && \hspace*{-1.5cm} 
 \frac{ \Gamma^\rmi{Born}_{1\to 3} }{  2 (g_1^2 + 3 g_2^2) }
 \; \to \; 
 \scat{1\to3}(\ala,\aQ,\aS) \, 
  \la{example_1to3_alt}
  \\ & \times & \!\!\! 
                \biggl\{ 
% \nn & & +
                     - \frac{
                        \E\cdot\P^{ }_\ala
                       }{ 
                         { s^{ }_{\ala\aQ} } - m_\bla^2 }
                     + \frac{
                       (\MM - m_\cS^2 )\,
                        \E\cdot(\P^{ }_\aQ + 2 \P^{ }_\ala)
                       }{ 
                         ({ s^{ }_{\ala\aQ} - m_\bla^2 })
                         ({ s^{ }_{\aQ\aS} - m_{\tilde{\cS}}^2 }) }
%  \nn & &  
                 -      \frac{ 
                          \E\cdot(\P^{ }_\ala + \K )    
                             }{ { s^{ }_{\aQ\aS} - m_{\tilde{\cS}}^2 } }
                 -      \frac{ 
                         2 m_\cS^2 \, \E\cdot\P^{ }_\ala  
                             }{ ({ s^{ }_{\aQ\aS} - m_{\tilde{\cS}}^2 })^2 }
                       \biggr\} 
 \;. \nonumber  \hspace*{7mm}
\ea
The (would-be) masses of the decay products are denoted
by $m_\ala^2 \equiv \P_\ala^2$, $m_\cQ^2 \equiv \P_\aQ^2$, 
$m_\cS^2 \equiv \P_\aS^2$.

%%%%%%%%%%%%%%%%%%%%%%%%%%% SUBSECTION %%%%%%%%%%%%%%%%%%%%%%%%%%%%%%%%%%%
%
\subsection{Corresponding virtual corrections}
\la{ss:virtual}

If the energy entering 
one of the propagators in \eq\nr{example_1to3_alt} 
equals the on-shell energy of that line --- that is, 
if three energies are related to each other like in 
a $1\leftrightarrow 2$ process --- 
then the integrand diverges, and the value of the 
integral is ambiguous. 
In accordance with the 
KLN theorem~\cite{kln1,kln2},
such divergences are cancelled by virtual corrections to 
$1\leftrightarrow 2$ processes. Conversely, requiring the 
cancellation of divergences, 
we can reconstruct virtual corrections. 

A key observation for cancelling the divergences is  
that the thermal distribution functions appearing 
in \eqs\nr{calN1to3}, \nr{calN2to2}, \nr{calN3to1}
can always be factorized,
in fact in several ways: 
\ba
 \mathcal{N}^{ }_{a,b,c}
 \!\! & = & \!\!  
 \bigl[
    1
      + n^{ }_{\sigma_a\sigma_b}(\epsilon^{ }_a + \epsilon^{ }_b 
                                 - \mu^{ }_a - \mu^{ }_b ) 
      + n^{ }_{\sigma_c}(\epsilon^{ }_c - \mu^{ }_c) 
 \bigr]
 \bigl[
        1 + n^{ }_{\sigma_a}(\epsilon^{ }_a - \mu^{ }_a )
          + n^{ }_{\sigma_b}(\epsilon^{ }_b - \mu^{ }_b )
 \bigr]
 \;, \nn 
 \la{N_1to3_gen}
 \\
%%%%
 \mathcal{N}^{ }_{a;b,c}
 \!\! & = & \!\!  
 \bigl[
        1 + n^{ }_{\sigma_a\sigma_b}(\epsilon^{ }_b - \epsilon^{ }_a
                                    - \mu^{ }_a - \mu^{ }_b)
          + n^{ }_{\sigma_c}(\epsilon^{ }_c - \mu^{ }_c )
 \bigr]
 \bigl[     n^{ }_{\sigma_a}(\epsilon^{ }_a + \mu^{ }_a)
          - n^{ }_{\sigma_b}(\epsilon^{ }_b - \mu^{ }_b) 
 \bigr]
 \nn
 \la{N_2to2_t_gen}
 \\
%%%%
 \!\! & = & \!\! 
 \bigl[
        n^{ }_{\sigma_a}(\epsilon^{ }_a + \mu^{ }_a) 
      - n^{ }_{\sigma_b\sigma_c}(\epsilon^{ }_b + \epsilon^{ }_c
                                - \mu^{ }_b - \mu^{ }_c ) 
 \bigr]
 \bigl[
        1 + n^{ }_{\sigma_b}(\epsilon^{ }_b - \mu^{ }_b)
          + n^{ }_{\sigma_c}(\epsilon^{ }_c - \mu^{ }_c )\bigr]
 \;, \nn
 \la{N_2to2_s_gen}
 \\
%%%%
 \mathcal{N}^{ }_{a,b;c}
 \!\! & = & \!\!
 \bigl[
      1+ n^{ }_{\sigma_a\sigma_c}(\epsilon^{ }_a - \epsilon^{ }_c
                                 + \mu^{ }_a + \mu^{ }_c)
       + n^{ }_{\sigma_b}(\epsilon^{ }_b + \mu^{ }_b) 
 \bigr]
 \bigl[
            n^{ }_{\sigma_a}(\epsilon^{ }_a + \mu^{ }_a)
          - n^{ }_{\sigma_c}(\epsilon^{ }_c - \mu^{ }_c)
 \bigr]
 \nn 
 \la{N_3to1_t_gen}
 \\
%%%%
 \!\! & = & \!\!
 \bigl[
        n^{ }_{\sigma_a\sigma_b}(\epsilon^{ }_a + \epsilon^{ }_b
      + \mu^{ }_a + \mu^{ }_b)
      - n^{ }_{\sigma_c}(\epsilon^{ }_c - \mu^{ }_c) 
 \bigr]
 \bigl[
        1 + n^{ }_{\sigma_a}(\epsilon^{ }_a + \mu^{ }_a)
          + n^{ }_{\sigma_b}(\epsilon^{ }_b + \mu^{ }_b)
 \bigr]
 \;. \nn 
 \la{N_3to1_s_gen}
\ea
In the cases involving energy differences, another useful representation
can be obtained through the identity 
$
 n^{ }_{\sigma}(-x) = -1 - n^{ }_{\sigma}(x)
$.
Eq.~\nr{N_1to3_gen} can be put in two alternative forms by the exchanges 
$a\leftrightarrow c$ and $b\leftrightarrow c$, 
whereas \eqs\nr{N_2to2_t_gen} and \nr{N_3to1_t_gen} give
additional identities via the exchanges $b\leftrightarrow c$ 
and $a\leftrightarrow b$, respectively. 

All the factors in \eqs\nr{N_1to3_gen}--\nr{N_3to1_s_gen}
are of the type that appear in $1\leftrightarrow 2$ processes, 
defined in \eqs\nr{N_1to2_gen} and \nr{N_2to1_gen}.
It is such $1\leftrightarrow 2$ processes
that cancel the IR divergences of the 
$2\leftrightarrow 2$ and $1\leftrightarrow 3$ reactions. 

In order to make the statement concrete, 
let us denote thermally averaged phase space integrals 
for $1\leftrightarrow 2$ processes, interpreted 
as operators acting on the momenta $\P^{ }_a$ and $\P^{ }_b$
with the rule introduced in \eq\nr{operator}, by 
\ba
 \scat{1\to2}(a,b) & \equiv & 
 \frac{1}{2} \int \! {\rm d}\Omega^{ }_{1\to2} 
          \, \mathcal{N}^{ }_{a,b} 
 \;, \\ 
%%%%
 {\rm d}\Omega^{ }_{1\to2} & \equiv & 
 \frac{1}{(2\pi)^6}
 \frac{{\rm d}^3\vc{p}^{ }_a}{2 \epsilon^{ }_{a}}
 \frac{{\rm d}^3\vc{p}^{ }_b}{2 \epsilon^{ }_{b}}
 \, (2\pi)^4 \delta^{(4)}(\K - \P^{ }_a
                       - \P^{ }_b )
 \;, \\[2mm] 
%%%%
 \mathcal{N}^{ }_{a,b} 
 & \equiv & 
 \bar{n}^{ }_{\sigma_a}(\epsilon^{ }_{a} - \mu^{ }_a)
 \,\bar{n}^{ }_{\sigma_b}(\epsilon^{ }_{b} - \mu^{ }_b)
% \nn 
% & - & 
 - 
   n^{ }_{\sigma_a}(\epsilon^{ }_{a} - \mu^{ }_a)
 \,n^{ }_{\sigma_b}(\epsilon^{ }_{b} - \mu^{ }_b)
 \;, \la{N_1to2_gen} \\[2mm]
%%%%%%%%%%%%%%%%%%%%%%%%%%%%%%%%%%%%%%%%%%%
 \scat{2\to1}(-a;b) & \equiv & 
 \frac{1}{2} \int \! {\rm d}\Omega^{ }_{2\to1} 
          \, \mathcal{N}^{ }_{a;b} 
 \;, \\ 
%%%%
 {\rm d}\Omega^{ }_{2\to1} & \equiv & 
 \frac{1}{(2\pi)^6}
 \frac{{\rm d}^3\vc{p}^{ }_a}{2 \epsilon^{ }_{a}}
 \frac{{\rm d}^3\vc{p}^{ }_b}{2 \epsilon^{ }_{b}}
 \, (2\pi)^4 \delta^{(4)}(\K + \P^{ }_a 
                       - \P^{ }_b )
 \;, \\[2mm] 
%%%%
 \mathcal{N}^{ }_{a;b} 
 & \equiv & 
 n^{ }_{\sigma_a}(\epsilon^{ }_{a} + \mu^{ }_a)
 \,\bar{n}^{ }_{\sigma_b}(\epsilon^{ }_{b} - \mu^{ }_b)
% \nn 
% & - & 
 - 
 \bar{n}^{ }_{\sigma_a}(\epsilon^{ }_{a} + \mu^{ }_a)
 \,n^{ }_{\sigma_b}(\epsilon^{ }_{b} - \mu^{ }_b)
 \;. \hspace*{5mm} \la{N_2to1_gen}
%%%%%%%%%%%%%%%%%%%%%%%%%%%%%%%%%%%%%%%%%%%%%%%
\ea
Furthermore we denote
\be
 \scat{1\leftrightarrow2}(a,b) 
 \; \equiv \; 
 \scat{1\to2}(a,b) + 
 \scat{2\to1}(-a;b) + 
 \scat{2\to1}(-b;a) 
 \;. \la{scat1lr2}
\ee 
% as well as $s^{ }_{ab} \; \equiv \; (\P^{ }_a + \P^{ }_b)^2$. 

%%%%%%%%%%%%%%%%%%%%% PARAGRAPH %%%%%%%%%%%%%%%%%%%%%%%%%%%%%%%%%%%%%%%%%%%
%
%\paragraph{Single pole.}
\subsubsection*{Single pole}

Let us start by considering a first-order pole in 
a variable $s^{ }_{ab}$. 
The corresponding ``residue'' is denoted by 
\be
 \widehat\Theta^{(1)}_{ab}(\P^{ }_a,\P^{ }_b,\P^{ }_c)
 \; \equiv \; 
 \lim_{s^{ }_{ab}\to m^2_{d}} 
 (s^{ }_{ab} -  m^2_{d})\,\Theta(\P^{ }_a,\P^{ }_b,\P^{ }_c)
 \;, \la{residue1}
\ee
where $ \Theta(\P^{ }_a,\P^{ }_b,\P^{ }_c) $ is the matrix element
squared appearing in \eq\nr{master}. 

There are various ways to establish what kind of 
$1\leftrightarrow 2$ processes are needed in order to cancel 
this pole. Probably the most economical is to consider 
a master sum-integral, as explained in appendix~A.2. In this case, 
the $2\leftrightarrow 2$ and $1\leftrightarrow 3$ processes and
the virtual corrections to $1\leftrightarrow 2$ processes
that cancel their poles are generated simultaneously from 
a single structure that respects the KLN theorem. 
Alternatively, one could take the cancellation
of divergences as the construction principle, and then make
use of the factorization formulae in 
\eqs\nr{N_1to3_gen}--\nr{N_3to1_s_gen}, however 
the implementation of the latter strategy is tedious.  

We adopt a notation where 
$\Gamma^\rmi{Born}_{1\leftrightarrow 2}$ denotes the interaction rate
corresponding to a Born-level $ 1\leftrightarrow 2 $ process, 
and $\Delta \Gamma^\rmi{Born}_{1\leftrightarrow 2}$
the virtual corrections to this process. Our algorithm reconstructs, 
strictly speaking, {\em only the IR sensitive part of the full} 
$\Delta \Gamma^\rmi{Born}_{1\leftrightarrow 2}$.
Specifically, 
the virtual correction associated with a residue 
like in \eq\nr{residue1} reads 
\ba
 \Delta \Gamma^\rmi{Born}_{1\leftrightarrow 2}
 & \supset & 
  \scat{1\leftrightarrow 2}(d,c)
  \, B(\P^{ }_{d};a,b)
 \; \widehat\Theta^{(1)}_{ab}(\P^{ }_a,\P^{ }_b,\K - \P^{ }_d)
 \;. \la{virtual_bubble}
\ea
Here we have defined a ``bubble'' operator, $B$,  
acting on the momenta $\P^{ }_a$ and $\P^{ }_b$, 
and having $\P^{ }_d$ as an input parameter, as 
\ba 
 B(\P^{ }_d;a,b)\, \Phi(\P^{ }_a,\P^{ }_b)
 & \equiv & 
 \int_{\vc{p}^{ }_a} 
 \frac{\frac{1}{2} + n^{ }_{\sigma_a}(\epsilon^{ }_a - \mu^{ }_a)}
      {2\epsilon^{ }_a}
 \,
 \frac{\Phi(\P^{ }_a,\P^{ }_d-\P^{ }_a)
 }{(\P^{ }_d - \P^{ }_a)^2 - m_b^2}
%%%%
 \nn 
 & + & 
 \int_{\vc{p}^{ }_a} 
 \frac{\frac{1}{2} + n^{ }_{\sigma_a}(\epsilon^{ }_a + \mu^{ }_a)}
      {2\epsilon^{ }_a}
 \,
 \frac{\Phi(-\P^{ }_a,\P^{ }_d+\P^{ }_a)
 }{(\P^{ }_d + \P^{ }_a)^2 - m_b^2}
%%%%
 \nn 
 & + & 
 \int_{\vc{p}^{ }_b} 
 \frac{\frac{1}{2} + n^{ }_{\sigma_b}(\epsilon^{ }_b - \mu^{ }_b)}
      {2\epsilon^{ }_b}
 \,
 \frac{\Phi(\P^{ }_d-\P^{ }_b,\P^{ }_b)
 }{(\P^{ }_d - \P^{ }_b)^2 - m_a^2}
%%%%
 \nn 
 & + & 
 \int_{\vc{p}^{ }_b} 
 \frac{\frac{1}{2} + n^{ }_{\sigma_b}(\epsilon^{ }_b + \mu^{ }_b)}
      {2\epsilon^{ }_b}
 \,
 \frac{\Phi(\P^{ }_d+\P^{ }_b,-\P^{ }_b)
 }{(\P^{ }_d + \P^{ }_b)^2 - m_a^2}
 \;, \la{bubble}
\ea
where 
$
 \int_{\vc{p}}
 \equiv \int \! \frac{ {\rm d}^{3-2\epsilon}\vc{p} }{ (2\pi)^{3-2\epsilon} }
$.
Let us stress again that, 
taken literally, \eq\nr{bubble} is not well-defined on its 
own, due to the poles that the integrands contain, 
however it is well-defined
in connection with \eq\nr{master}, because then the poles cancel. 

%%%%%%%%%%%%%%%%%%%%% PARAGRAPH %%%%%%%%%%%%%%%%%%%%%%%%%%%%%%%%%%%%%%%%%%%
%
%\paragraph{Multiple poles in a single variable.}
\subsubsection*{Multiple poles in a single variable}

If some process can be mediated by different particle types
(e.g.\ Higgs and $Z^0$ bosons), then it is possible that 
interference terms contain
two different poles in a single kinematic variable. In this
case, we can partial fraction the dependence on that variable, 
\be
 \frac{1}{(s^{ }_{ab}-m^2_{d}) (s^{ }_{ab}-m^2_{e})}
 = 
 \frac{1}{m^2_{d} - m^2_{e}}
 \biggl( 
 \frac{1}{s^{ }_{ab}-m^2_{d}}
 - 
 \frac{1}{s^{ }_{ab}-m^2_{e}}
 \biggr)
 \;, 
\ee
and then make use of \eq\nr{virtual_bubble}.

A related situation is met if there is a second order pole 
in a single variable. 
The corresponding residue is denoted by 
\be
 \widehat\Theta^{(2)}_{ab}(\P^{ }_a,\P^{ }_b,\P^{ }_c)
 \; \equiv \; 
 \lim_{s^{ }_{ab}\to m^2_{d}} 
 (s^{ }_{ab} -  m^2_{d})^2\,\Theta(\P^{ }_a,\P^{ }_b,\P^{ }_c)
 \;. \la{residue2}
\ee
The simplest procedure in this case is to make use of the result
for a single pole, just taking a mass derivative thereof: 
\ba
 \Delta \Gamma^\rmi{Born}_{1\leftrightarrow 2}
 & \supset & 
%% (m^2_{d})^n_{ }
 \frac{{\rm d}}{{\rm d}m^2_{d}}
 \biggl\{ 
 \scat{1\leftrightarrow2}(d,c)
 \, B(\P^{ }_{d};a,b)
 \; \widehat\Theta^{(2)}_{ab}(\P^{ }_a,\P^{ }_b,\K-\P^{ }_d)
%% \frac{1}{
%%            (m^2_{d})^n_{ }
%%         }
 \biggr\} 
 \;. \hspace*{5mm} \la{virtual_bubble_derivative}
\ea
Here we have assumed a notation like that introduced in
\eq\nr{example_1to3_alt}, whereby $m_d^2$ appears as 
a pole location. 
We note that the mass derivative acts both on 
$
 \scat{1\leftrightarrow2}(d,c)
$, 
and on the coefficient function 
$B(\P^{ }_{d};a,b)\,\widehat\Theta^{(2)}_{ab}$, 
which is a function of $m^{2}_{d}$ 
via its dependence on $\P^{ }_{d}$.

%%%%%%%%%%%%%%%%%%%%% PARAGRAPH %%%%%%%%%%%%%%%%%%%%%%%%%%%%%%%%%%%%%%%%%%%
%
%\paragraph{Single poles in separate variables.}
\subsubsection*{Single poles in separate variables}

The most complicated case is when there are poles 
in two separate kinematic variables, 
as can happen in interference terms. Say, if the variables
$s^{ }_{ab}$ and $s^{ }_{bc}$ can go on-shell, 
then the corresponding residue is defined as 
\be
 \widetilde\Theta^{ }_{ab;bc}(\P^{ }_a,\P^{ }_b,\P^{ }_c)
 \; \equiv \; 
 \lim_{s^{ }_{ab}\to m^2_{d}} 
 \lim_{s^{ }_{bc}\to m^2_{e}} 
 (s^{ }_{ab} -  m^2_{d})\,
 (s^{ }_{bc} -  m^2_{e})\,
 \Theta(\P^{ }_a,\P^{ }_b,\P^{ }_c)
 \;, \la{residue3}
\ee
where $\Theta$ is the function in \eq\nr{master}. 

It is important to realize that if a residue like 
in \eq\nr{residue3} exists in a $1\to 3$ matrix element, then
there is necessarily also another decay channel, namely one
in which the particles $d$, $b$, $e$ are in the final state, 
whereas the propagators of $a$ and $c$ appear in the matrix
element squared. Concretely, noting that 
the direction of the line $b$ gets inverted in the second ``cut'', 
poles of this type must appear in the combination
\ba
 \Gamma^\rmi{Born}_{1\to 3} & \supset & 
 \scat{1\to3}(a,b,c) \, 
 \frac{\widetilde{\Theta}^{ }_{ab;bc}(\P^{ }_a,\P^{ }_b,\P^{ }_c)}
 {[\,(\P^{ }_a + \P^{ }_b)^2 - m_d^2\,]\,
  [\,(\P^{ }_c + \P^{ }_b)^2 - m_e^2\,]}
%%%%%%%%%%%%%
 \nn 
 & + & 
 \scat{1\to3}(d,-b,e) \, 
 \frac{\widetilde{\Theta}^{ }_{ab;bc}
  (\P^{ }_d+\P^{ }_b,-\P^{ }_b,\P^{ }_e+\P^{ }_b)}
 {[\,(\P^{ }_d + \P^{ }_b)^2 - m_a^2\,]\,
  [\,(\P^{ }_e + \P^{ }_b)^2 - m_c^2\,]}
 \;. \la{structure_master}
\ea
Once again, a convenient way to prove that interference terms necessarily 
appear in this form goes through the consideration of master sum-integrals, 
as shown in appendix~A.1. We note that  
if the particle~$b$ is neutral
(i.e.\ carries no chemical potential), 
and $a=d$, $c=e$, 
% and $ \widetilde{\Theta}^{ }_{ab;bc}$ respects the symmetry
% $ 
%    \widetilde{\Theta}^{ }_{ab;bc}
%       (\P^{ }_a+\P^{ }_b,-\P^{ }_b,\P^{ }_c+\P^{ }_b) 
%  = \widetilde{\Theta}^{ }_{ab;bc}
%       (\P^{ }_a,\P^{ }_b,\P^{ }_c)
% $, 
then \eq\nr{structure_master} can be reduced to a single structure. 

The virtual corrections cancelling the 
poles of \eq\nr{structure_master}
can now be expressed as 
\ba
 \Delta \Gamma^\rmi{Born}_{1\leftrightarrow 2}
 & \supset & 
   \scat{1\leftrightarrow2}(d,c)
 \, C(\P^{ }_d\,,\,\P^{ }_c\, ; \,a,b,e)
 \; \widetilde\Theta^{ }_{ab;bc}(\P^{ }_a,\P^{ }_b,\P^{ }_c)
 \nn 
%%%%%%
 & + & 
    \scat{1\leftrightarrow2}(a,e)
 \, C(\P^{ }_a\,,\,\P^{ }_e \,;\, d,-b,c)
 \; \widetilde\Theta^{ }_{ab;bc}
 (\P^{ }_d + \P^{ }_b,-\P^{ }_b,\P^{ }_e + \P^{ }_b)
%%%%%%
 \;, \hspace*{5mm} \la{virtual_triangle}
\ea
where we have defined 
a ``cubic'' or ``triangle'' operator, $C$, which is similar to the 
bubble operator in \eq\nr{bubble}, but acts now  
on three momenta, 
with two incoming momenta (those related to $\scat{1\leftrightarrow2}$)
appearing as parameters. The incoming momenta are
constrained by their sum equalling $\K$
(cf.\ \eq\nr{outer_int}), which permits
to simplify some of the structures appearing:  
\ba 
 && \hspace*{-1.5cm} 
 C(\P^{ }_d,\P^{ }_e ; a,b,c)\, \Phi(\P^{ }_a,\P^{ }_b,\P^{ }_c)
%%%%
 \nn[2mm]
 & \equiv & 
 \int_{\vc{p}^{ }_a} 
 \frac{\frac{1}{2} + n^{ }_{\sigma_a}(\epsilon^{ }_a - \mu^{ }_a)}
      {2\epsilon^{ }_a}
 \,
 \frac{\Phi(\P^{ }_a\,,\,\P^{ }_d-\P^{ }_a\,,\,\K-\P^{ }_a)
 }{ [\, (\P^{ }_d - \P^{ }_a)^2 - m_b^2 \,]\, 
    [\, (\K - \P^{ }_a)^2 - m_c^2 \,]}
%%%%
 \nn 
 & + & 
 \int_{\vc{p}^{ }_a} 
 \frac{\frac{1}{2} + n^{ }_{\sigma_a}(\epsilon^{ }_a + \mu^{ }_a)}
      {2\epsilon^{ }_a}
 \,
 \frac{\Phi(-\P^{ }_a\,,\,\P^{ }_d+\P^{ }_a\,,\,\K+\P^{ }_a)
 }{ [\, (\P^{ }_d + \P^{ }_a)^2 - m_b^2 \,]\, 
    [\, (\K + \P^{ }_a)^2 - m_c^2 \,]}
%%%%
 \nn 
 & + & 
 \int_{\vc{p}^{ }_b} 
 \frac{\frac{1}{2} + n^{ }_{\sigma_b}(\epsilon^{ }_b - \mu^{ }_b)}
      {2\epsilon^{ }_b}
 \,
 \frac{\Phi(\P^{ }_d-\P^{ }_b\,,\,\P^{ }_b\,,\,\P^{ }_e+\P^{ }_b)
 }{ [\, (\P^{ }_d - \P^{ }_b)^2 - m_a^2 \,]\, 
    [\, (\P^{ }_e + \P^{ }_b)^2 - m_c^2 \,]}
%%%%
 \nn 
 & + & 
 \int_{\vc{p}^{ }_b} 
 \frac{\frac{1}{2} + n^{ }_{\sigma_b}(\epsilon^{ }_b + \mu^{ }_b)}
      {2\epsilon^{ }_b}
 \,
 \frac{\Phi(\P^{ }_d+\P^{ }_b\,,\,-\P^{ }_b\,,\,\P^{ }_e-\P^{ }_b)
 }{ [\, (\P^{ }_d + \P^{ }_b)^2 - m_a^2 \,]\, 
    [\, (\P^{ }_e - \P^{ }_b)^2 - m_c^2 \,]}
%%%%
 \nn 
 & + & 
 \int_{\vc{p}^{ }_c} 
 \frac{\frac{1}{2} + n^{ }_{\sigma_c}(\epsilon^{ }_c - \mu^{ }_c)}
      {2\epsilon^{ }_c}
 \,
 \frac{\Phi(\K - \P^{ }_c\,,\,\P^{ }_c-\P^{ }_e\,,\,\P^{ }_c)
 }{ [\, (\K - \P^{ }_c)^2 - m_a^2 \,]\, 
    [\, (\P^{ }_e - \P^{ }_c)^2 - m_b^2 \,] }
%%%%
 \nn 
 & + & 
 \int_{\vc{p}^{ }_c} 
 \frac{\frac{1}{2} + n^{ }_{\sigma_c}(\epsilon^{ }_c + \mu^{ }_c)}
      {2\epsilon^{ }_c}
 \,
 \frac{\Phi(\K + \P^{ }_c\,,\,-\P^{ }_c-\P^{ }_e\,,\,-\P^{ }_c)
 }{ [\, (\K + \P^{ }_c)^2 - m_a^2 \,]\, 
    [\, (\P^{ }_e + \P^{ }_c)^2 - m_b^2 \,]}
 \;. \la{triangle}
\ea
A rather tedious analysis shows that \eq\nr{virtual_triangle} indeed cancels  
all divergences of \eq\nr{structure_master}; alternatively, 
\eq\nr{virtual_triangle} can be obtained by considering the cuts of 
a corresponding master sum-integral, as discussed in appendix~A.2. 

It is important to stress that the two terms in \eq\nr{virtual_triangle}
{\em do not} correspond one-to-one 
to the two terms in \eq\nr{structure_master}. 
Rather, to cancel the poles of the first term in \eq\nr{structure_master}, 
parts of both terms of \eq\nr{virtual_triangle} are needed. 
Nevertheless, since both
terms of \eq\nr{structure_master} are guaranteed to appear in a correct
computation of $\Gamma^\rmi{Born}_{1\to 3}$, we can effectively adopt
an algorithm which derives the first term of \eq\nr{virtual_triangle}
from the first term of \eq\nr{structure_master}, and similarly
for the second terms. This does produce the correct
$\Delta \Gamma^\rmi{Born}_{1\leftrightarrow2}$ for
\eq\nr{virtual_triangle}. 

%%%%%%%%%%%%%%%%%%%%% PARAGRAPH %%%%%%%%%%%%%%%%%%%%%%%%%%%%%%%%%%%%%%%%%%%
%
%\paragraph{Example}
\subsubsection*{Example}

For the case in \eq\nr{example_1to3_alt}, there are poles corresponding
to the variables $s^{ }_{\ala\aQ}$ and $s^{ }_{\aQ\aS}$, 
located at $m_\bla^2$ and $m_{\tilde{\cS}}^2$, respectively. 
We also note that the interference term respects the symmetry 
described below \eq\nr{structure_master}, 
so that only one of the structures 
in \eq\nr{virtual_triangle} needs to be included. 
This then leads to 
\ba
 && \hspace*{-0.8cm} 
 \frac{ \Delta \Gamma^\rmi{Born}_{1 \leftrightarrow 2} }
      { 2(g_1^2 + 3 g_2^2) }
 \; \supset \;  % \biggl\{ 
 \nn
%%%%%%
 & - &    
          \scat{1\leftrightarrow2}(\bla,\aS)  
          \,\Bigl[ 
            \, B(\P^{ }_{\bla} \,;\, \ala,\aQ)
            \, \E\cdot\P^{ }_{\ala}
      +  (m_\cS^2 - M^2) 
               \, C(\P^{ }_{\bla},\P^{ }_{\aS} \,;\, \ala,\aQ,\bS) 
               \, \E \cdot ( \P^{ }_{\aQ} + 2 \P^{ }_{\ala} )
          \, \Bigr]
 \nn[1mm]
%%%%%%
 & - &    
          \scat{1\leftrightarrow2}(\ala,\bS)  
          \, B(\P^{ }_{\bS} \,;\, \aQ,\aS)
          \, \E\cdot ( \P^{ }_{\ala} + \K ) 
 \nn
%%%%%%
 & - & 
          2 m_\cS^2  \frac{{\rm d}}{{\rm d}m_{\tilde{\cS}}^2 }
          \Bigl[ \, 
            \scat{1\leftrightarrow2}(\ala,\bS)  
          \, B(\P^{ }_{\bS} \,;\, \aQ,\aS)
          \, \E\cdot\P^{ }_{\ala} 
          \, \Bigr]
%%%%
 % \, \biggr\} 
 \;. 
 \la{ex_virtual}
\ea

%%%%%%%%%%%%%%%%%%%%%%%%%%%%%% SECTION %%%%%%%%%%%%%%%%%%%%%%%%%%%%%%%%%%%
%
\section{IR regularization} 
\la{se:IR}

Having found a representation for 
$2\leftrightarrow 2$ and $1\leftrightarrow 3$ scatterings
(cf.\ \eq\nr{master}) as well as the virtual corrections that
render these expressions finite (cf.\ \eq\nr{ex_virtual}), 
the next task is to carry out the integrals in practice. 
However, the two sets are ill-defined separately. 
Therefore we introduce {\em generic masses} as intermediate regulators 
that render both sets finite, and check in the end 
that the results are stable if the 
masses are taken to their physical values.  

For the following, it is helpful to assume that only 
first order poles are present. Second order poles, like the one
appearing in \eq\nr{example_1to3_alt}, are to be viewed as
parametric derivatives, 
e.g.\ 
$
 1 / ({ s^{ }_{\aQ\aS} - m_{\tilde{\cS}}^2 })^2 
 = 
 \partial^{ }_{m_{\tilde{\cS}}^2}
 1 / ({ s^{ }_{\aQ\aS} - m_{\tilde{\cS}}^2 })
$, 
as was already done in \eq\nr{ex_virtual}.

If we partial fraction the energy dependence of 
the $2\leftrightarrow 2$ and $1\leftrightarrow 3$ matrix elements 
squared, it appears in products of fractions like
\be
 \frac{1}{\pm \epsilon^{ }_1 \pm \epsilon^{ }_2 \pm \epsilon^{ }_3}
 \;. \la{frac1}
\ee
Let now $\epsilon^{ }_3$ represent an energy variable that becomes
singular in the sense of \eq\nr{residue1}. 
In the $2\leftrightarrow 2$ and $1\leftrightarrow 3$ scatterings, 
these fractions always appear pairwise, such that when put together 
we end up having combinations like 
\be
 \frac{2 \epsilon^{ }_3}{ 
 (\epsilon^{ }_1 \pm \epsilon^{ }_2)^2 - \epsilon_3^2
 }
 \;, \la{frac4}
\ee
where the numerator cancels against a similar term from the 
integration measure. This reflects the fact that $\epsilon^{ }_{1,2}$
are energies of external state particles, whereas $\epsilon_3^{ }$
is that of a virtual state. 
In the virtual corrections, 
the same would-be divergence appears as 
\be
 \frac{2\epsilon^{ }_1}{   
 (\epsilon^{ }_2 \pm \epsilon^{ }_3)^2 - \epsilon^2_1 
 }
 \;, \quad
  \frac{2\epsilon^{ }_2}{   
 (\epsilon^{ }_1 \pm \epsilon^{ }_3)^2 - \epsilon^2_2 
 }
 \;. \la{frac3} 
\ee 
Now $\epsilon^{ }_1$ or $\epsilon^{ }_2$ 
does not appear as an integration variable, 
but rather represents the energy of a virtual particle, 
$\epsilon_1^2 = (\vc{p}^{ }_2 \pm \vc{p}^{ }_3)^2 + m_1^2$ or 
$\epsilon_2^2 = (\vc{p}^{ }_1 \pm \vc{p}^{ }_3)^2 + m_2^2$.

We define integrals over  
\eqs\nr{frac4} and \nr{frac3} as {\em principal values}. 
Proceeding first with the virtual corrections, the principal value
concerns the integration over the directions of a momentum 
like $\vc{p}^{ }_2$ or $\vc{p}^{ }_1$. 
As shown in \se\ref{ss:phasespace_virtual}
and appendix~C, these angular averages can be carried out
analytically, and the results are non-singular 
as long as masses are finite, apart from integrable (logarithmic)
singularities that influence final energy integrations.

Returning then to $2\leftrightarrow 2$ and $1\leftrightarrow 3$
scatterings, the situation is more complicated. In many cases we can take 
$
 s^{\pm}_{12} = 
 (\epsilon^{ }_1 \pm \epsilon^{ }_2)^2 - (\vc{p}^{ }_1 \pm \vc{p}^{ }_2)^2
$
as an integration variable in \eq\nr{frac4}, and then the singularity is 
located at the position where $s^{\pm}_{12} = m_3^2$. However, 
in interference terms, two separate poles appear. Even though one of
the singularities can still be easily localized, the other one is
more challenging. The principal value integration {\it per se} can be 
dealt with in connection with azimuthal averaging (cf.\ \eq\nr{calF}), 
however a remnant divergence can manifest itself in energy variables. 

We note that, if the singularities of the energy
variables cannot be localized analytically, 
it is possible to take care of them numerically. This goes 
by implementing principal value integration as a limit, {\it viz.}
\be
 \frac{1}{
           (\epsilon^{ }_1 \pm \epsilon^{ }_2)^2  - \epsilon_3^2
         }
 \longrightarrow
 \mathbbm{P}
  \frac{1}{
          (\epsilon^{ }_1 \pm \epsilon^{ }_2)^2  - \epsilon_3^2
         }
 \; \equiv \; 
 \lim_{\delta\to 0}
 \re
 \biggl[
    \frac{1}{
          (\epsilon^{ }_1 \pm \epsilon^{ }_2)^2  - \epsilon_3^2 
            + 2 i \delta}
 \biggr] 
 \;. \la{frac2}
\ee
In general $\delta$ has to be set to a small value for reliable results, 
e.g.\ $\delta \sim (10^{-4}T)^2$. 

Finally, it is appropriate to stress that actual poles do not always
appear. To this end, consider a general structure appearing in 
\eq\nr{structure_master}, 
\be
 \Gamma^\rmi{Born}_{1\to 3}  \; \supset \;   
 \scat{1\to3}(a,b,c) \, 
 \frac{\widetilde{\Theta}^{ }_{ab;bc}(\P^{ }_a,\P^{ }_b,\P^{ }_c)}
 {[\,(\P^{ }_a + \P^{ }_b)^2 - m_d^2\,]\,
  [\,(\P^{ }_c + \P^{ }_b)^2 - m_e^2\,]}
 \;. 
\ee
The allowed ranges of the kinematic invariants for the $1\to3$ process
(``Dalitz plots'') and
its crossings can be established as usual
(cf.\ \se\ref{ss:phasespace_real} and  
appendices~\ref{app:1to3}--\ref{app:3to1_s}).
In particular,  
$1\to 3$ channels are realized if $M > m^{ }_a + m^{ }_b + m^{ }_c$, 
and poles exist in the propagators if
\be
 m^{ }_d \in \bigl( m^{ }_a + m^{ }_b, M - m^{ }_c \bigr)
 \;, \quad
 m^{ }_e \in \bigl( m^{ }_b + m^{ }_c, M - m^{ }_a \bigr)
 \;, \la{ris1}
\ee
respectively. The $2\leftrightarrow 2$ channels are always open, 
and poles exist if 
\ba
 m^{ }_d < \mbox{min}\bigl(|m^{ }_a - m^{ }_b|,|M - m^{ }_c| \bigr)
 & \mbox{or} & 
 m^{ }_d > \mbox{max}\bigl( m^{ }_a + m^{ }_b, M + m^{ }_c \bigr)
 \;, \\ 
 m^{ }_e < \mbox{min}\bigl(|m^{ }_b - m^{ }_c|,|M - m^{ }_a| \bigr)
 & \mbox{or} & 
 m^{ }_e > \mbox{max}\bigl( m^{ }_b + m^{ }_c, M + m^{ }_a \bigr)
 \;. \la{ris2}
\ea
The $3\to 1$ channels can be realized in three different ways, 
and poles are met if 
\ba
 m^{ }_a > m^{ }_b + m^{ }_c + M 
 & \Rightarrow & 
 m^{ }_d \in \bigl( M + m^{ }_c, m^{ }_a - m^{ }_b \bigr)
 \;, \quad
 m^{ }_e \in \bigl( m^{ }_b + m^{ }_c, m^{ }_a - M \bigr)
 \;, \hspace*{7mm} \\ 
%%%%%
 m^{ }_b > m^{ }_a + m^{ }_c + M 
 & \Rightarrow & 
 m^{ }_d \in \bigl( M + m^{ }_c, m^{ }_b - m^{ }_a \bigr)
 \;, \quad
 m^{ }_e \in \bigl( M + m^{ }_a, m^{ }_b - m^{ }_c \bigr)
 \;, \hspace*{7mm} \\ 
%%%%%
 m^{ }_c > m^{ }_a + m^{ }_b + M 
 & \Rightarrow & 
 m^{ }_d \in \bigl( m^{ }_a + m^{ }_b , m^{ }_c - M  \bigr)
 \;, \quad
 m^{ }_e \in \bigl( M + m^{ }_a , m^{ }_c - m^{ }_b \bigr)
 \;. \hspace*{7mm} \la{ris3}
%%%%%
\ea
For the example in \eq\nr{example_1to3_alt}, 
with 
$a\to \ala$, 
$b\to \aQ$, 
$c\to \aS$, 
$d\to \bla$, 
$e\to \bS$, 
and setting subsequently $m_{\tilde{\cS}}\to m_\cS$ and 
$m^{ }_\bla \to m^{ }_\ala$, and assuming that 
$m_\cS$ and $M$ are macroscopic 
whereas $m^{ }_\ala$ and $\mQ^{ }$ are small, 
poles can only exist in the $2\leftrightarrow 2$ channels, and then only if 
\be
 \mQ^{ } > 2 m^{ }_\ala 
 \quad (\,\mbox{there are poles}\,)
 \;. \la{ir_cond}
\ee
In the physical world, both of these masses vanish in an unresummed 
computation. The results depend continuously on them, 
so that the massless limit is well-defined. 
If we take $\mQ\to 0$ first, \eq\nr{ir_cond} indicates that 
there are {\em no poles} in any of the real scatterings. 

Even if unproblematic in principle, 
the case in which poles do appear is discussed from a different
perspective in \se\ref{sss:ris}. 

%%%%%%%%%%%%%%%%%%%%%%%%%%%%%% SECTION %%%%%%%%%%%%%%%%%%%%%%%%%%%%%%%%%%%
%
\section{Phase space integrals} 
\la{se:phasespace}

%%%%%%%%%%%%%%%%%%%%%%%%%%% SUBSECTION %%%%%%%%%%%%%%%%%%%%%%%%%%%%%%%%%%%
%
\subsection{Phase space integrals for $t$-channel 
$2\leftrightarrow 2$ reactions}
\la{ss:phasespace_real}

Among the reactions in \eq\nr{master}, 
we start by considering $2\leftrightarrow 2$ processes, 
which in many cases are physically the most important ones. 
A corresponding analysis of $1\to 3$ 
and $3\to 1$ processes can be found in appendix~\ref{app:further}.

For purposes of the practical integration, 
it is convenient to label the initial and
final-state four-momenta, energies, masses and chemical potentials 
with different symbols 
($\K^{ }_a$ vs.\ $\P^{ }_a$, 
$E^{ }_a$ vs.\ $\epsilon^{ }_a$, 
$M^{ }_a$ vs.\ $m^{ }_a$,
and $\nu^{ }_a$ vs.\ $\mu^{ }_a$, respectively). 
Here on-shell energies are denoted by
\be
 E^{ }_{a(\vc{k})} \; \equiv \; \sqrt{\vc{k}^2 + M_a^2}
 \;, \quad
 \epsilon^{ }_{a(\vc{p})} \; \equiv \; \sqrt{\vc{p}^2 + m_a^2}
% \;,  \quad
% \omega \; \equiv \; \sqrt{\vc{k}^2 + M^2}
 \;, \la{on-shell}
\ee
so that 
$
 \K^{ }_a \; = \; (E^{ }_{a(\vc{k}_a)},\vc{k}^{ }_a)  
$
and   
$
 \P^{ }_a \; = \; (\epsilon^{ }_{a(\vc{p}_a)},\vc{p}^{ }_a)
$.
% and 
% $
% \K \; \equiv \; (\omega,\vc{k}) 
% $.
The momentum label can often be left out from the 
energies without a danger of confusion.

With the given momenta and channels, 
Mandelstam invariants are defined as usual, 
\ba
 \scat{1\to3}: && \s{ij} = (\P^{ }_i  + \P^{ }_j)^2 \;, \la{sij} \\ 
 \scat{2\to2}: && s = (\P^{ }_1 + \P^{ }_2)^2 \;, \quad
                  t = (\P^{ }_1 - \K^{ }_1)^2 \;, \quad
                  u = (\P^{ }_2 - \K^{ }_1)^2 \;, \\ 
 \scat{3\to1}: && s = (\K^{ }_1 + \K^{ }_2)^2 \;, \quad
                  t = (\P^{ }_1 - \K^{ }_1)^2 \;, \quad
                  u = (\P^{ }_1 - \K^{ }_2)^2 \;. \la{mstam_3}
\ea

For $2\leftrightarrow 2$ processes, we introduce two parametrizations, 
the $t$- and $s$-channel ones~\cite{bb2}. 
In principle both of them can be used 
for any matrix element squared. 
However, if only one propagator appears in the integrand, the kinematic
variable should be chosen to correspond to that variable, because
then we can easily identify the pole location (if one appears), 
and implement the corresponding principal value integration 
(cf.\ \se\ref{se:IR}). In the interference terms, 
where two propagators appear, 
it is convenient to choose 
the parametrization according to which is the ``most singular''
kinematic invariant,  in order to understand the behaviour  
in the massless limit (cf.\ \se\ref{sss:regge}). 
The most singular variable is identified by putting auxiliary masses
such as those introduced in \eq\nr{example_1to3_alt} to zero. 
Furthermore, if $u$ appears as a singular variable, we can make use of 
the substitution $\P^{ }_1\leftrightarrow\P^{ }_2$,
in order to rename it into $t$. Thus, the most singular variable
can always be chosen as $t$ or $s$.

With this prescription, and numbering the momenta 
and chemical potentials corresponding
to $\scat{2\to2}(-a^{ }_1;b^{ }_1,b^{ }_2)$ as 
$
 \K^{ }_1 \equiv -\P^{ }_{a_1^{ }}
$ ,
$
 \P^{ }_i \equiv \P^{ }_{b_i^{ }}
$, 
$
 \nu^{ }_1 \equiv - \mu^{ }_{a_1^{ }}
$, 
$
 \mu^{ }_i \equiv \mu^{ }_{b_i^{ }}
$,  
so that the sign flips associated with
initial-stage particles have already been 
included when showing $\K^{ }_1$ and $\nu^{ }_1$, 
the crossings of \eq\nr{master} yield the following 
$t$-channel rate for the example of \nr{example_1to3_alt}
(the case of $s$-channel scatterings is discussed in
appendix~\ref{app:2to2_s}):
\ba
 && \hspace*{-1cm}
 \frac{ \Gamma^\rmi{Born}_{2\leftrightarrow 2(t)} }
 { 2 (g_1^2 + 3 g_2^2) }
 \; \to \; 
 \scat{2\to2}(-\aQ;\ala,\aS) \, 
%  2 (g_1^2 + 3 g_2^2)
%  \nn[2mm] & \times & 
                \biggl\{ 
% \nn & & +
                      - \frac{\E\cdot\P^{ }_1}{t - m_\bla^2}
                      + \frac{
                        ( m_\cS^2 - \MM )\,
                        \E\cdot( \K^{ }_1 - 2 \P^{ }_1 )
                       }{ 
                        ({ t - m_\bla^2 })\, ({ u - m_{\tilde{\cS}}^2 }) }
                       \biggr\} 
%%%%%%%%%%%%%%%%
 \nn[2mm] 
 & + & 
 \scat{2\to2}(-\ala;\aQ,\aS) \, 
%  2 (g_1^2 + 3 g_2^2)
%  \nn[2mm] & \times & 
                \biggl\{ 
% \nn & & +
                      \frac{ \E\cdot\K^{ }_1 }{t - m_\bla^2} 
                      + 
                      \frac{
                        ( \MM - m_\cS^2 )\,
                        \E\cdot( \P^{ }_1 - 2 \K^{ }_1 )
                       }{ 
                        ({ t - m_\bla^2 }) \, ({ s - m_{\tilde{\cS}}^2 }) }
                       \biggr\} 
%%%%%%%%%%%%%%%%
  \la{example_2to2_t} \\[2mm] 
 & + & 
   \bigl[\, \scat{2\to2}(-\aQ;\aS,\ala) + \scat{2\to2}(-\aS;\aQ,\ala)
 \,\bigr]\, 
%  2 (g_1^2 + 3 g_2^2)
%  \nn[2mm] & \times & 
                \biggl\{ 
% \nn & & +
                 -      \frac{ 
                         \E\cdot ( \P^{ }_2 + \K ) 
                             }{ { t - m_{\tilde{\cS}}^2 } }
                 -      \frac{ 
                         2 m_\cS^2 \, \E\cdot\P^{ }_2   
                             }{ ({ t - m_{\tilde{\cS}}^2 })^2 }
                       \biggr\} 
 \;.\nonumber
\ea
Here $\E$ can be $\K$, or the medium
four-velocity, $\U\equiv(1,\vc{0})$,
or a linear combination thereof. 

The key idea now is to have $t$ as an integration variable, 
so that most of the poles are easily resolved. 
To achieve this we introduce a four-momentum $\mathcal{Q}$ such 
that $t = \mathcal{Q}^2$, and rephrase 
the integration measure from \eq\nr{dO2to2} as 
\ba
 & & \hspace*{-1.2cm}
 \int \! {\rm d}\Omega^{t}_{2\leftrightarrow 2}
 \; \equiv \; 
 \int \! \frac{{\rm d}^3\vc{p}_1^{ }
             \,{\rm d}^3\vc{p}_2^{ } 
             \,{\rm d}^3\vc{k}_1^{ }
             \, {\rm d}^4\mathcal{Q} }
         {8 (2\pi)^9 \epsilon^{ }_{1(\vc{p}_1)}
          \,\epsilon^{ }_{2(\vc{p}_2)}
          \,E^{ }_{1(\vc{k}_1)}}
  \, (2\pi)^4\, 
 \delta^{(4)}_{ }
    \bigl(\P^{ }_1 - \K^{ }_1 - \mathcal{Q}\bigr)\,
 \delta^{(4)}_{ }\bigl(\mathcal{Q} + \P^{ }_2 - \K \bigr)
 \nn 
 & = & 
 \frac{1}{8(2\pi)^5}
 \int\! \frac{{\rm d}^3\vc{p}_1^{ }
              \, {\rm d} q^{ }_0
              \, {\rm d}^3\vc{q} }
             {\epsilon^{ }_{1(\vc{p}_1)} 
              \, \epsilon^{ }_{2(\vc{q-k})}
              \, E^{ }_{1(\vc{q-p}_1)}
             }
 \, 
 \delta\bigl(\epsilon^{ }_{1(\vc{p}_1)} 
              - E^{ }_{1(\vc{q-p}_1)} - q^{ }_0\bigr) 
 \delta\bigl( q^{ }_0 + \epsilon^{ }_{2(\vc{q-k})} - \omega \bigr)
 \;, \hspace*{5mm} \la{dPhi22_t}  
\ea
where we have integrated over $\vc{p}^{ }_2$ and $\vc{k}^{ }_1$.
The Dirac-$\delta$'s fix two angles as 
\be
 \vc{q}\cdot \vc{p}_1^{ }  = 
 q^{ }_{0\,} \epsilon^{ }_{1({p}^{ }_1)} + 
 \frac{M_1^2 - m_1^2 - t}{2} 
 \;, \quad
 \vc{q}\cdot\vc{k} = 
 q^{ }_{0\,} \omega + 
 \frac{m_2^2 - M^2 - t}{2}
 \;, \la{angles_t}
\ee
where $p^{ }_i \; \equiv \; |\vc{p}_i^{ }|$.
The other Mandelstam variables can be expressed as 
\be
 u %% = (\P^{ }_1 - \K^{ })^2 
   = m_1^2 + M^2 + 2\,(\vc{k}\cdot\vc{p}^{ }_1 
                   - \omega \epsilon^{ }_{1({p}^{ }_1)})
 \;, \quad
 %% s = m_1^2 + m_2^2 + M_1^2 + M^2 - t - u
 s = m_2^2 + M_1^2 - t - 2\,(\vc{k}\cdot\vc{p}^{ }_1 
                   - \omega \epsilon^{ }_{1({p}^{ }_1)})
 \;. \la{mandelstam_2to2_t}
\ee
If products like $1/(u^m s^n)$ appear, with $m,n > 0$, 
the dependence on $\vc{k}\cdot\vc{p}^{ }_1$ should be 
partial fractioned to appear as powers of inverse 
first-order polynomials. 

The azimuthal angle between $\vc{k}$ and $\vc{p}^{ }_1$ is unconstrained
as of now.
We note that this angle does not 
appear inside the thermal distribution functions in this parametrization
(cf.\ \eq\nr{N_2to2_t}). Choosing $\vc{q}$ as the $z$-axis, we may write 
\ba
 \vc{k} \cdot \vc{p}^{ }_1 = 
 \underbrace{
 k p^{ }_1 
 \, 
 \cos\theta^{ }_{\vc{q},\vc{k}} 
 \cos\theta^{ }_{\vc{q},\vc{p}_1} }_{\equiv a}
 + 
 \underbrace{ 
 k p^{ }_1 
 \, 
 \sin\theta^{ }_{\vc{q},\vc{k}}
 \sin\theta^{ }_{\vc{q},\vc{p}_1} }_{\equiv b}
 \, \cos\varphi
 \;, 
\ea
where 
$ \cos\theta^{ }_{\vc{q},\vc{k}} $ 
and
$ \cos\theta^{ }_{\vc{q},\vc{p}_1} $
are fixed through \eq\nr{angles_t}.
Let us define a generating function, incorporating
the possibility of a numerical principal value 
regularization {\it \`a la} \eq\nr{frac2}, as 
\ba
 \mathcal{F}^{ }_{\vc{q};\vc{k},\vc{p}_1;z} & \equiv & 
 \lim_{\delta\to 0}
 \re\, 
 \biggl\langle 
    \frac{1}{z - \vc{k}\cdot\vc{p}^{ }_1  + i \delta  }
 \biggr\rangle^{ }_{\!\varphi}
 \la{calF}  
% \\[1mm] 
%%%%%%%%%%%
% & \equiv & 
 \; \equiv \; 
 \lim_{\delta\to 0}
 \re 
 \int_{-\pi}^{\pi} \! 
 \frac{ {\rm d}\varphi }{2\pi} 
 \, 
 \frac{1}{z - a - b \cos\varphi  + i \delta}
 \hspace*{5mm} \nn[1mm] 
%%%%%%%%%%%
 & = & 
% \; =  \;
 \lim_{\delta\to 0}
   \re
   \frac{\sign(\,z - a\,)}
        {\sqrt{ 
          (z - a + i \delta)^2 - b^2 
        }}
 \;. \hspace*{5mm} \la{azimuthal_t_res}
\ea
This average 
vanishes if $|z-a| < |b|$ (for $\delta\to 0$), 
i.e.\ when poles are actually crossed. Nevertheless 
the function is singular when approaching this domain
from the outside, 
i.e.\ $|z-a|\to |b|^+_{ }$, so that a regularization is still needed, 
either numerically through~$\delta$, or analytically, 
by resolving the singular domains of the outer integrations. 

Through a series in $1/z$, the azimuthal averages
of positive powers of $\vc{k}\cdot\vc{p}^{ }_1$ are readily extracted
from \eq\nr{azimuthal_t_res}. In this case we can put
$\sign(\,z-a\,)\to 1$, yielding
\be
 \bigl\langle \,\vc{k}\cdot\vc{p}^{ }_1 
 \bigr\rangle^{ }_{\!\varphi} = a \;, \quad
 \bigl\langle (\vc{k}\cdot\vc{p}^{ }_1)^2
 \bigr\rangle^{ }_{\!\varphi} = 
   a^2 + \frac{b^2}{2} \;, \quad 
 \bigl\langle (\vc{k}\cdot\vc{p}^{ }_1)^3
 \bigr\rangle^{ }_{\!\varphi} = 
 a \biggl( a^2 + \frac{3 b^2}{2} \biggr)
 \;. 
 \la{azimuthal_t_res2}
\ee
Derivatives with respect to $z$ yield averages of negative powers, 
however they are not needed for
\eq\nr{example_2to2_t}, unless we use a $t$-channel
parametrization for an $s$-channel process
(cf.\ \eq\nr{example_2to2_s}). 

With all angles resolved, 
the remaining integration measure can be worked out. 
As a first step, we may consider separately the two ``vertices''
(or energy conservation constraints) appearing in 
\eq\nr{dPhi22_t}, finding 
\ba
 && \hspace*{-1.3cm}
 \int \! {\rm d}\Omega^{t}_{2\leftrightarrow 2}
  =  
 \frac{1}{(4\pi)^3k}
 \int_{-\infty}^{\omega - m^{ }_2}\!{\rm d}q^{ }_0 
 \int^{ k + \sqrt{(q^{ }_0 - \omega)^2 - m_2^2} }
     _{|k - \sqrt{(q^{ }_0 - \omega)^2 - m_2^2}|}
     {\rm d}q  
 \, \biggl\{ 
 \theta(-t)\int_{ \epsilon_1^- }^{\infty} 
 \! {\rm d}\epsilon^{ }_1
 \nn 
 & +  & 
   \theta(t)\,
   \theta\bigl((m^{ }_1 - M^{ }_1)q^{ }_0\bigr)\, 
   \theta\bigl((m^{ }_1 - M^{ }_1)^2-t\bigr)
 \int^{ \epsilon_1^+ } 
     _{ \epsilon_1^- }
 \! {\rm d} \epsilon^{ }_{1}
 \biggr\} 
 \;,
 \hspace*{4mm} \la{dPhi22_t_prefinal}
\ea
where 
$\epsilon_1^\pm$ are from \eq\nr{2to2t_e1pm}.
Subsequently it is advantageous to replace the variables
$q^{ }_0,q$ through $t = q_0^2 - q^2, q^{ }_0$, yielding a Jacobian
\be
 {\rm d}q^{ }_0\, {\rm d}q = 
 \frac{ {\rm d}t\,{\rm d}q^{ }_0 }{2q}
 \;. 
\ee 
The integration range of $t$ can be established as 
$(-\infty,(m^{ }_2 - M)^2)$, and for the part $t > 0$ we find that
the sign of $q^{ }_0$ depends on $M - m^{ }_2$, such that
$\theta(t)\theta((M- m^{ }_2) q^{ }_0 )$ may be inserted in the 
integrand. Altogether, \eq\nr{dPhi22_t_prefinal} can thus be
converted into
\ba
 && \hspace*{-1.5cm} 
 \int \! {\rm d}\Omega^{t}_{2\leftrightarrow 2}
 \; = \;  
 \frac{1}{(4\pi)^3k}
 \biggl\{ 
   \int_{-\infty}^0 \! {\rm d}t 
   \int_{q_0^-}^{q_0^+} \! {\rm d}q^{ }_0 
   \int_{\epsilon_1^-}^{\infty} \! {\rm d}\epsilon^{ }_1
 \nn 
 & + & 
    \theta((M - m^{ }_2)(m^{ }_1 - M^{ }_1))
   \int_0^{\rmi{min$((M - m^{ }_2)^2,(m^{ }_1 - M^{ }_1)^2)$}} 
    \!\!\!  {\rm d}t \, 
   \int_{q_0^-}^{q_0^+} \! {\rm d}q^{ }_0 
   \int_{\epsilon_1^-}^{\epsilon_1^+}
                                          \! {\rm d}\epsilon^{ }_1
 \biggr\} \frac{1}{2q} 
 \;, \la{dPhi22_t_final} \hspace*{5mm} 
\ea
where $q = \sqrt{q_0^2 - t}$ and 
\ba
 q_0^{\pm} & \equiv & 
 \frac{ \omega (t+M_{ }^2-m_2^2)
                         \pm k\kallen(t,M_{ }^2,m_2^2)}{2M^2}
 \;, \la{q0pm} \\
 \epsilon_1^{\pm} & \equiv & 
 \frac{ q^{ }_0(t+m_1^2-M_1^2)
                         \pm q\kallen(t,m_1^2,M_1^2)}{2t}
 \;. \la{2to2t_e1pm}
\ea
Here the K\"all\'en function is defined as 
\be
 \kallen^{ } ( x,m_1^2,m_2^2 )
  \; \equiv \; 
 \sqrt{     x^2+m_1^4+m_2^4
              - 2 x ( m_1^2  + m_2^2 )
              - 2 m_1^2 m_2^2 
      }
 \;. \la{kallen}
\ee

Finally, 
the thermal distributions associated with 
$2\leftrightarrow 2$ scatterings, cf.\ \eq\nr{calN2to2}, 
can be factorized as in \eq\nr{N_2to2_t_gen}, 
which after the insertion of  
$q^{ }_0 = \epsilon^{ }_1 - E^{ }_1 = \omega - \epsilon^{ }_2$
from \eq\nr{dPhi22_t}
and renaming chemical potentials as mentioned above yields 
\ba
 \mathcal{N}^{ }_{\tau_1;\sigma_1\sigma_2}
 \!\! & = & \!\!  
 \bigl[
            n^{ }_{\tau_1\sigma_1}(q^{ }_0 - \mu^{ }_1 + \nu^{ }_1 )
          - n^{ }_{\sigma_2}(q^{ }_0 - \omega + \mu^{ }_2 )
 \bigr]
 \bigl[
            n^{ }_{\tau_1}(\epsilon^{ }_1 - q^{ }_0  - \nu^{ }_1 )
          - n^{ }_{\sigma_1}(\epsilon^{ }_1  - \mu^{ }_1 ) 
 \bigr]
 \;. \nn \la{N_2to2_t}
\ea
The first factor implies that the $q^{ }_0$-integral
is exponentially localized around modest $|q^{ }_0|$, even if the
boundaries from \eq\nr{q0pm} can obtain large values; 
likewise, 
the integration over~$\epsilon^{ }_1$ is exponentially convergent
at large~$\epsilon^{ }_1$.
% \footnote{%
%  This means that in practice 
%  the integration bounds $\epsilon_1^{\pm}$
%  can be replaced with
%  $\mbox{max}(\Lambda,q^{ }_0 + \Lambda)$ 
%  if they exceed this value, with $\Lambda$ chosen
%  so that $\exp(-\Lambda/T)$ matches the desired accuracy level.} 

We note that if $\tau^{ }_1\sigma^{ }_1 = +$, the first factor
in \eq\nr{N_2to2_t}
has a pole at $ q^{ }_0 - \mu^{ }_1 + \nu^{ }_1 = 0$, but the second
factor vanishes at the same point, lifting it. In order to avoid 
numerical issues with this, it is helpful to replace \eq\nr{N_2to2_t}
by an alternative representation  
if $ | q^{ }_0 - \mu^{ }_1 + \nu^{ }_1 | \ll T $, {\it viz}.
$
 \mathcal{N}^{ }_{\tau_1;\sigma_1\sigma_2} 
 = 
 \{\,
  1 +  
  n^{-1}_{\tau_1\sigma_1}(q^{ }_0 - \mu^{ }_1 + \nu^{ }_1 )
  \, [
        1 + n^{ }_{\sigma_2}(\omega - q^{ }_0  - \mu^{ }_2 ) 
     ]
 \} 
 \, [ 
      1 +   n^{ }_{\tau_1}(\epsilon^{ }_1 - q^{ }_0  - \nu^{ }_1 )
 ] \, 
            n^{ }_{\sigma_1}(\epsilon^{ }_1  - \mu^{ }_1 ) 
$.

%%%%%%%%%%%%%%%%%%%%%%%%%%% SUBSECTION %%%%%%%%%%%%%%%%%%%%%%%%%%%%%%%%%%%
%
\subsection{Phase space integrals for virtual corrections} 
\la{ss:phasespace_virtual}

The virtual corrections, discussed 
in \se\ref{ss:virtual}, have the structure of a $1\leftrightarrow 2$
phase space average, which we call an ``outer integral'', 
convoluted with a bubble or triangle function, 
which we call an ``inner integral''
(cf.\ \eqs\nr{virtual_bubble}, 
\nr{virtual_bubble_derivative}, 
\nr{virtual_triangle}).
Let us discuss these in turn. 

%%%%%%%%%%%%%%%%%%%%% PARAGRAPH %%%%%%%%%%%%%%%%%%%%%%%%%%%%%%%%%%%%%%%%%%%
%
\subsubsection*{Outer integration}

An outer integral like 
$
    \scat{1\leftrightarrow2}(d,c) 
$
in \eq\nr{virtual_bubble}
contains three different channels, corresponding to $1\to 2$ decays
and two different $2\to 1$ inverse decays, cf.\ \eq\nr{scat1lr2}.
Remarkably, the three channels can be combined into a single expression,
once we make use of the identity 
$
  n^{ }_{\sigma}(-x) = -\, \bar{n}^{ }_{\sigma}(x)
$ 
in the $2\to 1$ channels and substitute 
$\epsilon^{ }_i \to -\epsilon^{ }_i$ in one of them. 
For instance, if we choose $\epsilon^{ }_d$ as the integration variable, 
then 
\ba
 && \hspace*{-1.0cm} 
 \scat{1\leftrightarrow 2}(c,d) \, \Phi(\P^{ }_c,\P^{ }_d)
% \nn 
% & = &
 \; = \; 
 \frac{\theta(\kallensq(M^2,m_c^2,m_d^2))}{16\pi k} 
 \nn 
%%%%%%%%%%%%%%%%%%%%%
 & \times & \hspace*{-2mm} 
 \int_{\epsilon_d^-}^{\epsilon_d^+}
 \! {\rm d}\epsilon^{ }_d \, 
 \bigl[
   1
     + n^{ }_{\sigma_c}(\omega - \epsilon^{ }_d - \mu^{ }_c)
     + n^{ }_{\sigma_d}(\epsilon^{ }_d - \mu^{ }_d) 
 \bigr]
 \,
 \sign\bigl( \epsilon^{ }_d(\omega - \epsilon^{ }_d) \bigr)
 \, 
 \Phi\bigl( 
  \K - \P^{ }_d, \P^{ }_d
%  (\omega - \epsilon^{ }_d,\vc{k} - \vc{p}^{ }_d),
%  (\epsilon^{ }_d,\vc{p}^{ }_d)
 \bigr)
 \nn 
%%%%%%%%%%%%%%%%%%%%
 & \times & \hspace*{-2mm}
 \biggl\{ 
   1 + \epsilon\, 
       \ln\biggl[
             \frac{k^2\bmu^2}{M^2(\epsilon_d^+ - \epsilon^{ }_d)
                                 (\epsilon^{ }_d - \epsilon_d^-)}
          \biggr]
     + \rmO(\epsilon^2) 
 \biggr\} 
 \;, \la{outer_int}   
\ea
where $\kallen$ is from \eq\nr{kallen}, 
we have expressed the spacetime dimension as $D = 4-2\epsilon$, 
and the integration bounds are 
analogous to \eq\nr{q0pm}, {\it viz.} 
\be
   \epsilon^{\pm}_d = \frac{\omega\, (M^2 + m_d^2 - m_c^2)
   \pm k \kallen(M^2,m_c^2,m_d^2)}{2M^2}
 \;. \la{1to2_bounds}
\ee
Furthermore the direction of $\vc{p}^{ }_d$ is fixed 
similarly to the second constraint in \eq\nr{angles_t}, 
\be
   \vc{k}\cdot\vc{p}^{ }_d = 
   \omega \, \epsilon^{ }_d + \frac{m_c^2 - m_d^2 - M^2}{2}
 \;. \la{kdotpd}
\ee
The sign function in \eq\nr{outer_int} is 
negative in the $2\to 1$ channels,
in one because
of the inverted sign of $\epsilon^{ }_d$ and in the other because
of the negative sign of $\omega - \epsilon^{ }_d$.\footnote{%
 The integrand is exponentially localized close to 
 one of the integration boundaries, 
 so it makes sense to have more numerical resolution there. 
 } 

For completeness we have shown 
the part of $\rmO(\epsilon)$ in \eq\nr{outer_int}, 
because the inner integrals
normally contain $1/\epsilon$ divergences; 
$\bmu$ denotes the scale parameter of the $\msbar$ scheme. 
In this context is also appropriate to note that in $D$ dimensions, 
some of the coefficients in \eq\nr{example_1to3} contain additional
parts proportional to $D-4$, which had been omitted. If we worry about
the renormalization of the coupling associated with 
the vertex in \fig\ref{fig:1to3}(left), these terms should be 
included, but normally such renormalization effects are genuinely small
in a weakly coupled system, unlike the IR effects that we are
mostly interested in. 

%%%%%%%%%%%%%%%%%%%%% PARAGRAPH %%%%%%%%%%%%%%%%%%%%%%%%%%%%%%%%%%%%%%%%%%%
%
\subsubsection*{Inner integrations}

For the inner integrals, we need to consider $B$ from \eq\nr{bubble}
and $C$ from \eq\nr{triangle}. For generality we start with $C$, which
contains one more angular variable. 
Let us recall that all propagators within the thermal averages 
are assumed regularized as principal values. 

We may now express the integration measure in spherical coordinates as usual.
Assuming that the inner integral is UV finite (we return to this below), we
write 
\be
 \int^{ }_{\vc{p}_a} 
 = 
 \frac{1}{(2\pi)^2} 
 \int_{m_a}^{\infty} \! {\rm d}\epsilon^{ }_a \, 
 \epsilon^{ }_a \, p^{ }_a \, 
 \int_{-1}^{+1} \! {\rm d}\cos\theta 
 \,\biggl\{ \frac{1}{2\pi} \int_{0}^{2\pi} \! {\rm d}\varphi \biggr\} 
 \;, 
\ee
and similarly for 
$
 \int^{ }_{\vc{p}_b}
$
and
$
 \int^{ }_{\vc{p}_c}
$
in \eq\nr{triangle}. 

For the integrand, we consider two possible structures, 
defined in \eqs\nr{def_G} and \nr{def_H}. 
There is freedom in choosing the axis with respect to which $\theta$ is 
measured. This way, the angular integrals can be carried out explicitly,
as detailed in appendix~C. 

The two middle terms of \eq\nr{triangle}, associated with $\epsilon^{ }_b$, 
require a more careful look, as the dependence on the angles needs
to be partial fractioned, in order to bring the results in a form in which
we can make use of \eq\nr{def_H}. For this we may write
\ba
 & & \frac{1 }{ [\, (\P^{ }_d - \P^{ }_b)^2 - m_a^2 \,]\, 
                [\, (\P^{ }_e + \P^{ }_b)^2 - m_c^2 \,]}
 \nn 
 & = &  
 \biggl[
   \frac{1}{  (\P^{ }_d - \P^{ }_b)^2 - m_a^2 } 
 - 
   \frac{1}{  (\P^{ }_e + \P^{ }_b)^2 - m_c^2 } 
 \biggr]
 \frac{1}{m_a^2 + m_e^2 - m_c^2 - m_d^2 + 2 \K\cdot\P^{ }_b}
 \;, 
\ea
where we made use of $\P^{ }_d + \P^{ }_e = \K$.

Collecting together the contributions,
and making use of $\mathcal{H}$ from \eq\nr{def_H}, we obtain
\ba 
 && \hspace*{-0.8cm} 
 C(\P^{ }_d,\P^{ }_e \,;\, a,b,c)\, \Phi(\P^{ }_a,\P^{ }_b,\P^{ }_c)
%%%%
 \nn[2mm]
 & \simeq &\! 
 \int_{m_a}^{\infty} 
 \! \frac{{\rm d}\epsilon^{ }_a \, p^{ }_a}{(4\pi)^2}  \, 
 \biggl\{ 
 \nn &  
 + &\!
   \biggl[ \fr12 + n^{ }_{\sigma_a}(\epsilon^{ }_a - \mu^{ }_a)\biggr]
   \mathcal{H}^{ }_{\vc{p}_a;\vc{p}_d,\vc{k};
        \epsilon^{ }_a\epsilon^{ }_d
        + \frac{ m_b^2-m_a^2-m_d^2 }{2} 
        \,,\,
        \epsilon^{ }_a\omega 
        + \frac{ m_c^2 - m_a^2 - M^2 }{2}
   }
   {\Phi(\P^{ }_a,\P^{ }_d-\P^{ }_a,\K - \P^{ }_a)}
 \nn &  
 + &\!
   \biggl[ \fr12 + n^{ }_{\sigma_a}(\epsilon^{ }_a + \mu^{ }_a)\biggr]
   \mathcal{H}^{ }_{\vc{p}_a;\vc{p}_d,\vc{k};
        \epsilon^{ }_a\epsilon^{ }_d 
        + \frac{ m_d^2+m_a^2-m_b^2 }{2} 
        \,,\,
        \epsilon^{ }_a\omega 
        + \frac{ M^2 + m_a^2 - m_c^2 }{2}        
   }
   {\Phi(-\P^{ }_a,\P^{ }_d+\P^{ }_a,\K + \P^{ }_a)}
 \biggr\} 
%%%%
 \nn 
 & + &\! 
 \int_{m_b}^{\infty} 
 \! \frac{{\rm d}\epsilon^{ }_b \, p^{ }_b}{(4\pi)^2}  \, 
 \biggl\{ 
 \nn &  
 - &\!
   \biggl[ \fr12 + n^{ }_{\sigma_b}(\epsilon^{ }_b - \mu^{ }_b)\biggr]
   \mathcal{H}^{ }_{\vc{p}_b;\vc{p}_d,\vc{k};
       \epsilon^{ }_b\epsilon^{ }_d
      + \frac{ m_a^2-m_b^2-m_d^2 }{2} 
      \,,\,
      \epsilon^{ }_b\omega 
      + \frac{ m_a^2+m_e^2-m_c^2-m_d^2 }{2}
   }
   {\Phi(\P^{ }_d-\P^{ }_b,\P^{ }_b,\P^{ }_e+\P^{ }_b)}
 \nn & 
 - &\!
   \biggl[ \fr12 + n^{ }_{\sigma_b}(\epsilon^{ }_b + \mu^{ }_b)\biggr]
   \mathcal{H}^{ }_{\vc{p}_b;\vc{p}_d,\vc{k};
       \epsilon^{ }_b\epsilon^{ }_d
      +\frac{ m_d^2 + m_b^2-m_a^2}{2} 
      \,,\,
      \epsilon^{ }_b\omega 
      + \frac{ m_c^2+m_d^2 - m_a^2 -m_e^2 }{2}
   }
   {\Phi(\P^{ }_d+\P^{ }_b,-\P^{ }_b,\P^{ }_e-\P^{ }_b)}
 \nn &  
 - &\! 
   \biggl[ \fr12 + n^{ }_{\sigma_b}(\epsilon^{ }_b - \mu^{ }_b)\biggr]
   \mathcal{H}^{ }_{\vc{p}_b;\vc{p}_e,\vc{k};
       \epsilon^{ }_b\epsilon^{ }_e
      +\frac{ m_e^2+m_b^2-m_c^2}{2} 
      \,,\,
      \epsilon^{ }_b\omega
      + \frac{ m_a^2+m_e^2-m_c^2-m_d^2 }{2}
   }
   {\Phi(\P^{ }_d-\P^{ }_b,\P^{ }_b,\P^{ }_e+\P^{ }_b)}
 \nn &  
 - &\!
   \biggl[ \fr12 + n^{ }_{\sigma_b}(\epsilon^{ }_b + \mu^{ }_b)\biggr]
   \mathcal{H}^{ }_{\vc{p}_b;\vc{p}_e,\vc{k};
        \epsilon^{ }_b\epsilon^{ }_e
        + \frac{m_c^2-m_b^2-m_e^2 }{2} 
        \,,\,
      \epsilon^{ }_b\omega
      + \frac{ m_c^2+m_d^2 - m_a^2 - m_e^2}{2}
   }
   {\Phi(\P^{ }_d+\P^{ }_b,-\P^{ }_b,\P^{ }_e-\P^{ }_b)}
   \biggr\} 
%%%%
 \nn 
 & + &\! 
 \int_{m_c}^{\infty} 
 \! \frac{{\rm d}\epsilon^{ }_c \, p^{ }_c}{(4\pi)^2}  \, 
 \biggl\{ 
 \nn &  
 + &\!
   \biggl[ \fr12 + n^{ }_{\sigma_c}(\epsilon^{ }_c - \mu^{ }_c)\biggr]
   \mathcal{H}^{ }_{\vc{p}_c;\vc{p}_e,\vc{k};
        \epsilon^{ }_c\epsilon^{ }_e
        + \frac{m_b^2-m_c^2-m_e^2 }{2} 
        \,,\,
        \epsilon^{ }_c\omega
        + \frac{ m_a^2 - m_c^2 - M^2 }{2} 
   }
   {\Phi(\K-\P^{ }_c,\P^{ }_c-\P^{ }_e,\P^{ }_c)}
 \nn &  
 + &\!
   \biggl[ \fr12 + n^{ }_{\sigma_c}(\epsilon^{ }_c + \mu^{ }_c)\biggr]
   \mathcal{H}^{ }_{\vc{p}_c;\vc{p}_e,\vc{k};
        \epsilon^{ }_c\epsilon^{ }_e
        + \frac{m_e^2+m_c^2-m_b^2}{2}
        \,,\,
        \epsilon^{ }_c\omega 
        + \frac{ M^2 + m_c^2 - m_a^2}{2} 
   }
   {\Phi(\K+\P^{ }_c,-\P^{ }_c-\P^{ }_e,-\P^{ }_c)}
 \biggr\} 
 \;. \nn \la{triangle_resolved}
\ea
We have used the sign $\,\simeq\,$ because the integrals 
are, in general, UV divergent; 
below \eq\nr{bubble_resolved} 
we turn to how the divergences can be subtracted.

For the integral $B$ from \eq\nr{bubble}, where only one propagator
appears, the results can be obtained from \eq\nr{triangle_resolved}
by appropriately dropping structures and renaming variables:  
\ba 
 && \hspace*{-1.5cm} 
 B(\P^{ }_d \,;\, a,b)\, \Phi(\P^{ }_a,\P^{ }_b)
%%%%
 \nn[2mm]
 & \simeq &\! 
 \int_{m_a}^{\infty} 
 \! \frac{{\rm d}\epsilon^{ }_a \, p^{ }_a}{(4\pi)^2}  \, 
 \biggl\{ 
 \nn &  
 - &\!
  \biggl[ \fr12 + n^{ }_{\sigma_a}(\epsilon^{ }_a - \mu^{ }_a)\biggr]
   2 \mathcal{G}^{ }_{\vc{p}_a;\vc{p}_d,\vc{k};
       \epsilon^{ }_a\epsilon^{ }_d
      + \frac{m_b^2-m_a^2-m_d^2 }{2} 
   }
   \,\Phi(\P^{ }_a,\P^{ }_d-\P^{ }_a)
 \nn &  
 + &\!
   \biggl[ \fr12 + n^{ }_{\sigma_a}(\epsilon^{ }_a + \mu^{ }_a)\biggr]
   2 \mathcal{G}^{ }_{\vc{p}_a;\vc{p}_d,\vc{k};
        \epsilon^{ }_a\epsilon^{ }_d 
        + \frac{ m_d^2 +m_a^2-m_b^2}{2} 
   }
   \,\Phi(-\P^{ }_a,\P^{ }_d+\P^{ }_a)
 \biggr\} 
%%%%
 \nn 
 & + &\! 
 \int_{m_b}^{\infty} 
 \! \frac{{\rm d}\epsilon^{ }_b \, p^{ }_b}{(4\pi)^2}  \, 
 \biggl\{ 
 \nn &  
 - &\!
  \biggl[ \fr12 + n^{ }_{\sigma_b}(\epsilon^{ }_b - \mu^{ }_b)\biggr]
   2 \mathcal{G}^{ }_{\vc{p}_b;\vc{p}_d,\vc{k};
         \epsilon^{ }_b\epsilon^{ }_d 
         + \frac{m_a^2-m_b^2-m_d^2 }{2} 
   }
   \,\Phi(\P^{ }_d-\P^{ }_b,\P^{ }_b)
 \nn &  
 + &\!
   \biggl[ \fr12 + n^{ }_{\sigma_b}(\epsilon^{ }_b + \mu^{ }_b)\biggr]
   2 \mathcal{G}^{ }_{\vc{p}_b;\vc{p}_d,\vc{k};
       \epsilon^{ }_b\epsilon^{ }_d 
       + \frac{ m_d^2+m_b^2-m_a^2}{2} 
   }
   \,\Phi(\P^{ }_d+\P^{ }_b,-\P^{ }_b)
 \biggr\} 
 \;. \hspace*{5mm} \la{bubble_resolved}
\ea
Here $\mathcal{G}$ is from \eq\nr{def_G}. 
%\footnote{% 
% In addition to UV divergences, 
% there can be mild (integrable) singularities within
% the integration ranges. For instance, 
% for the integral over $\epsilon^{ }_a$ in \eq\nr{bubble_resolved}, 
% these are at 
% $
%   [{\epsilon^{ }_d\, (m_a^2 + m_d^2 - m_b^2)
%   \pm p^{ }_d \kallen(m_d^2,m_a^2,m_b^2)}]/({2m_d^2})
% $ 
% if $m^{ }_d > m^{ }_a + m^{ }_b$ or $m^{ }_a > m^{ }_d + m^{ }_b$, 
% and at 
% $
%   [{\epsilon^{ }_d\, (m_b^2 - m_a^2 - m_d^2)
%   \pm p^{ }_d \kallen(m_d^2,m_a^2,m_b^2)}]/({2m_d^2})
% $ 
% if $m^{ }_b > m^{ }_a + m^{ }_d$.
% }

%%%%%%%%%%%%%%%%%%%%% PARAGRAPH %%%%%%%%%%%%%%%%%%%%%%%%%%%%%%%%%%%%%%%%%%%
%
\subsubsection*{UV subtraction}

The nature of the integrals in \eqs\nr{triangle_resolved}, 
\nr{bubble_resolved} depends on the functions $\Phi$ which, 
in turn, depend on the spin structure of the non-equilibrium particle
considered. For the example in \eq\nr{example_1to3_alt}, 
which leads to \eq\nr{ex_virtual}, 
$\Phi$ contains the helicity projectors $\E\cdot \P^{ }_i$.
In this case, or if $\Phi$ is independent of the integration 
momenta, only the function $B$ needs a subtraction. In case of a quadratic
momentum dependence, $C$ would need a subtraction as well. However, even
if $C$ were finite, we endorse adopting a subtraction procedure, 
specified as follows. 

The most straightforward implementation of a subtraction is to 
omit the vacuum contributions from \eq\nr{bubble_resolved}, and 
to compute them separately. That is, we may write
\be
 B(\P^{ }_d \,;\, a,b) = 
 B^{ }_0(\P^{ }_d \,;\, a,b) + 
 B^{ }_\T(\P^{ }_d \,;\, a,b)
 \;, 
\ee
where $B^{ }_\T$ is obtained by dropping the temperature-independent
factors $\frac{1}{2}$ from \eq\nr{bubble_resolved}:
\ba
 B^{ }_\T(\P^{ }_d \,;\, a,b)
 & \equiv &  
 B^{ }(\P^{ }_d \,;\, a,b)
 \bigr|^{ }_{\frac{1}{2} + n^{ }_{\sigma_i} \to n^{ }_{\sigma_i}}
 \;, \hspace*{5mm} \la{bubble_resolved_T}
 \\ 
 B^{ }_0(\P^{ }_d \,;\, a,b)
 & \equiv &  
 B^{ }(\P^{ }_d \,;\, a,b)
 \bigr|^{ }_{\frac{1}{2} + n^{ }_{\sigma_i} \to \frac{1}{2} }
 \;.
\ea
The part $B^{ }_\T$ vanishes at zero temperature. 
By returning back to \eq\nr{bubble},
it is possible to verify that the vacuum factors, 
in turn, amount to a usual vacuum integral,\footnote{%  
 What is meant here is that the factors 
 in \eq\nr{bubble} combine into 
 \ba
  && 
  \frac{1}{4\epsilon^{ }_a}
  \biggl[
    \frac{1}{(\epsilon^{ }_d - \epsilon^{ }_a)^2 - \epsilon_b^2 } 
  + 
    \frac{1}{(\epsilon^{ }_d + \epsilon^{ }_a)^2 - \epsilon_b^2 } 
  \biggr]
  + 
  \frac{1}{4\epsilon^{ }_b}
  \biggl[
    \frac{1}{(\epsilon^{ }_d - \epsilon^{ }_b)^2 - \epsilon_a^2 } 
  + 
    \frac{1}{(\epsilon^{ }_d + \epsilon^{ }_b)^2 - \epsilon_a^2 } 
  \biggr]
  \nn 
  & = & 
  \frac{1}{2\epsilon^{ }_a}
    \frac{1}{(\epsilon^{ }_d + \epsilon^{ }_a)^2 - \epsilon_b^2 } 
  + 
  \frac{1}{2\epsilon^{ }_b}
    \frac{1}{(\epsilon^{ }_d - \epsilon^{ }_b)^2 - \epsilon_a^2 } 
    \stackrel{p^{ }_0 \leftrightarrow i p^{ }_\rmiii{E} }{=} 
  - \int_{-\infty}^{\infty} 
  \! \frac{{\rm d} p^{ }_\rmiii{E} }{2\pi}
  \frac{1}{[p_\rmiii{E}^2 +  \epsilon_a^2 ]
           [(z^{ }_d - p^{ }_\rmiii{E})^2 + \epsilon_b^2 ]}
  \biggr|^{ }_{z_d \to - i \epsilon_d}
  \hspace*{-5mm}, \hspace*{6mm} \la{B0_details}
 \ea
 where the integral has been defined after Wick rotation 
 to Euclidean signature.
 Rotating back, the principal value (cf.\ \se\ref{se:IR})
 corresponds
 to the real part of a usual Feynman integral, 
 as indicated in \eq\nr{B0}.
 \la{fn:B0}}  
\be
 B^{ }_0(\P^{ }_d \,;\, a,b)\, \Phi(\P^{ }_a,\P^{ }_b) 
 \to 
 \re \int_{\P} 
  \biggl\{  
    \frac{  i\, \Phi(\P, \P^{ }_d - \P)
         }{[\P^2 - {m}_a^2][( \P^{ }_d - \P)^2 - {m}_b^2]} 
  \biggr\}
 \;. \la{B0}
\ee

Now, the vacuum integral can be worked out with standard methods. 
If the function $\Phi$ in \eq\nr{B0} depends on momenta, the result 
can be reduced to integrals with trivial numerators by making
use of Passarino-Veltman relations, 
however the details depend
on the helicity structure considered. For the case of \eq\nr{ex_virtual}, 
where the momentum dependence is linear, {\it viz.}\
$
 \Phi(\P^{ }_a,\P^{ }_b) = c^{ }_a \P^{ }_a + c^{ }_b \P^{ }_b
$, 
we get
\be
 \int_{\P} 
  \biggl\{  
    \frac{  i\, \Phi(\P, \P^{ }_d - \P)
         }{[\P^2 - {m}_a^2][( \P^{ }_d - \P)^2 - {m}_b^2]} 
  \biggr\}
 = 
 \int_{\P} 
  \biggl\{  
    \frac{  i\, [ c^{ }_b  \P^{ }_d + (c^{ }_a - c^{ }_b) \P  ]
         }{[\P^2 - {m}_a^2][( \P^{ }_d - \P)^2 - {m}_b^2]} 
  \biggr\}
\ee
where  
\ba
 && \hspace*{-1.2cm}
 \int_{\P} 
  \frac{i\P}{[\,\P^2 - {m}_a^2\,][\,(\P^{ }_d - \P)^2 - {m}_b^2\,]}
 \la{B1} \\
 & = & 
 \frac{\P^{ }_d}{2 m_d^2}
 \int_{\P} 
 \biggl\{ 
  \frac{i( m_d^2 + {m}_a^2 - {m}_b^2)}
       {[\,\P^2 - {m}_a^2\,][\,(\P^{ }_d - \P)^2 - {m}_b^2\,]}
   + \frac{i}{\P^2 - {m}_b^2} - \frac{i}{\P^2 - {m}_a^2}
 \biggr\} 
 \;. \nonumber 
\ea
The integrals on the second line of \eq\nr{B1} can be evaluated,  
\ba
 \re \int_{\P} 
           \frac{i}{\P^2 - {m}^2}
 & = &
  - \frac{{m}^2}{(4\pi)^2}
   \biggl( \frac{1}{\epsilon} + \ln\frac{\bmu^2}{m^2} +1 \biggr)
 \;, \la{A0_expl} \\
%%%
 \re \int_{\P} 
  \frac{i}{[\,\P^2 - {m}_a^2\,]
           [\,(\P^{ }_d - \P)^2 - {m}_b^2\,]}
 & = & 
  - \frac{1}{(4\pi)^2}
   \biggl[ \frac{1}{\epsilon}
 \la{B0_expl} \\ 
 & + & \int_0^1 \! {\rm d}x\, 
 \ln\biggl|\, 
  \frac{\bmu^2}{ {m}^{2}_a\, x  +  {m}^{2}_b (1-x) 
 - m_d^2\, x(1-x) } 
 \,\biggr|
 \;  
 \biggr]
 \;, \hspace*{5mm} \nonumber 
\ea
where terms of $\rmO(\epsilon)$ have been omitted. The integral over $x$
could be carried out~\cite{hv}, 
however the integral representation is a practical tool. 
The are mild singularities within the integration range if 
$m^{ }_d > m^{ }_a + m^{ }_b$, located at 
$
 x = [m_b^2 + m_d^2 - m_a^2 \pm \kallen(m_a^2,m_b^2,m_d^2)]/(2 m_d^2)
$.

The presence of $1/\epsilon$-divergences in \eqs\nr{A0_expl} and \nr{B0_expl} 
reflects the fact that we are considering a bare expression. In a full 
computation, the $1/\epsilon$-divergences of virtual corrections cancel
against the renormalization of the parameters that appear in the 
leading-order $1\leftrightarrow 2$ process. For simplicity, we assume
in the following that this renormalization has been taken care of, 
specifically that the $1\leftrightarrow 2$ process is expressed
in terms of $\msbar$ scheme couplings and masses, evaluated at
the scale $\bmu = 2\pi T$. Under this assumption, the 
$1/\epsilon$ divergences can be dropped from
\eqs\nr{A0_expl} and \nr{B0_expl} and the scale can be 
set to $\bmu = 2\pi T$. 
It is appropriate to stress, however, that were we considering
the bare rate, before renormalization, then we should recall
that the outer integral of \eq\nr{outer_int} contains a part
of $\rmO(\epsilon)$, as well as possible overall coefficients
proportional to $D-4$, 
which would give a finite contribution
once multiplied by the $1/\epsilon$-divergence. 

A splitup into thermal and vacuum parts is also possible for 
the triangle integral $C$, even if it does not always contain UV divergences:
\ba
 C^{ }_\T(\P^{ }_d,\P^{ }_e \,;\, a,b,c)
 & \equiv &  
 C^{ }(\P^{ }_d,\P^{ }_e \,;\, a,b,c)
 \bigr|^{ }_{\frac{1}{2} + n^{ }_{\sigma_i} \to n^{ }_{\sigma_i}}
 \;, \hspace*{5mm} \la{triangle_resolved_T}
 \\ 
 C^{ }_0(\P^{ }_d,\P^{ }_e \,;\, a,b,c)
 & \equiv &  
 C^{ }(\P^{ }_d,\P^{ }_e \,;\, a,b,c)
 \bigr|^{ }_{\frac{1}{2} + n^{ }_{\sigma_i} \to \frac{1}{2} }
 \;.
\ea  
The vacuum limit from \eq\nr{triangle} 
can be expressed as\hspace*{0.1mm}\footnote{%
 This can be established like in footnote~\ref{fn:B0}, 
 but now there are six terms to be combined.
 However, as shown in ref.~\cite{hv},
 a representation keeping channels with different physical
 interpretations separate 
 is preferable,  cf.\ \eq\nr{C0_rep2}.
 We have checked numerically 
 that evaluating the vacuum part of \eq\nr{triangle},
 with angular integrals expressed like in \eq\nr{triangle_resolved},
 is indeed consistent with \eq\nr{C0_rep2}.
 }
\ba
 && \hspace*{-1.5cm} 
 C^{ }_0( \P^{ }_d, \P^{ }_e \,;\, a,b,c)
 \, \Phi(\P^{ }_a,\P^{ }_b,\P^{ }_c)
 \nn[2mm] 
 & = &  
 \re 
 \int^{ }_{\P}  
  \frac{i \,\Phi(\P^{ }\,,\, \P^{ }_d-\P^{ }\,,\,\K-\P^{ })
 }{ [\, \P^2_{ } - {m}_a^2\,]\,
    [\, ( \P^{ }_d - \P^{ })^2 - {m}_b^2 \,]\, 
    [\, (\K - \P^{ })^2 - {m}_c^2 \,]}
 \;. 
\ea
In the case of a linear dependence like in \eq\nr{ex_virtual}, we can write
\be
 \Phi = c^{ }_a \P + c^{ }_b ( \P^{ }_d - \P) + c^{ }_c (\K - \P)
 = (c^{ }_a - c^{ }_b - c^{ }_c)\,\P 
 + c^{ }_b\,  \P^{ }_d + c^{ }_c\, \K 
 \;. 
\ee
The first structure points in a direction spanned by 
the latter two vectors, 
\be
 \int^{ }_{\P} \frac{i \P}{[...][...][...]}
 = 
 \alpha^{ }_b\,  \P^{ }_d + \alpha^{ }_c\, \K 
 \;, 
\ee
where
\be
 \left( 
  \begin{array}{c} 
   \alpha^{ }_b \\ 
   \alpha^{ }_c 
  \end{array} 
 \right)
 = 
 \frac{1}{\kallensq( m_d^2, m_e^2,\MM)}
 \left( 
  \begin{array}{cc}  
        - 2 \MM & m_d^2 - m_e^2 + \MM \\ 
   m_d^2 - m_e^2 + \MM & -2 m_d^2 
  \end{array}
 \right)
 \left(  
  \begin{array}{c}
    \mathcal{I}^{ }_b \\ 
    \mathcal{I}^{ }_c 
  \end{array} 
 \right)
 \;, 
\ee
with 
\ba
 \mathcal{I}^{ }_b & \equiv & 
 \int^{ }_{\P}  
     \frac{i\, (2 \P\cdot \P^{ }_d )}{[...][...][...]}
 \; = \; 
 \int^{ }_{\P} \biggl\{ 
     \frac{i ( m_d^2 + m_a^2 - m_b^2) }{[...][...][...]}
 \nn & & 
   + \frac{i}{[\P^2 - {m}_b^2][( \P^{ }_e-\P)^2 - {m}_c^2 ]}
   - \frac{i}{[\P^2 - {m}_a^2][(\K-\P)^2 - {m}_c^2 ]}
 \biggr\} 
 \;, \la{Ib} \\[2mm] 
%%%
 \mathcal{I}^{ }_c & \equiv & 
 \int^{ }_{\P}  
     \frac{i\, (2 \P\cdot\K )}{[...][...][...]}
 \; = \; 
 \int^{ }_{\P} \biggl\{ 
     \frac{i (\MM + m_a^2 -m_c^2)}{[...][...][...]}
 \nn & & 
   + \frac{i}{[\P^2 - {m}_b^2][(\P^{ }_e-\P)^2 - {m}_c^2 ]}
   - \frac{i}{[\P^2 - {m}_a^2][(\P^{ }_d-\P)^2 - {m}_b^2 ]}
 \biggr\} 
 \;. \la{Ic}
\ea
The Feynman representation for the first rows of 
\eqs\nr{Ib} and \nr{Ic} reads 
\ba
 && \hspace*{-1.2cm} 
 \re \int^{ }_{\P} 
  \frac{i 
 }{ [\, \P^2_{ } - {m}_a^2\,]\,
    [\, (\P^{ }_d - \P^{ })^2 - {m}_b^2 \,]\, 
    [\, (\K - \P^{ })^2 - {m}_c^2 \,]}
 \la{C0_rep} \\
%%%%%%%%%%%%%%%
 & = & 
 \frac{1}{(4\pi)^2}
 \int_0^1 \! {\rm d}y \! \int_0^y \! {\rm d}x \, \mathbbm{P} 
 \frac{1}{
   {m}_a^2 x 
 + [ {m}_c^2  
 - \MM x ](1-y) 
 + [ {m}_b^2  
 - m_d^2\, x 
 - m_e^2\, (1-y)
   ] (y-x)
 }
 \nn[2mm] 
%%%%%%%%%%%%%%%
 & = & 
 \frac{1}{(4\pi)^2(c+2 b\alpha)}
 \int_0^1 \! {\rm d}y \, 
 \biggl\{ 
    \frac{1}{y - y^{ }_1}
    \ln \biggl| 
          \frac{b y^2 + (c+e) y + a + d + f}
               {b y_1^2 + (c+e) y^{ }_1 + a + d + f}
        \biggr|
 \nn[2mm] 
%%%%%%%%%%%%%%%
 & - &  
    \frac{1}{y - y^{ }_2}
    \ln \biggl| 
          \frac{(a + b + c) y^2 + (d+e) y + f}
               {(a + b + c) y_2^2 + (d+e) y^{ }_2 + f}
        \biggr|
  \, + \, 
    \frac{1}{y - y^{ }_3}
    \ln \biggl| 
          \frac{a y^2 + d y + f}
               {a y_3^2 + d y^{ }_3 + f}
        \biggr|
 \; \biggr\}
 \;, \la{C0_rep2}
\ea
where we have followed ref.~\cite{hv}, 
denoting 
$
 a \equiv m_e^2
$, 
$
 b \equiv m_d^2 
$, 
$
 c \equiv M^2 - m_d^2 - m_e^2
$, 
$
 d \equiv m_b^2 - m_c^2 - m_e^2 
$, 
$
 e \equiv m_a^2 - m_b^2 + m_e^2 - M^2
$, 
$
 f \equiv m_c^2 
$,
$
 \alpha \equiv [m_d^2 + m_e^2 - M^2  \pm \kallen(m_d^2,m_e^2,M^2)]
 / (2 m_d^2) 
$, 
$
 y^{ }_0 \equiv - (d + e \alpha) / (c + 2 b \alpha)
$, 
$
 y^{ }_1 \equiv y^{ }_0 + \alpha
$, 
$
 y^{ }_2 \equiv y^{ }_0 / ( 1 - \alpha)
$, 
$
 y^{ }_3 \equiv - y^{ }_0 / \alpha
$. 
The kinematics of the outer integration guarantees that $\alpha$ is real
(cf.\ \eq\nr{virtual_triangle}); the sign in front of $\kallen$ is best
chosen so that no large cancellation takes place in the numerator of 
$\alpha$. 
If the arguments of the logarithms have real zeros, 
which happens for 
$\lambda(m_a^2,m_b^2,m_d^2) \ge 0$, 
$\lambda(m_a^2,m_c^2,M^2) \ge 0$, 
$\lambda(m_b^2,m_c^2,m_e^2) \ge 0$, 
respectively, 
we may express the 
corresponding integral in terms of (real parts of) 
four dilogarithms~\cite{hv}, otherwise 
the integral representations are quite efficient. 

%%%%%%%%%%%%%%%%%%%%%%%%%%% SECTION %%%%%%%%%%%%%%%%%%%%%%%%%%%%%%%%%%%
%
\section{Temperature and chemical potential induced IR divergences}
\la{se:resum}

After the inclusion of the virtual corrections
from \se\ref{ss:virtual}
(implemented through \se\ref{ss:phasespace_virtual}), 
there are no poles in any of the 
integrands that we have considered, and consequently 
logarithmic and double-logarithmic IR divergences have been lifted. 
However, the Bose and Fermi distributions
introduce new scales, the temperature~$T$ and the chemical 
potentials~$\mu^{ }_i$. These lead to possible new sources of
divergences, namely that 
\bi

\item[(a)]
a vacuum decay rate is boosted by $\sim (T /m^{ }_i)^n, (T / M)^n$,
and diverges if $m^{ }_i, M \ll T$;  

\item[(b)]
the Bose distribution is $\sim T / ( m^{ }_i - \mu^{ }_i)$ 
at small momenta, and diverges if $\mu^{ }_i \to m_i^{ }$.

\ei
Such divergences imply that the naive perturbative series
needs to be re-organized, or ``resummed''.
Unfortunately it is difficult to automate 
this treatment, as it requires 
a carefully tuned subtraction-addition step,
in order to avoid double counting when implementing 
the resummation that eliminates the divergence. 
Here we illustrate the procedure for the example 
of \fig\ref{fig:1to3}, and briefly mention other
widely discussed applications. 

%%%%%%%%%%%%%%%%%%%%%%% SUBSECTION %%%%%%%%%%%%%%%%%%%%%%%%%%%%%%%
%
\subsection{$2\leftrightarrow 2$ scatterings via soft $t$-channel exchange}
\la{sss:regge}

We start by considering what may be referred to as 
the high-energy small-angle limit, 
parametrically $\lambda^2 \ll -t \ll s$, where
$\lambda^2\sim m^{2}_i,M^2$ and $s\sim (\pi T)^2$.
In vacuum the structure of scattering amplitudes 
has been studied in great detail in this ``Regge'' domain. 
If $\lambda^2 \ll (\pi T)^2$, the thermally averaged scattering 
rates in general diverge, as we now review.

If we expand the integrand in \eq\nr{example_2to2_t}
in a Laurent series in $q^{ }_0,q$, the leading term may come with 
a negative power of $q^{ }_0,q$ and a positive
power of $\epsilon^{ }_1$. The latter integral is 
weighted by the thermal distribution functions, 
and therefore turns into
a positive power of the temperature. To compensate for the dimensions, 
the domain of small $q^{ }_0,q$ then leads to an IR divergence. 

In order to isolate this divergence, we may
work out the angles from \eq\nr{angles_t}, the azi\-muthal 
average from \eq\nr{azimuthal_t_res}, as well as all kinematic variables, 
in the massless limit. To facilitate this task, it is 
essential that the parametrization through $\mathcal{Q}$ has been 
so chosen that the divergences are associated with small values
of $q^{ }_0,q$. Denoting the massless value of $\epsilon^{ }_1$ by $p^{ }_1$,
we may approximate the integrand by taking 
\be
 q^{ }_0,q \; \ll \; p^{ }_1
 \;. 
\ee 
In this limit \eq\nr{angles_t} implies that 
\be
% \vc{q}\cdot\vc{p}^{ }_1 \approx q^{ }_0 p^{ }_1
% \;, \quad
% \vc{q}\cdot\vc{k} \approx \frac{q^2 + 2 q^{ }_0 k - q_0^2}{2}
% \;, \quad
 \cos\theta^{ }_{\vc{q},\vc{p}_1} \approx \frac{q^{ }_0}{q}
 \;, \quad
 \cos\theta^{ }_{\vc{q},\vc{k}}
 \approx 
 \frac{q^{ }_0}{q} + \frac{q^2 - q_0^2}{2 q k}
 \;. \la{IR_angles_t}
\ee
We remark that 
$k$ is often approximated as 
$\sim p^{ }_1 \gg q^{ }_0,q$, however
we have included the correction $(q^2-q_0^2)/(2qk)$ in 
$ \cos\theta^{ }_{\vc{q},\vc{k}} $,
given that the leading term 
$ 
 \cos\theta^{ }_{\vc{q},\vc{k}} \simeq q^{ }_0 / q
$ 
cancels against   
$ 
 \cos\theta^{ }_{\vc{q},\vc{p}_1}
$,  
when estimating the magnitude of inverse powers of $u$ and $s$
in \eqs\nr{iu} and \nr{iuiu}. 
In the massless limit, 
the integration 
measure from \eq\nr{dPhi22_t_prefinal} becomes
\be
 \lim_{m^{ }_i,M\to 0}
 \int \! {\rm d}\Omega^{t}_{2\leftrightarrow 2}
  =  
 \frac{1}{(4\pi)^3k}
 \int_{-\infty}^{k}
      \! {\rm d}q^{ }_0 
 \int^{ 2 k - q^{ }_0 }
     _{| q^{ }_0 |}
     \! {\rm d}q  
 \int^{\infty}_{(q^{ }_0 + q)/2}
     \! {\rm d}p^{ }_1
 \;, \la{dPhi22_t_massless}
\ee
implying that $t < 0$, i.e.\ $|q^{ }_0| < q$, so that the 
angles in \eq\nr{IR_angles_t} are well-defined.  
% (In the massless limit, a representation in terms of $q^{ }_0,q$ is
% often more transparent than in terms of $t,q^{ }_0$.) 

For estimating the kinematic variables $u,s$ from 
\eq\nr{mandelstam_2to2_t}, we need to consider the azi\-muthal
average from \eqs\nr{azimuthal_t_res}, \nr{azimuthal_t_res2}. 
Inserting subsequently 
$
 \cos\theta^{ }_{\vc{q},\vc{p}_1}
$ 
and 
$
 \cos\theta^{ }_{\vc{q},\vc{k}}
$, yields
\ba
 && u \;\approx\; -s \;\approx\;
 \frac{p^{ }_1 (q_0^2 - q^2) (2k - q^{ }_0)}{q^2}
 \;, \la{u} \\ 
 && u^2 \;\approx\; -u s \;\approx\; s^2 \;\approx\;
 \frac{p^{2}_1 (q_0^2 - q^2)^2 [ 3 (2k - q^{ }_0)^2 - q^2 ]}{2 q^4}
 \;, \la{uu} \\
 && \frac{1}{u} \;\approx\; - \frac{1}{s} \;\approx\;  
 \frac{q}{p^{ }_1(q_0^2 - q^2)}
 \;, \la{iu} \\ 
 && \frac{1}{u^2} \;\approx\; -\frac{1}{u s} 
                  \;\approx\; \frac{1}{s^2} \;\approx\;
 \frac{q(2k-q^{ }_0)}{p^{2}_1(q_0^2 - q^2)^2}
 \;. \la{iuiu}
\ea
% We note, furthermore, that
% thermal distribution functions are non-singular 
% at small $q^{ }_0$.\footnote{% 
%  According to \eq\nr{N_2to2_t}, 
%  if $\tau^{ }_1 = \sigma^{ }_1$ and $\nu^{ }_1 = \mu^{ }_1$, 
%  the first structure is $\sim T/q^{ }_0$  
%  whereas the latter term is $\sim q^{ }_0 n'_{\tau_1}(p^{ }_1)$.
%  Otherwise both terms are $\sim O(1)$. 
%  }

Now, 
\eqs\nr{dPhi22_t_massless}--\nr{uu} imply that integrands like 
$
  s/t, u/t \sim p^{ }_1 (2k - q^{ }_0)/q^2
$ 
lead to 
a logarithmically IR-divergent integral, whereas
$ 
  s^2/t^2, u^2/t^2 \sim 
  p^{2}_1 [ 3 (2k - q^{ }_0)^2 - q^2 ] / q^4
$ 
lead to a quadratic divergence. 
In contrast, 
$
  t/s,t/u\sim q/p^{ }_1
$ 
and 
$
 t^2/s^2,t^2/u^2 \sim q(2k-q^{ }_0)/p_1^2
$
indicate that if we choose a $t$-channel parametrization for
a process that is really of $u$ or $s$-channel nature, then 
the IR sensitivity is transmitted from $q$ to $p^{ }_1$, 
and is therefore not clearly resolved. 

It should be stressed that, as long as masses are non-zero, 
these IR sensitivities are not divergences in a literal sense.  
Their presence simply implies that if we consider the limit 
$m^{ }_i,M \ll \pi T$, the thermally averaged cross sections become
anomalously large. The way to handle the large terms goes through 
Hard Thermal Loop (HTL) resummation~\cite{ht1,ht2,ht3,ht4}, which
gives a thermal mass $\sim g^2T^2$ to the would-be massless modes. 
% Unfortunately this physics cannot be deduced from \eq\nr{example_1to3_alt},
% since some of the mass corrections do not lead to corresponding
% $2\leftrightarrow 2, 1\leftrightarrow 3$ processes from which
% they could be reconstructed. In other words, 
% focussing on a specific problem, 
% the full mass corrections need to be determined separately. 
Subsequently the Born-level result can be rectified through
a subtraction-addition step, with the addition part involving the 
HTL-resummed result and the subtraction part its unresummed form, 
eliminating double counting.
Adapting a clever trick developed in ref.~\cite{sch}, whereby
thermal light-cone observables can be analytically continued to static ones, 
HTL-resummed results have been worked up to the NLO level 
for certain interaction rates characterizing
massless or nearly massless probe particles~\cite{qhat,nlo_width}.

For the concrete example of \fig\ref{fig:1to3}, HTL resummation affects
the lepton propagator, whereas the scalar propagator only experiences
a mass correction.\footnote{% 
 In general, HTL computations include vertex corrections as well. 
 There is none for our example with a Yukawa vertex,
 and more generally they 
 are unimportant when considering ``hard'' momenta, $k\sim \pi T$.
 } 
Let us denote the 
lepton spectral function by
\be
 \rho^{ }_\ell (q^{ }_0,\vc{q}) \; \equiv \;
 \Bigl( q^{ }_0 \, \hat{\rho}^{ }_0 (q^{ }_0,q)
 ,\vc{q}\, \hat{\rho}^{ }_s (q^{ }_0,q)
 \, \Bigr)
 \;, \la{rhoell}
\ee
where the temporal and spatial parts read
($q^{ }_0 \equiv \re q^{ }_0 + i 0^+_{ }$)~\cite{weldon,qed2}
\ba
 \hat{\rho}^{ }_\mu(q^{ }_0,q) & = & 
 \im \Biggl\{ 
 \frac{1 - \frac{\delta^{ }_{\mu,0}  \mellT^2 L}{2q^{ }_0} 
         + \frac{\delta^{ }_{\mu,s}  \mellT^2 (1-q^{ }_0 L)}{2 q^2}  }
 {q_0^2 - q^2 - \mellT^2
  + \frac{\mellT^4 
 [
  (q L)^2 - (1 - q^{ }_0 L)^2 
 ]
         }{4 q^2} 
 }
 \Biggr\}
 \;, \quad
 L \equiv \frac{1}{2q} \ln \frac{q^{ }_0 + q}{q^{ }_0 - q}
 \;. \la{rhomu}
\ea
For $|q^{ }_0| < q$,
the overall imaginary part originates from $\im L = -\pi/(2q)$, 
whereas for $|q^{ }_0| > q$, 
it originates from a zero of the denominator.   
The thermal lepton mass reads
\be
 \mellT^2 
 \; \equiv \; 
 \frac{g_1^2 + 3 g_2^2}{4}
  \int_{\vc{p}} 
  \frac{
         2 \nB^{ }(p)
         + \nF^{ }(p+\mu^{ }_\ell)
         + \nF^{ }(p-\mu^{ }_\ell)  
  }{p} 
 \; = \;
 \frac{g_1^2 + 3 g_2^2}{16}
 \biggl( T^2 + \frac{\mu_\ell^2}{\pi^2} \biggr) 
 \;. \la{mell}
\ee

The interaction rate $\Gamma^\rmii{HTL}_{ }$ is determined
by making use of \eq\nr{rhoell}. Integrating over angles, the remaining
measure can be cast in the form of \eq\nr{dPhi22_t_prefinal},
with $\vc{q}\cdot \vc{k}$ fixed
according to \eq\nr{angles_t} and the
integral over $\epsilon^{ }_1$  
represented by \eq\nr{mell}.  
In the HTL spectral function, 
it is the spacelike domain, with $\theta(-t)$, 
which corresponds to $2\leftrightarrow 2$ scatterings. 

Subsequently, we have to subtract the part
already included in \eq\nr{example_2to2_t}. 
% Specifically, 
% this concerns the first terms on its first and second rows. 
This subtraction can be obtained by ``re-expanding'' \eq\nr{rhomu} to 
$\rmO(g^2 T^2)\sim \rmO(\mellT^2)$. With a slight abuse of notation, 
we however keep a lepton mass in the denominator, renaming it into
$m_\bla^2$, whereby it serves as an intermediate
IR regulator in the sense of \se\ref{se:IR}.  This then leads to 
\ba
 \Delta\Gamma^\rmii{HTL}_{2\leftrightarrow 2}
 & = & 
 -\frac{1}{(2\pi)^2k}
   \int_{-\infty}^0 \! {\rm d}t 
   \int_{q_0^-}^{q_0^+} \! {\rm d}q^{ }_0 
 \, 
 \Bigl[ 
   1 - \nF^{ }(q^{ }_0 - \mu^{ }_\ell)
     + \nB^{ }(\omega - q^{ }_0 - \muS^{ })
 \Bigr] 
 \, \Lambda(q^{ }_0,\omega)
 \hspace*{8mm} 
 \nn
%%%%%%%
 & \times & 
 \mathbbm{P}
 \biggl\{ 
   \vare^{ }_0 q^{ }_0 \,
   \biggl[
     \hat{\rho}^{ }_0(q^{ }_0,q) - 
     \frac{\pi \mellT^2}{4 q^{ }_0 q (t - m_\bla^2 )} 
   \biggr]
%  \nn 
%%%%%%%%
% & & \hspace*{1.5cm}
   - \frac{\vec{\vare}\cdot\vc{k}\, \vc{q}\cdot\vc{k} }{k^2} \, 
   \biggl[
    \hat{\rho}^{ }_s(q^{ }_0,q) - 
    \frac{\pi \mellT^2 q^{ }_0}{4 q^3 (t - m_\bla^2)} 
   \biggr]
 \biggr\} 
 \;, \nn
  \la{2to2_htl} 
\ea
where 
$\E \, \equiv \, (\vare^{ }_0,\vec{\vare}\,)$,
and the integration bounds are a special case of \eq\nr{q0pm}, 
{\it viz.} 
\be
  q_0^{\pm} \; \equiv \; 
 \frac{ \omega (M_{ }^2+t-m_\cS^2)
                         \pm k\kallen(M_{ }^2,t,m_\cS^2)}{2M^2}
 \;. 
 \la{q0pm_htl}
\ee

Let us mention that if we choose $m^{ }_\bla = \mellT^{ }$
in \eq\nr{2to2_htl}, then the numerical value of \eq\nr{2to2_htl} is 
quite small. This implies that using the thermal value 
$\mellT^{ }$ in the unresummed computation of \se\ref{ss:phasespace_real}
would yield a reasonable
approximation to the resummed result. That said, 
the error would {\em not be} parametrically of higher order; 
\eq\nr{2to2_htl} is of higher order only as far as the domain 
$q^{ }_0,q \gg \mellT^{ }$ is concerned, 
where the subtractions are effective. 

In \eq\nr{2to2_htl}, 
we have introduced a function $\Lambda(q^{ }_0,\omega)$
that necessitates further elaboration. 
In order to correctly implement
HTL resummation for 
$ 
 q^{ }_0,q \sim \mellT^{ } \ll \pi T
$,
we require 
$
 \lim_{q^{ }_0\to 0}
 \Lambda(q^{ }_0,\omega) = 1
$.
What we do outside of this domain, 
where the effect is formally of higher order, 
is a matter of choice. 
A frequent logic is to set 
$\Lambda \to \Lambda^\star_{ }$, where 
\be
 \Bigl[ 
   1 - \nF^{ }(q^{ }_0 - \mu^{ }_\ell)
     + \nB^{ }(\omega - q^{ }_0 - \muS^{ })
 \Bigr] 
 \, \Lambda^\star_{ }(q^{ }_0,\omega)
 \; \equiv \;  
 \nF^{ }(\mu^{ }_\ell) + \nB^{ }(\omega - \muS^{ })
 \;. \la{choice_Lam}
\ee
A benefit of this choice is that the remaining integrals 
can be solved analytically in the so-called ultrarelativistic 
regime ($m_\cS^{},M \ll k \sim \pi T$). 
On the other hand, 
a numerical evaluation 
with general masses and momenta 
is more straightforward by setting
$\Lambda\to 1$. With the latter choice, 
the magnitude of \eq\nr{2to2_htl}
will be illustrated in \fig\ref{fig:example_subtractions}.

%%%%%%%%%%%%%%%%%%%%%%% SUBSECTION %%%%%%%%%%%%%%%%%%%%%%%%%%%%%%%
%
\subsection{$2\leftrightarrow2$ scatterings off soft bosons}
\la{sss:off}

Another possible source for an IR divergence 
in $2\leftrightarrow 2$ scatterings is when 
one of the external scatterers becomes soft.
Technically, this is the case when the argument
of a bosonic thermal distribution vanishes in \eq\nr{N_2to2_t}. 
Let us recall that chemical potentials associated with bosons
should be smaller than the particle masses, otherwise we are 
driven to Bose-Einstein condensation. In cosmology, chemical 
potentials are in any case small compared with the temperature. 
In most computations it is then sufficient to expand to 
first order in chemical potentials. In this case, the domain
of a small bosonic energy exhibits a quadratic pole,
$\nB'(q^{ }_0) = -\beta \nB^{ }(q^{ }_0)[ 1 + \nB^{ }(q^{ }_0)] 
 \approx - T/q_0^2 $ for $q^{ }_0 \ll T$, 
leading to a logarithmic divergence if the mass has been omitted. 
Such a divergence can arise for instance from  
the domain $q^{ }_0\sim \omega$ inside the thermal distribution 
$n^{ }_{\sigma_2}$ in \eq\nr{N_2to2_t}~\cite{cptheory}. 
It has also been pointed out that a similar effect can 
arise even without chemical potentials, if we consider 
virtual corrections to $1\leftrightarrow 2$ scatterings
and pick up an enhancement $\sim (T/q^{ }_0)^2$ from
two Bose distributions~\cite{interpolation}. Both of these
effects are related to kinematic endpoints for
scattering off soft Higgs bosons, 
and can be regularized simply by keeping the thermal Higgs mass finite. 

In the presence of chemical potentials, 
some care is needed with virtual lines as well. For instance, as 
discussed below 
\eq\nr{N_2to2_t}, $q^{ }_0$ may cross zero despite the particle 
having a mass, and then $n^{ }_{\tau_1\sigma_1}$ has a would-be 
pole at $q^{ }_0 = \mu^{ }_1 - \nu^{ }_1$ if 
$\tau^{ }_1\sigma^{ }_1 = +$, which is lifted by the second factor. 
If we expand in chemical potentials, 
the pole is of second order,  
but it is still lifted. In order to account for this, 
it is helpful to make use of 
the  representation given below \eq\nr{N_2to2_t}. 

%%%%%%%%%%%%%%%%%%%%%%% SUBSECTION %%%%%%%%%%%%%%%%%%%%%%%%%%%%%%%
%
\subsection{Small-angle $1+n\leftrightarrow2+n$ reactions}
\la{sss:lpm}

Even if at first sight
simpler than $2\leftrightarrow 2$ and $1\leftrightarrow 3$ scatterings, 
it turns out that the IR structure of
$1\leftrightarrow 2$ reactions is more complicated, 
as the phase space is strongly constrained by masses.
This implies that a subclass of $1+n\leftrightarrow 2+n$ reactions, 
with $n\ge 1$, can be as important as the $1\leftrightarrow 2$ reactions, 
and needs to be incorporated through
Landau-Pomeranchuk-Migdal (LPM) resummation~\cite{gelis3,amy1,bb1}. 
The procedure requires 
the inclusion of ``asymptotic'', 
i.e.\ large-momentum thermal mass corrections~\cite{weldon}. 
The full information cannot be easily deduced 
from a matrix element squared like \eq\nr{example_1to3_alt}, and
therefore requires an effort specific to the application in question. 
Profitting from ideas 
put forward in ref.~\cite{sch}, LPM-resummed results have been 
extended up to NLO in some cases~\cite{gm1,gm2,gm3}. 

Similarly to \se\ref{sss:regge}, 
LPM resummation requires a subtraction-addition step, manifestly avoiding
the danger of double counting. 
It is important here that as LPM resummation
is normally implemented after making use of  
kinematic simplifications pertinent to the ultrarelativistic regime, 
the subtraction should adhere to the same simplifications, 
guaranteeing that the resummation only has a negligible (higher-order)
effect in domains where it is not justified. It is an 
open problem, deserving further study, how LPM resummation could be 
smoothly connected to kinematic domains beyond the ultrarelativistic one. 

For the concrete example of \fig\ref{fig:1to3}, 
even though we do not discuss LPM resummation itself here, 
for the reason just mentioned, 
the {\em subtraction term} 
can be deduced from the same HTL computation that led to 
\eq\nr{2to2_htl}. The difference to \eq\nr{2to2_htl}
is that in $1\leftrightarrow 2$
reactions, the lepton is on-shell and timelike, 
and therefore we need to pick up 
the pole part from \eq\nr{rhomu}.\footnote{%
 Mathematically speaking, the denominator has multiple zeros, 
 however only the trivial one at $q^{ }_0 \approx \sqrt{q^2 + \mellT^2}$
 plays a role in the large-momentum regime pertinent to LPM resummation.  
 } 
Subsequently we again need to re-expand $\hat{\rho}^{ }_\mu$
up to $\rmO(g^2T^2)\sim \rmO(\mellT^2)$ to obtain the terms to 
be subtracted. 
However, great care is needed here: 
in \eq\nr{ex_virtual}, we have reconstructed only a part of the virtual
corrections to $1\leftrightarrow 2$ scatterings, namely those that have
a counterpart on the side of $2\leftrightarrow 2$ and 
$1\leftrightarrow 3$ scatterings. This subset does {\em not} include the 
corrections that modify the lepton mass but do nothing else. 
Therefore, only terms from the 
numerator of \eq\nr{rhomu} play a role in the re-expansion. 
Renaming
the mass in the denominator to $m_\bla^2$, which is viewed as an 
IR regulator like in \eq\nr{2to2_htl}, this leads to 
\be
 \Delta \Gamma^\rmii{HTL}_{1 \leftrightarrow 2}
 \; = \; 
 - 
 \scat{1\leftrightarrow2}(\bla,\aS)  
 \,
 \mellT^2 
 \, \mathbbm{P}  
 \biggl\{ 
  \biggl(
    \vare^{ }_0 
    - \frac{\vec{\vare}\cdot\vc{p}^{ }_\bla\, \epsilon^{ }_\bla }
           {p_\bla^2} 
  \biggr) 
  \frac{1}{p^{ }_\bla}
  \ln \biggl|
        \frac{\epsilon^{ }_\bla - p^{ }_\bla}
             {\epsilon^{ }_\bla + p^{ }_\bla} 
      \biggr|
  - \frac{2 \vec{\vare}\cdot\vc{p}^{ }_\bla }{p_\bla^2}
 \biggr\} 
 \;. \la{1to2_htl}
\ee
Here the integration boundaries and angle are from
\eqs\nr{1to2_bounds} and \nr{kdotpd}, and the expression
in curly brackets represents an angular average 
of the type in \eq\nr{def_G}. 
Eq.~\nr{1to2_htl} is closely related to and indeed
subtracts most of the first term on the first line of 
\eq\nr{ex_virtual}. 

The numerical magnitude of \eq\nr{1to2_htl} will be 
illustrated in \fig\ref{fig:example_subtractions}. 
We note that
the IR divergences
of \eqs\nr{2to2_htl} and \nr{1to2_htl}
cancel against each other when $m^{ }_\bla\to 0$~\cite{interpolation}; 
the IR divergence of \eq\nr{1to2_htl} originates from the region
of small $\epsilon^{ }_\bla$, where the phase
space distributions in 
$
  \scat{1\leftrightarrow2}(\bla,\aS)  
$ 
take the same form as on the right-hand side of \eq\nr{choice_Lam}.
% The cancellation can be understood from the fact that both 
% subtractions originate as limits of the first term
% in \eq\nr{example_1to3_alt} 
% and its corresponding virtual correction in \eq\nr{ex_virtual}.

%%%%%%%%%%%%%%%%%%%%%%% SUBSECTION %%%%%%%%%%%%%%%%%%%%%%%%%%%%%%%
%
\subsection{How about real intermediate states?}
\la{sss:ris}

One issue frequently discussed in the Boltzmann equation
literature is that of ``real intermediate states''
(cf.,\ e.g.,\ ref.~\cite{kw}). Consider a reaction
in which the Mandelstam variable entering a line can 
coincide with the mass-squared of that particle.
This case emerges, for instance, if we consider the 
process in \fig\ref{fig:1to3}(left), 
and let the scalar decay into 
a $t\bar{t}$ pair, assuming that with thermal modifications it could 
be possible to have $M - m^{ }_\ala > m^{ }_\cS > 2 m^{ }_t$. 
The general conditions for the appearance
of real intermediate states were listed in 
\eqs\nr{ris1}--\nr{ris3}. 

If the propagator in question appears quadratically in the 
cross section, 
the thermal phase space average looks potentially divergent.
In our procedure, as specified in \se\ref{se:IR}, this is regularized
by treating the quadratic propagator as a mass derivative of a principle
value, rendering the average integrable. 
There is necessarily also 
a $1\leftrightarrow 2$ reaction 
in which the resonant intermediate state appears
as a real particle. 
Virtual corrections to these $1\leftrightarrow 2$
processes (a closed $t\bar{t}$ loop in the above example)
reflect the same singularity. Specifically, 
after angular averaging, 
the remaining energy integrals,  
from \eqs\nr{outer_int} and \nr{bubble_resolved},  
cross a hypersurface where the argument of a logarithm vanishes. 
Even though this is again integrable, 
an efficient numerical procedure normally requires
dividing the integration domain into subregions, 
so that the singularities appear at their boundaries. 

To summarize, real intermediate states
(or, in the language of \se\ref{se:IR}, actual poles)
in the matrix elements squared lead to no IR 
singularities in our framework. However, their appearance makes it more 
tedious to render the practical integrations efficient. 

%%%%%%%%%%%%%%%%%%%%%%%%%%% SECTION %%%%%%%%%%%%%%%%%%%%%%%%%%%%%%%%%%%%%%

\section{Description of algebraic and numerical procedures}
\la{se:code}

Let us summarize in a procedural fashion an algorithm implementing
the ingredients introduced in \ses\ref{se:2to2} and \ref{se:phasespace}.

The general philosophy is to start from a $1\rightarrow 3$ amplitude, 
like in \eq\nr{example_1to3_alt}, 
describing how a ``non-equilibrium'' particle decays into ``plasma'' particles
(the process does not need to be kinematically allowed in practice). 
Let us stress again that even if we refer to a decay, 
inverse processes are always included as well. 
The non-equilibrium particle represents 
a slow variable which may fall out of equilibrium or never enter it
in the first place, 
whereas the plasma particles are fast ones, 
with their density matrices fully characterized by
a temperature and a handful of chemical potentials. 
There are really two algorithms, an algebraic~[{\sc a}]
and a numerical one~[{\sc n}].
Their ingredients are: 

% \pagebreak

\bi

\item{\sc input parameters describing plasma particles [n]:}

These parameters include the temperature;
various chemical potentials 
(lepton and baryon asymmetries, gauge field backgrounds); 
effective couplings and masses 
like $g^{ }_1, g^{ }_2, m^{ }_{\bla}, m_{\tilde{\cS}}^{ }$ 
in \eq\nr{example_1to3_alt},
which may 
incorporate ``hard loop corrections'' from fast processes, 
whereby these are in general functions of the temperature; 
masses $m^{ }_a$, chemical potentials $\mu^{ }_a$, 
and statistics $\sigma^{ }_a$, associated with 
the final state of the $1\to 3$ process; 
a specification for negative index choices, 
according to \eq\nr{prop}. 

\item{\sc input parameters describing the non-equilibrium particle [n]:}

Momentum $k$, mass $M$, and energy $\omega = \sqrt{k^2 + M^2}$; 
polarization state, e.g.\ through the four-momentum $\E$ 
in \eq\nr{example_1to3_alt}; 
Standard Model quantum numbers of the vertex 
to which the non-equilibrium particle couples, 
e.g.\ through its possible dependence on the  
lepton flavour in \eq\nr{example_1to3_alt}. 

\item{\sc definition of auxiliary functions [n]:}

The basic functions appearing are 
$n^{ }_{\sigma}(\epsilon)$ from \eq\nr{n_sigma}; 
the kinematic (K\"all\'en) function from \eq\nr{kallen}; 
azimuthal averages, given by 
\eqs\nr{azimuthal_t_res}, \nr{azimuthal_t_res2}; 
angular averages for virtual corrections, 
given by $\mathcal{G}$ and $\mathcal{H}$
from \eqs\nr{def_G}, \nr{def_H}. 

\item{\sc input of $1\to 3$ decay rate [a]:}

The key dynamical information enters through a function 
$
 \Theta^{ }_{ }(\P^{ }_1,\P^{ }_2,\P^{ }_3)
$
(e.g.\ \eq\nr{example_1to3_alt}), 
which effectively represents a matrix element squared for  
a Born-level decay rate,
$
   \Gamma^\rmi{Born}_{1\to 3} 
   = 
   \scat{1\to3}(a^{ }_1,a^{ }_2,a^{ }_3)
          \, \Theta^{ }_{ }(\P^{ }_1,\P^{ }_2,\P^{ }_3)
$, 
with $\P^{ }_i \equiv \P^{ }_{a_i}$.

\item{\sc reflection of $1\to 3$ decay rate into other channels [a]:}

The full (Born-level) interaction rate, 
incorporating the other channels and inverse processes,  
is obtained from the $1\to 3$ decay rate according to \eq\nr{master}.
Subsequently, we can make use of permutations of momenta, 
to label the four-momenta according to the conventions 
that we have adopted.
% in \ses\ref{ss:phasespace_real}, 
% \ref{app:1to3}--\ref{app:3to1_s}, respectively. 
For the $1\to 3$ process there are 
$3! = 6$ permutations, whereas
for the $2\leftrightarrow 2$ and $3\to 1$ cases the identification
of one initial- or one final-state particle, respectively, 
leaves over a two-fold freedom that can be used to 
rename variables. An automated implementation of this step 
requires a software with wildcard pattern matching capabilities, 
such as {\sc form}~\cite{form}. 

\item{\sc identification of virtual corrections [a]:}

The $1\to 3$ rate permits for the automatic identification of
IR sensitive virtual corrections to $1\leftrightarrow 2$ processes, 
by searching for poles
in the matrix elements squared,  
determining the corresponding residues, 
and attaching these to the proper thermal average (cf.\ \se\ref{ss:virtual}). 

\item{\sc integration measure and thermal distributions for given channel [a]:}

Proceeding to the integration, 
two angles are fixed by energy-momentum conservation,  
e.g.\ from \eq\nr{angles_t};
all kinematic invariants are fixed in terms of the chosen
integration variables, e.g.\ from \eq\nr{mandelstam_2to2_t}; 
thermal distributions can be represented in 
a factorized form, e.g.\ from \eq\nr{N_2to2_t}; 
integration measure can be inserted, 
e.g.\ from \eq\nr{dPhi22_t_final}. 

% \pagebreak

\item{\sc azimuthal average [a]:}

A key ingredient of the parametrization is that 
one the remaining integrals, over the azimuthal angle, 
can be carried out analytically. The dependence on the 
azimuthal angle originates through a single scalar product, 
for instance $\vc{k}\cdot\vc{p}^{ }_1$ in 
\eq\nr{calF}, which does not appear in thermal 
distributions.
After the full expression has been partial fractioned into terms 
containing positive powers of this variable, or inverse powers of 
first-order polynomials, 
the azimuthal average can be inserted, 
from \eqs\nr{azimuthal_t_res}, \nr{azimuthal_t_res2}.
For virtual corrections, the full angular averages can be resolved, 
from~appendix~C.

\item{\sc phase space integrals [n]:}

The remaining at most three-dimensional integration can 
be carried out numerically, e.g.\ from \eq\nr{dPhi22_t_final}. 
In virtual corrections the integration is normally two-dimensional.

% \item{\sc handling of chemical potentials [n]:}
% 
% We consider two options for treating the dependence of the 
% results on chemical potentials. If the bosonic chemical potentials are 
% smaller than the particle masses with which they are associated, 
% they can be included exactly. In cosmology, however, chemical potentials
% are small in any case. To handle this situation, the thermal distributions, 
% e.g.\ from \eq\nr{N_2to2_t}, can be expanded to first order in chemical
% potentials, avoiding the possibility of encountering spurious
% ill-defined expressions. 
  
\item{\sc tests [n]:}

The procedure contains some redundancies, 
which permit to crosscheck for its correct implementation. 
For $2\leftrightarrow 2$ and $3\to 1$ processes, 
if we override the permutations used to select the ``optimal''
parametrization, 
we should get the same result 
for both $t$- and $s$-channel parametrizations.
For $1\leftrightarrow 3$ processes, all 3! relabellings of 
the final-state momenta should lead to the same result. 

\ei

Let us conclude by 
showing example plots for the case of \eq\nr{example_1to3_alt}, 
for the plasma input parameters
$
 g^{ }_1=1/3
$, 
$ 
 g^{ }_2=2/3
$, 
$
 \muL = 10^{-3} T % \, \forall a
$, 
% $
%  \muB = 0
% $, 
$
 \muY = 2\times 10^{-2} T
$, 
where $\muY$ refers to the hypercharge chemical potential. 
The properties of the decay products are set as
$
 \sigma^{ }_{\ala} = -1, 
 m^{ }_\ala = 0.1 T, 
 \mu^{ }_\ala = \muL - \muY / 2
$;
$
 \sigma^{ }_{\cQ} = +1, 
 m^{ }_\cQ = 0.01 T, 
 \mu^{ }_\cQ = 0 
$; 
$
 \sigma^{ }_{\cS} = +1
$, 
$ 
 m^{ }_\cS = T 
$, 
$
 \mu^{ }_\cS = \muY / 2
$. 
The auxiliary masses are set to 
$ m^{ }_\bla = m^{ }_\ala$
and 
$ m_{\tilde{\cS}}^{ } = m_\cS^{ }$.
For the non-equilibrium particle we set 
$ 
 M=0.3T 
$ or 
$
 M=3T
$, 
% coupling it to the lepton flavour
% $ 
%  a=2
% $, 
choosing the helicity projection 
$
 \E = \K
$
or 
$
 \E = \U
$.
Recalling our convention of referring to the side 
of the non-equilibrium particle
as the initial state, 
$1\to3$ channels are open for the mass $M = 3T$,  
and $3\to 1$ channels for $M = 0.3T$. 
The $2\leftrightarrow 2$ scatterings are allowed in any case. 
The results are shown in \fig\ref{fig:example2to2}, including 
for comparison also the results originating from 
$1\leftrightarrow 2$ processes. For further illustration, 
in \fig\ref{fig:example_subtractions} we show 
the influence of thermal resummations
from \eqs\nr{2to2_htl} and \nr{1to2_htl}, whose effect is 
to replace the mass $m^{ }_\bla$ by the proper thermal 
lepton mass $\mellT^{ }$. Our final results are obtained
by summing together \figs\ref{fig:example2to2} and
\ref{fig:example_subtractions}; the outcome is shown
in \fig\ref{fig:imDELTA}, this time as a function of
$M/T$, for various fixed momenta $k/T$.
The conclusions that we draw
from these plots are discussed in \se\ref{se:concl}.

%%%%%%%%%%%%%%%%%%%%%%%%%%%%%%%%% FIGURE %%%%%%%%%%%%%%%%%%%%%%%%%%%%%%%%%
\begin{figure}[p]

\hspace*{-0.1cm}
\centerline{%
 \epsfysize=7.5cm\epsfbox{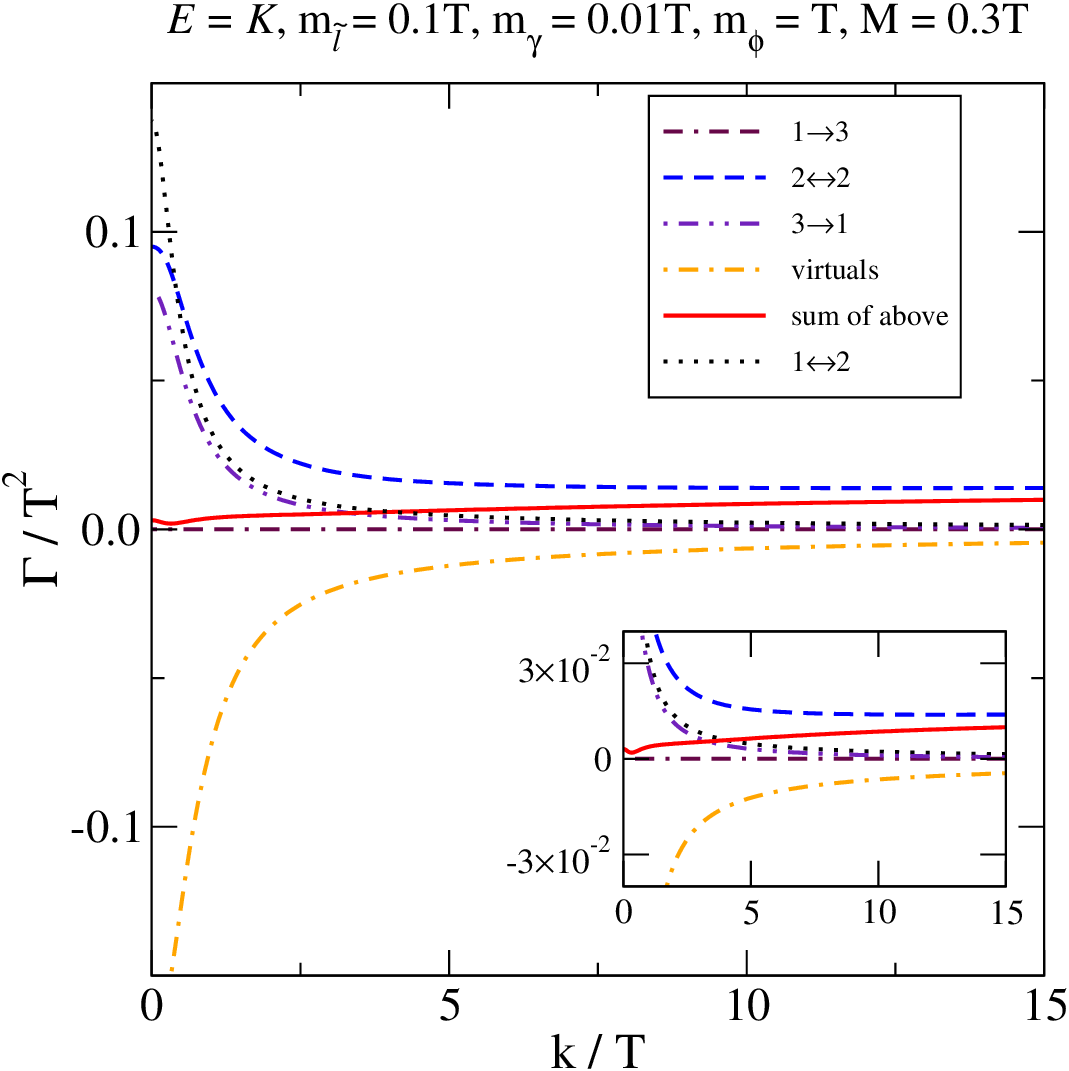}%
 \hspace{0.5cm}%
 \epsfysize=7.5cm\epsfbox{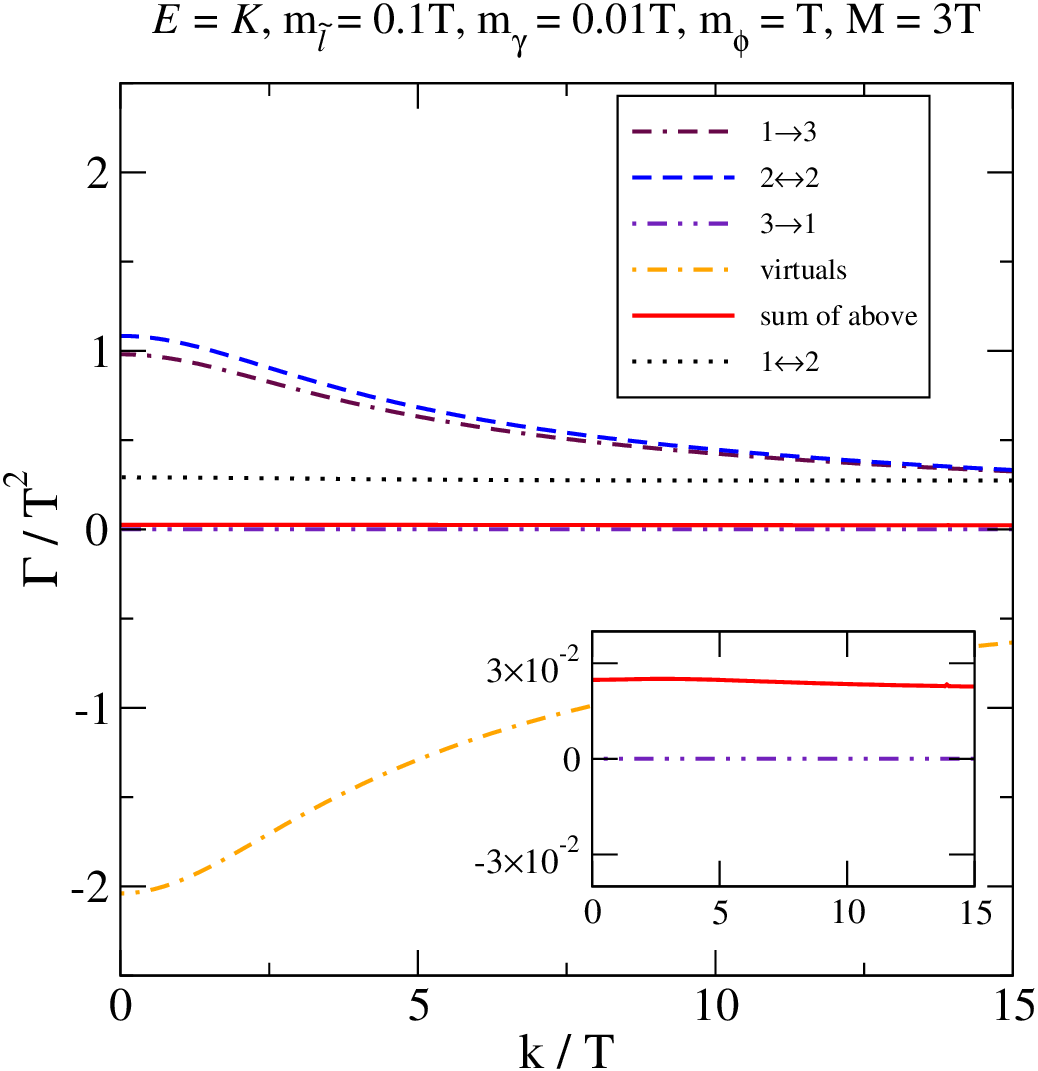}%
}

\vspace*{0.4cm}

\hspace*{-0.1cm}
\centerline{%
 \epsfysize=7.5cm\epsfbox{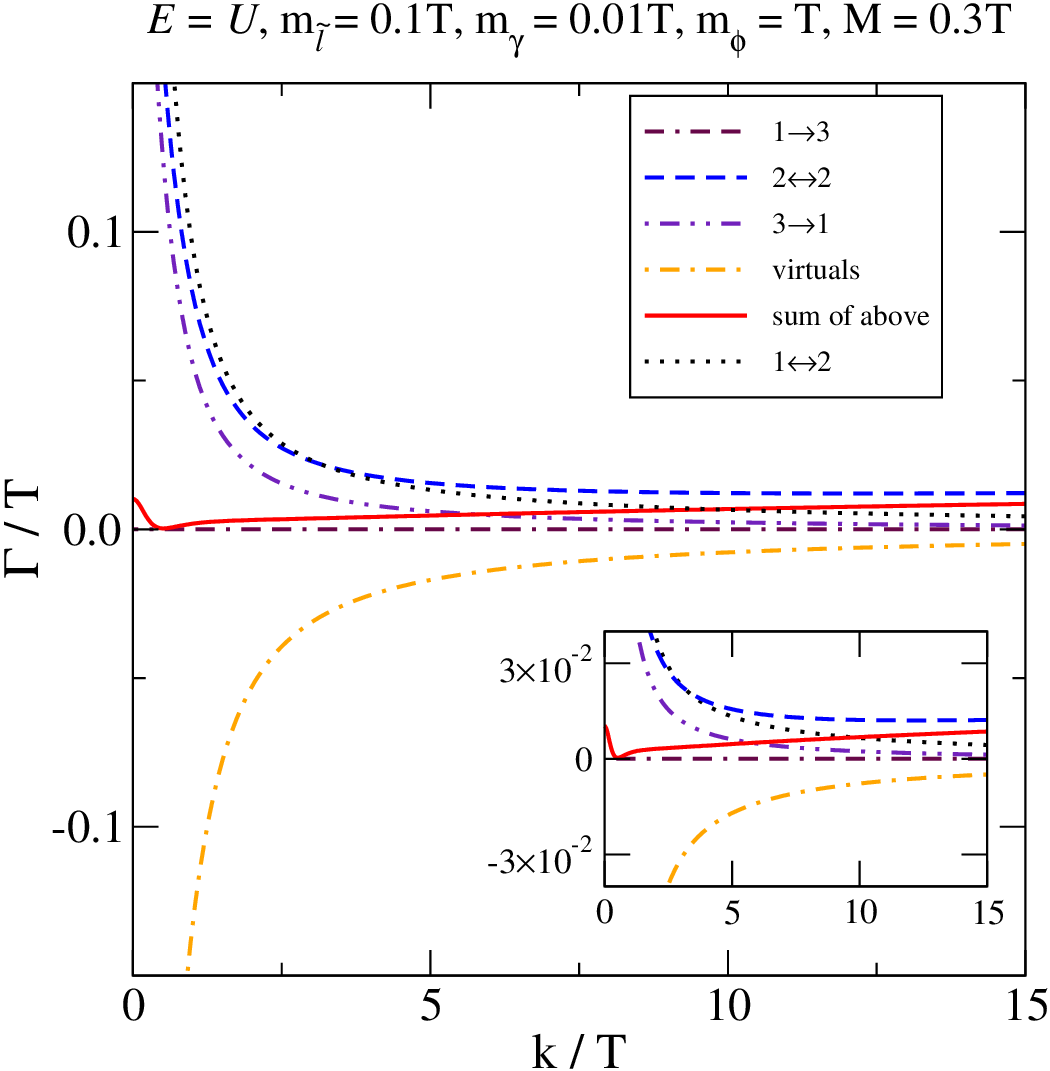}%
 \hspace{0.5cm}%
 \epsfysize=7.5cm\epsfbox{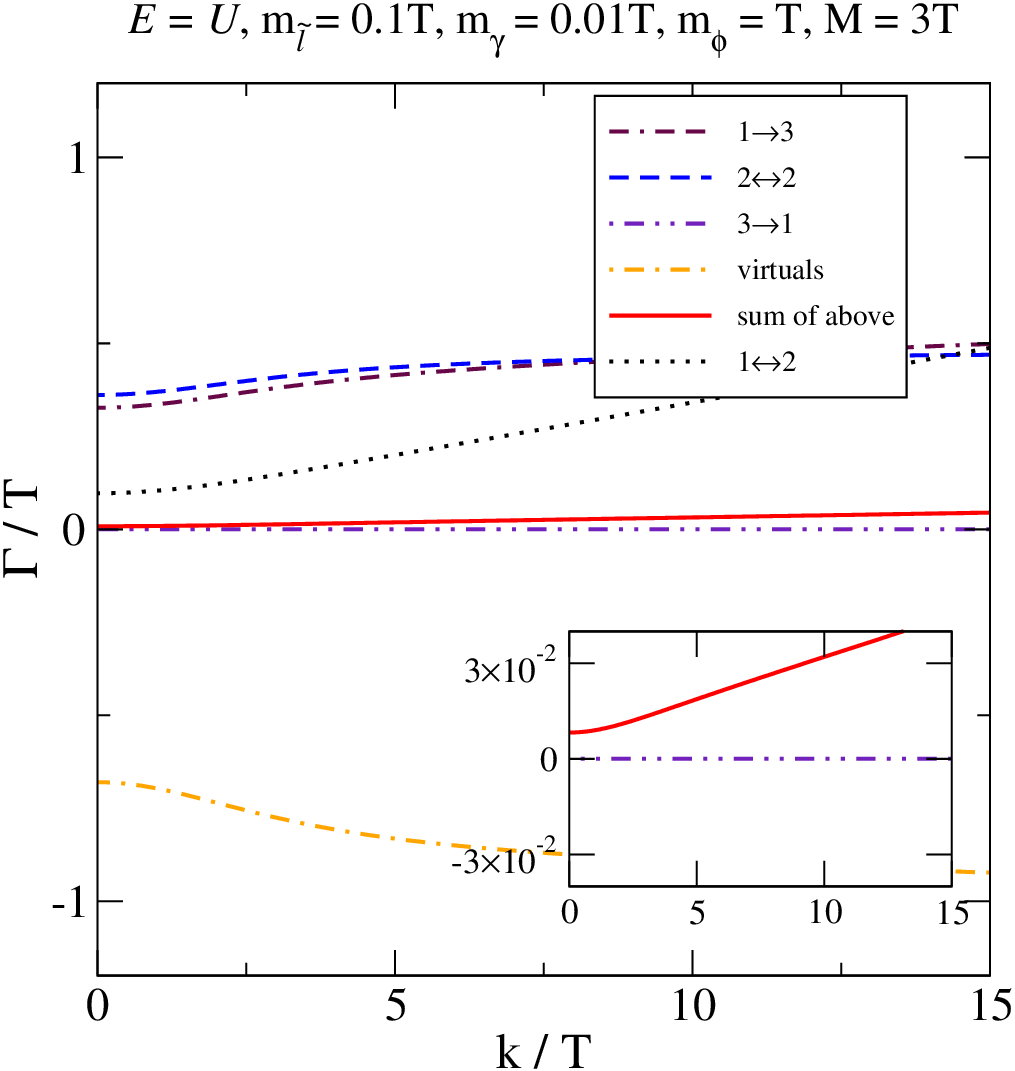}%
}

\caption[a]{\small
 Examples of  
 $
  \Gamma^\rmii{Born}_{2\leftrightarrow 2,1\leftrightarrow 3} 
 $,
 with the matrix element from \eq\nr{example_1to3_alt},
 and
 $
   \Delta\Gamma^\rmii{Born}_{1 \leftrightarrow 2 }
 $,
 from \eq\nr{ex_virtual}.    
 The sum of these two is compared with
 $
  \Gamma^\rmii{Born}_{1\leftrightarrow 2} 
 $, 
 with the matrix element 
 $
   \Theta^{ }_{ }(\P^{ }_\ala,\P^{ }_\aS) = 4 \E\cdot\P^{ }_\ala
 $.
 The mass is   
 $M = 0.3T$ (left) and $M = 3T$ (right), with  
 the other parameters explained at the end of 
 \se\ref{se:code}. 
 The top row shows results for $\E = \K$ (normalizing to $T^2$),  
 the bottom row for $\E = \U$ (normalizing to $T$). 
 The results are for $\mQ^{ } = 0.01T$, noting that real and 
 virtual corrections separately depend strongly on this IR regulator, 
 however the full result (solid line) 
 is independent of it. 
}

\la{fig:example2to2}
\end{figure}
%%%%%%%%%%%%%%%%%%%%%%%%%%%%%%%%%%%%%%%%%%%%%%%%%%%%%%%%%%%%%%%%%%%%%%%%%%%

%%%%%%%%%%%%%%%%%%%%%%%%%%%%%%%%% FIGURE %%%%%%%%%%%%%%%%%%%%%%%%%%%%%%%%%
\begin{figure}[p]

\hspace*{-0.1cm}
\centerline{%
 \epsfysize=7.5cm\epsfbox{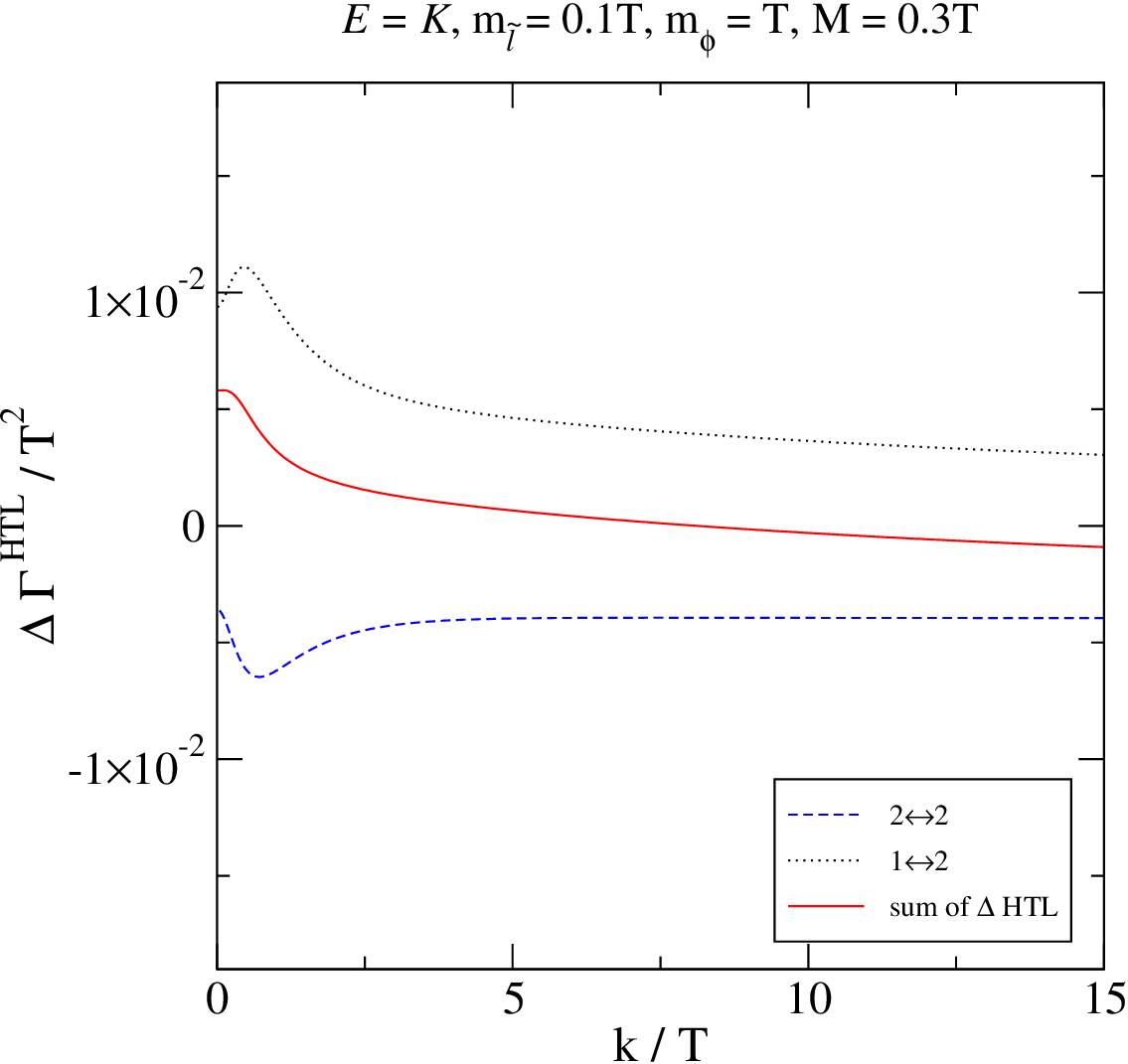}%
 \hspace{0.5cm}%
 \epsfysize=7.5cm\epsfbox{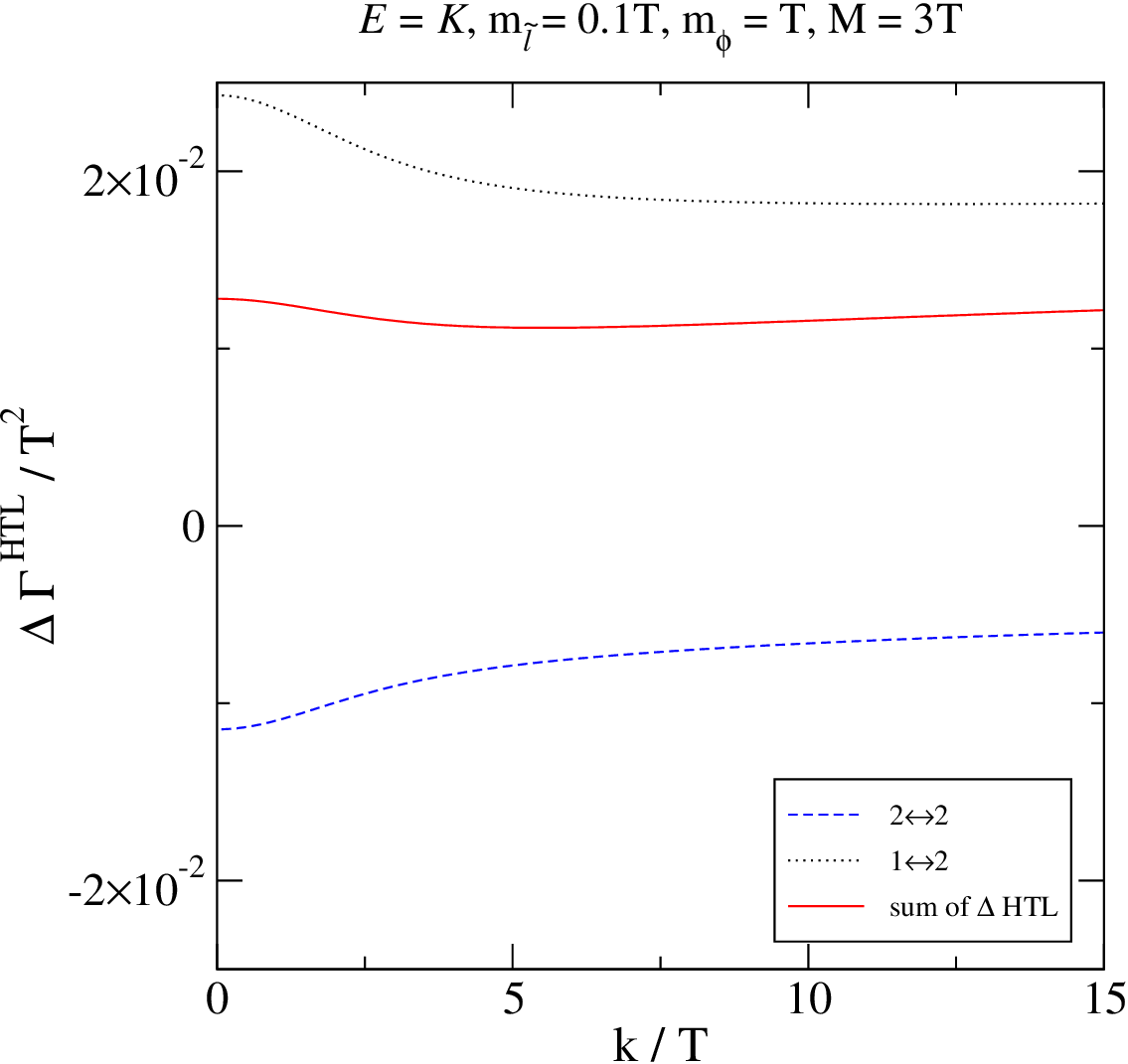}%
}

\vspace*{0.4cm}

\hspace*{-0.1cm}
\centerline{%
 \epsfysize=7.5cm\epsfbox{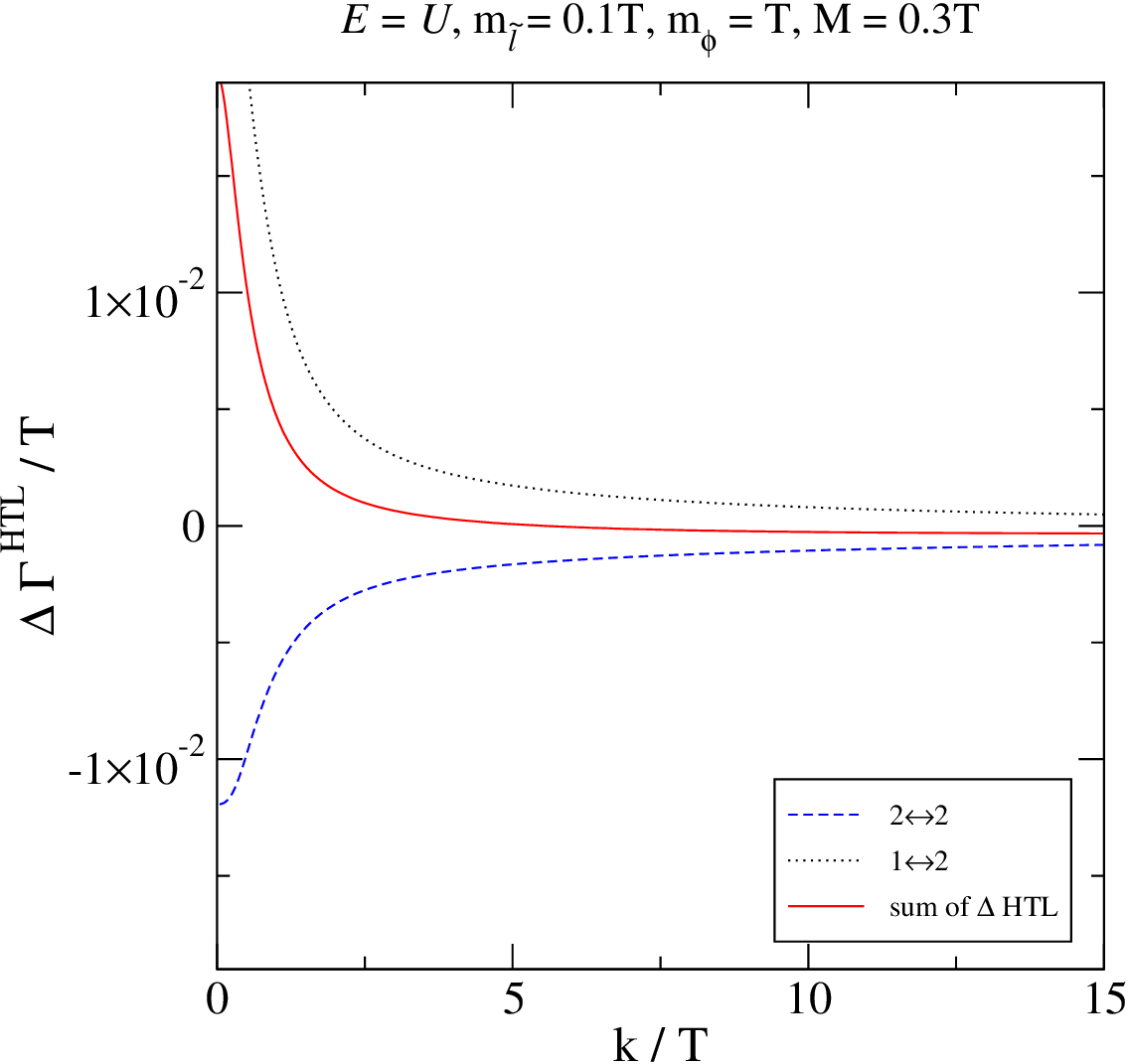}%
 \hspace{0.5cm}%
 \epsfysize=7.5cm\epsfbox{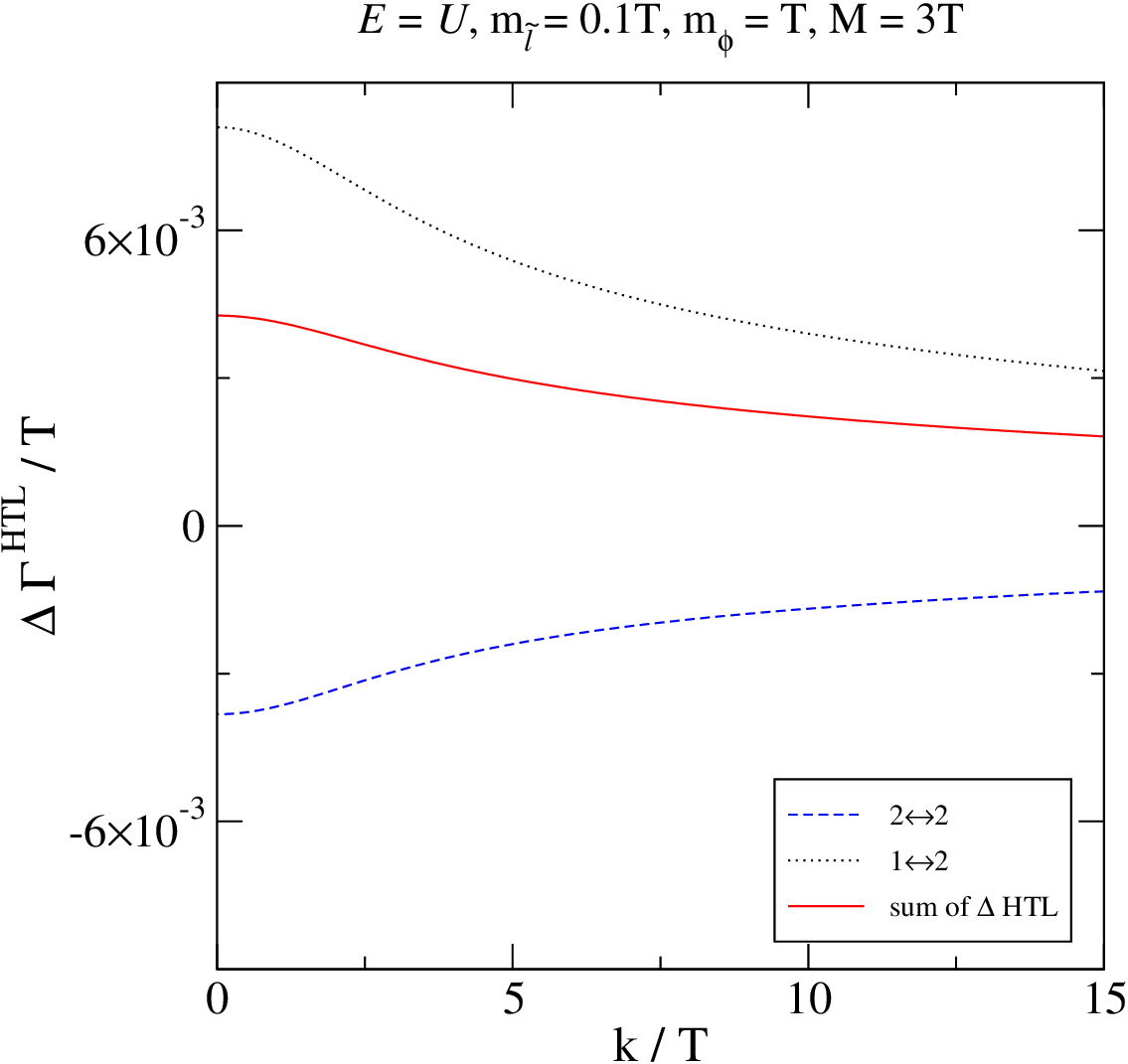}%
}

\caption[a]{\small
 Examples of a numerical evaluation of 
 $
  \Delta \Gamma^\rmii{HTL}_{2\leftrightarrow 2} 
 $,
 from \eq\nr{2to2_htl},  and
 $
  \Delta \Gamma^\rmii{HTL}_{1\leftrightarrow 2} 
 $, 
 from \eq\nr{1to2_htl}, 
 for $M = 0.3T$ (left) and $M = 3T$ (right).  
 The top row shows results for $\E = \K$ (normalizing to $T^2$),  
 the bottom row for $\E = \U$ (normalizing to $T$), with 
 the other parameters explained at the end of 
 \se\ref{se:code}. 
 The purpose of these corrections is to replace the auxiliary
 mass $m^{ }_\bla  = 0.1T $ through the physical thermal lepton mass 
 $\mellT^{ }  \approx 0.3T $, which parametrizes HTL propagators.
 The $2\leftrightarrow 2$ and 
 $1\leftrightarrow 2$ corrections depend strongly 
 on the IR regulator $m^{ }_\bla$, 
 however the full result (solid line) 
 is almost independent of it, as long as we stay away from the constrained 
 domain $M \in (m^{ }_\cS - m^{ }_\bla, m^{ }_\cS + m^{ }_\bla)$.
}

\la{fig:example_subtractions}
\end{figure}
%%%%%%%%%%%%%%%%%%%%%%%%%%%%%%%%%%%%%%%%%%%%%%%%%%%%%%%%%%%%%%%%%%%%%%%%%%%

%%%%%%%%%%%%%%%%%%%%%%%%%%%%%%%%% FIGURE %%%%%%%%%%%%%%%%%%%%%%%%%%%%%%%%%
\begin{figure}[p]

\hspace*{-0.1cm}
\centerline{%
 \epsfysize=7.5cm\epsfbox{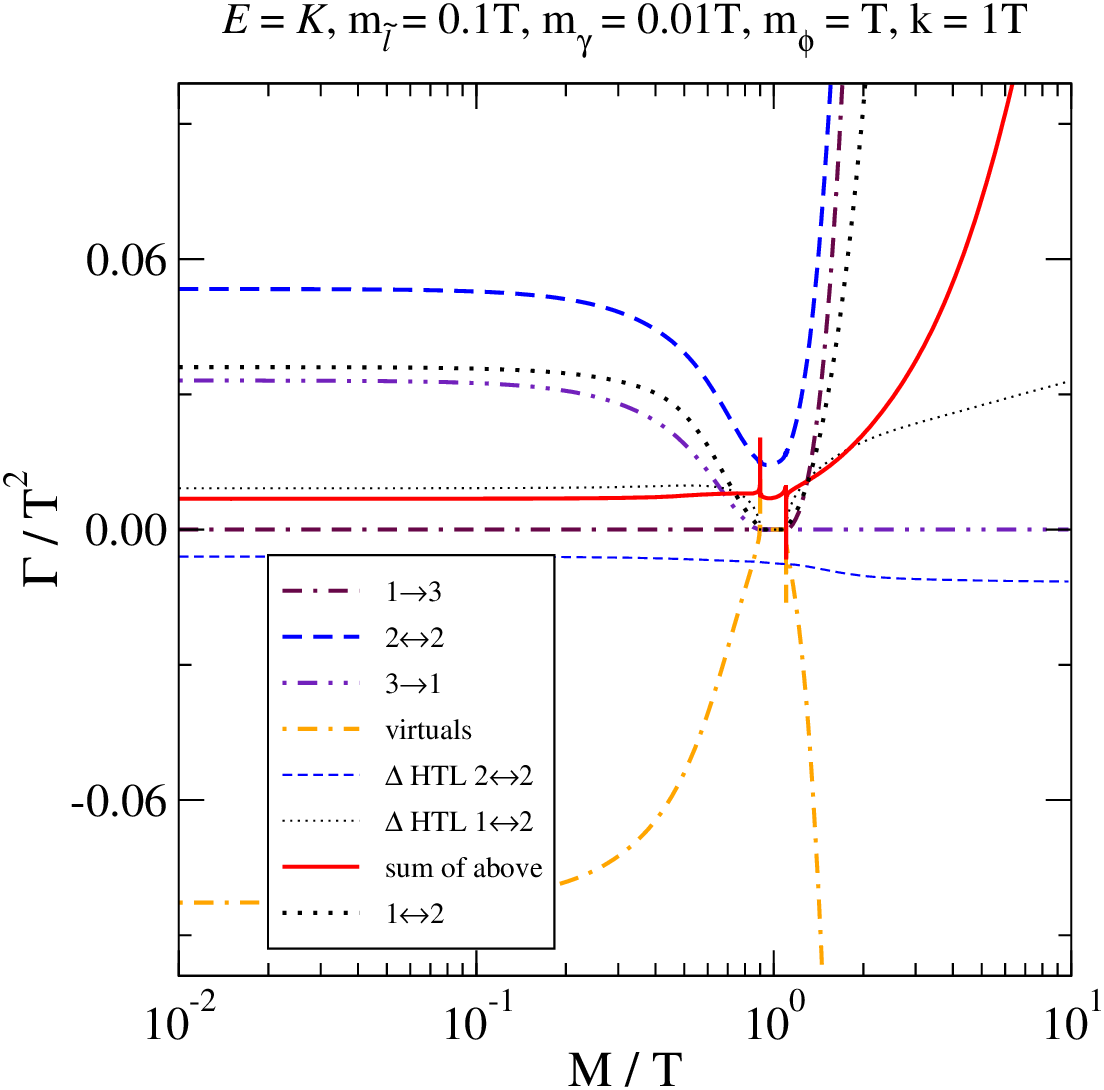}%
 \hspace{0.5cm}%
 \epsfysize=7.5cm\epsfbox{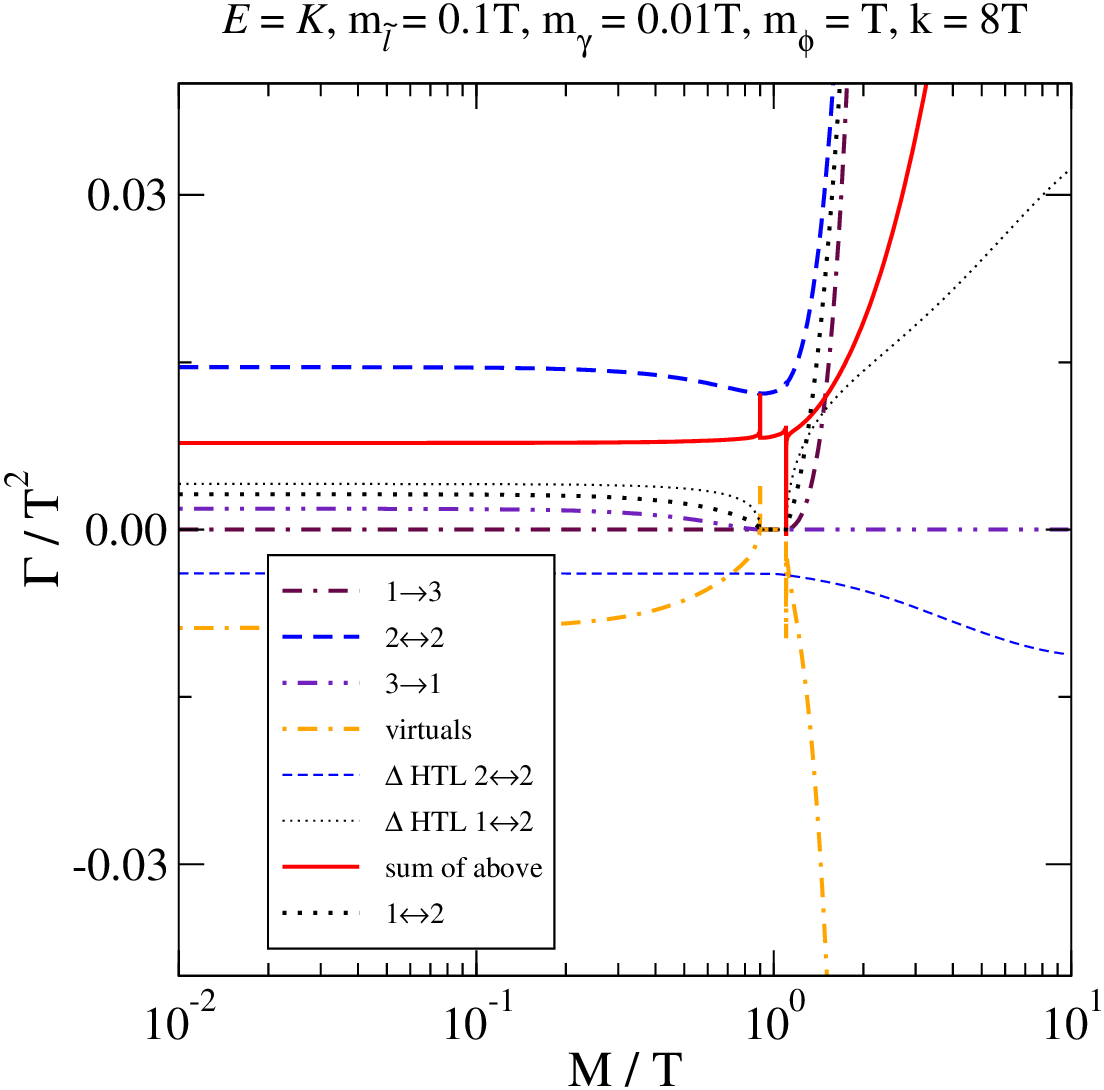}%
}

\vspace*{0.4cm}

\hspace*{-0.1cm}
\centerline{%
 \epsfysize=7.5cm\epsfbox{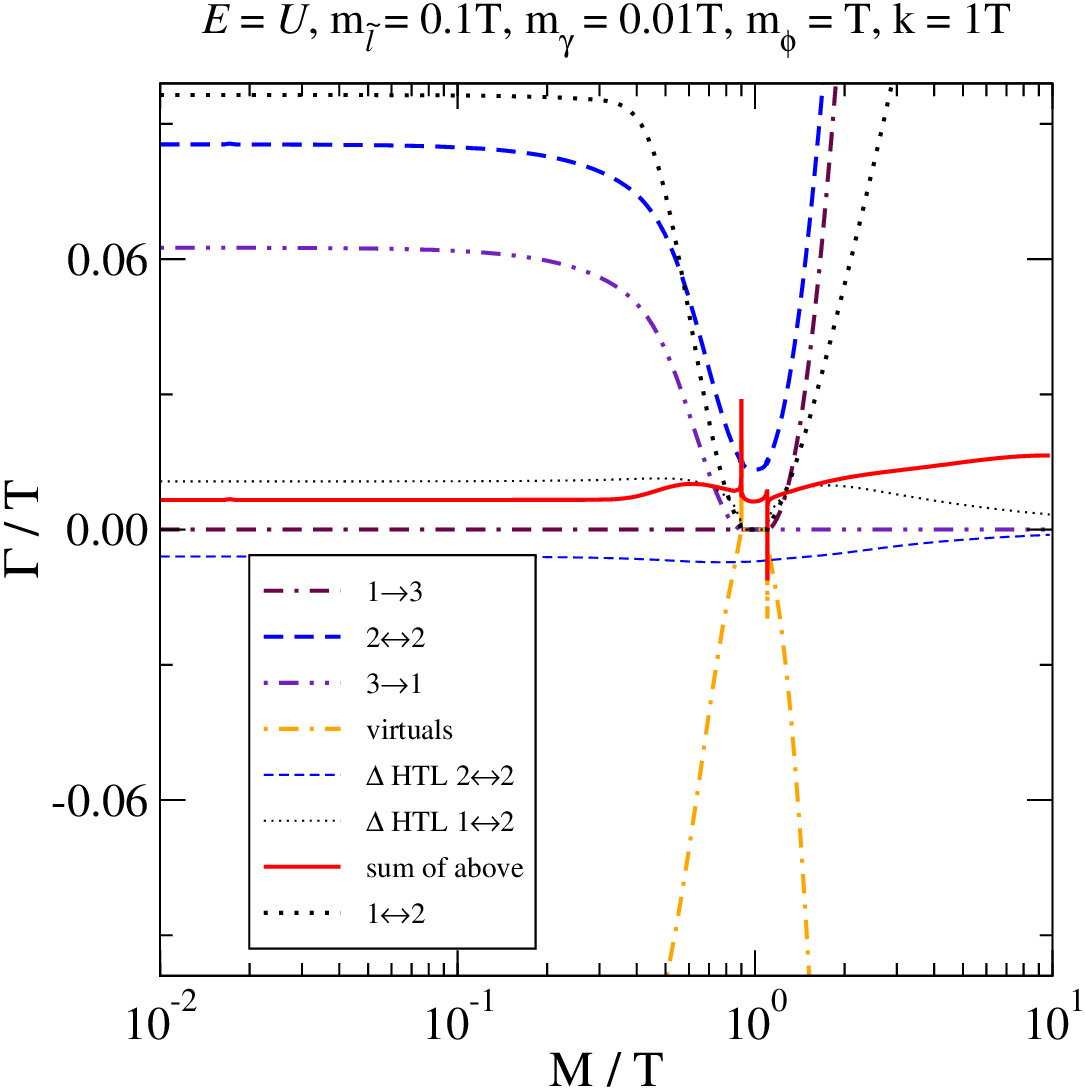}%
 \hspace{0.5cm}%
 \epsfysize=7.5cm\epsfbox{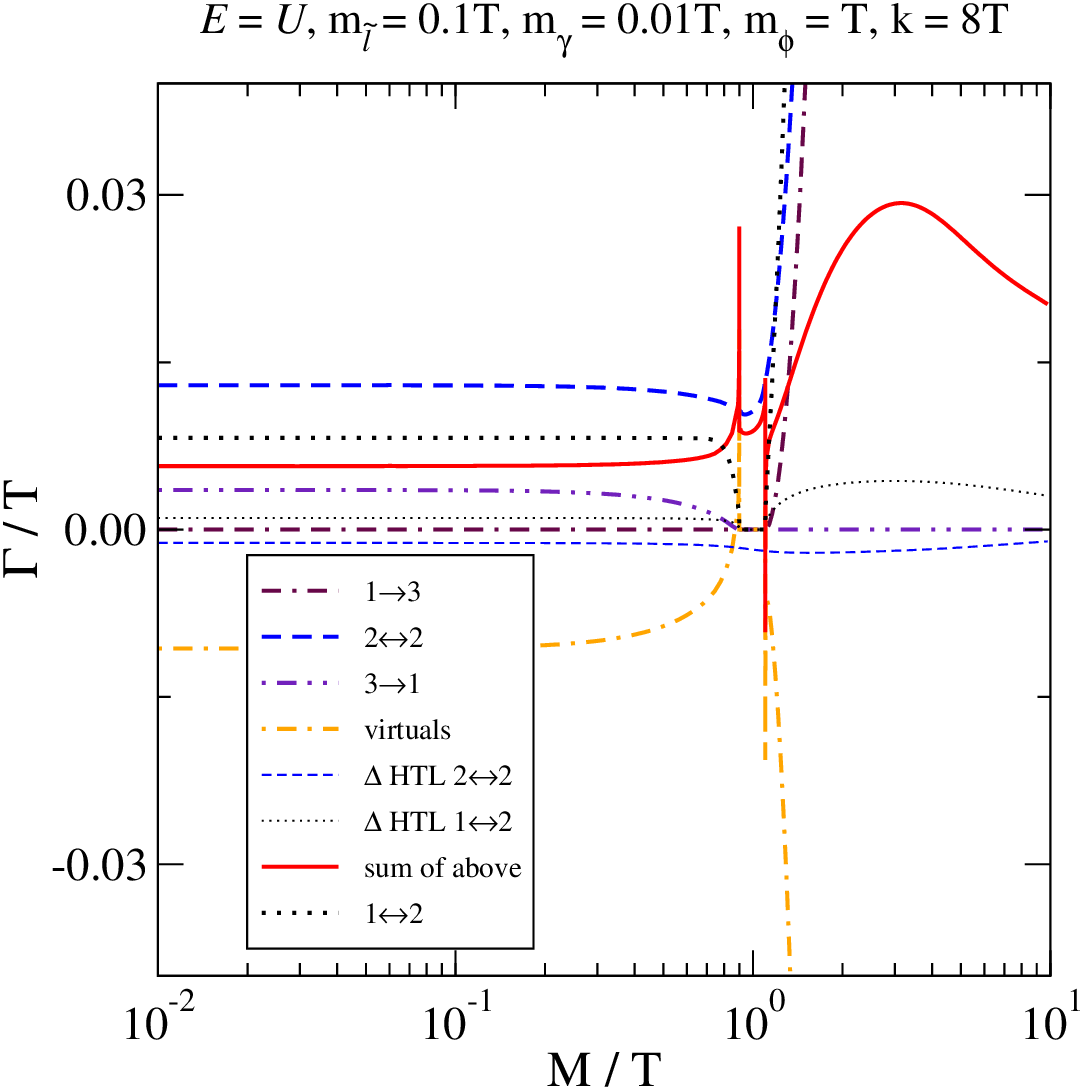}%
}

\caption[a]{\small
 Examples of 
 $
  \Gamma^\rmii{Born}_{2\leftrightarrow 2,1\leftrightarrow 3}
 + 
  \Delta\Gamma^\rmii{Born}_{1 \leftrightarrow 2 }
 + 
  \Delta\Gamma^\rmii{HTL}_{2\leftrightarrow 2,1\leftrightarrow 2}
 $ 
 from \eq\nr{full},
 as a function of $M$, for $k = T$ (left) and $k = 8T$ (right).  
 The masses are 
 $ 
  m^{ }_{\bla} = m^{ }_{\ala} = 0.1 T 
 $, 
 $
  m^{ }_{\cQ} = 0.01 T
 $,  
 $ 
  m^{ }_{\bS} =  m^{ }_{\cS} = T 
 $.
 The spikes at around $M = m^{ }_{\cS}\pm m^{ }_\bla$
 result from an incomplete cancellation 
 between the unresummed
 $
  \Gamma^\rmii{Born}_{2\leftrightarrow 2,1\leftrightarrow 3}
 + 
  \Delta\Gamma^\rmii{Born}_{1 \leftrightarrow 2 }
 $
 and the HTL subtraction-addition contribution
 $
  \Delta\Gamma^\rmii{HTL}_{2\leftrightarrow 2,1\leftrightarrow 2}
 $,  
 as the latter cannot remove threshold singularities 
 that originate from non-HTL structures, particularly
 the virtual double-pole corrections (last line of \eq\nr{ex_virtual}). 
 The spikes could be eliminated by sending $m^{ }_{\bla} \to 0$, 
 whereby the physical thermal lepton mass $\mellT^{ }  \approx 0.3T$ 
 is reinstated by the HTL contribution.
}

\la{fig:imDELTA}
\end{figure}
%%%%%%%%%%%%%%%%%%%%%%%%%%%%%%%%%%%%%%%%%%%%%%%%%%%%%%%%%%%%%%%%%%%%%%%%%%%

Finally, even if we are able to produce accurate results for 
several test cases, it is appropriate to acknowledge that numerical
integrations become less efficient in certain limits. Roughly speaking, 
the integrations are simple if all masses and momenta are of order $T$, 
whereas large scale hierarchies are challenging to handle. 
Examples are
the non-relativistic limit $M \gg T$, where a huge cancellation
between real and virtual corrections demands exquisite numerical 
precision, 
and the ultrarelativistic regime $M \ll T$, particularly with $k \gg T$, 
where the literal integration ranges can be broad but the integrands 
are strongly localized. 
Moreover, particle spectra leading to 
real poles can be costly, if principal value
integrations are regularized numerically rather than analytically 
(cf.\ \ses\ref{se:IR} and \ref{sss:ris}). We cannot exclude 
numerical inaccuracies if parameters are pushed to 
domains which happen to have eluded our tests. 

%%%%%%%%%%%%%%%%%%%%%%%%%%% SECTION %%%%%%%%%%%%%%%%%%%%%%%%%%%%%%%%%%%%%%

\section{Conclusions and outlook}
\la{se:concl}

We have described a method to represent 
and evaluate thermal $2\leftrightarrow 2$
and $1\leftrightarrow 3$ scattering rates, 
including a way to regularize and 
subtract the poles that appear in the matrix elements squared. 
Choosing a language in which the side of the non-equilibrium particle
is called the initial state 
(even if both processes and inverse processes are always included), 
the idea
is to give as input a $1\to 3$ matrix element squared, which
displays maximal symmetries, as all thermalized particles are
in the final state (the process does not need to be kinematically
allowed). The $2\leftrightarrow 2$ and $3\to 1$
scattering rates are obtained by crossing relations, 
and the virtual corrections to $1\leftrightarrow 2$ rates 
that cancel their poles are 
automatically identified (cf.\ \se\ref{ss:virtual}). 
Vacuum contributions can be pulled apart, so that 
the final step is to carry out an exponentially 
convergent three-dimensional numerical integral. The results
have been worked out for general chemical potentials, and can
therefore be applied not only to cosmology but, potentially, 
to dense astrophysical environments as well. 

In our framework, the full interaction rate can be represented as 
\be
  \Gamma \,\approx\, 
  \Gamma^\rmi{Born}_{2\leftrightarrow 2,1\leftrightarrow 3}
 + 
  \Delta\Gamma^\rmi{Born}_{1 \leftrightarrow 2 }
 + 
  \Delta\Gamma^\rmii{HTL}_{2\leftrightarrow 2,1\leftrightarrow 2}
 + 
  \Gamma^\rmii{LPM}_{1+n \leftrightarrow 2+n }
 \;. \la{full}
\ee
The first part, 
$
  \Gamma^\rmi{Born}_{2\leftrightarrow 2,1\leftrightarrow 3}
$, 
captures  
$2\leftrightarrow 2$
and $1\leftrightarrow 3$ rates, 
and the second, 
$
   \Delta\Gamma^\rmi{Born}_{1 \leftrightarrow 2 }
$, 
the virtual corrections that cancel their poles. 
The third part, 
$
  \Delta\Gamma^\rmii{HTL}_{2\leftrightarrow 2,1\leftrightarrow 2}
$, 
represents a subtraction-addition step implementing HTL resummation
(cf.\ \ses\ref{sss:regge} and \ref{sss:lpm}), 
which replaces auxiliary masses, 
used as an intermediate IR regulator, 
through physical thermal masses. The last part, 
$
  \Gamma^\rmii{LPM}_{1+n \leftrightarrow 2+n } 
$,
with $n\ge 0$,  
sums together $ 1+n \leftrightarrow 2+n $ processes. In the present
paper, 
$
  \Gamma^\rmii{LPM}_{1+n \leftrightarrow 2+n } 
$ 
has been approximated through its lowest-order term
($n=0$), 
$
  \Gamma^\rmi{Born}_{1 \leftrightarrow 2 } 
$, 
given that 
the corresponding formalism
has not been generalized to the case that some particles
cease to be ultrarelativistic 
(in \figs\ref{fig:example2to2}--\ref{fig:imDELTA} 
we have chosen $m^{ }_{\cS} = T$ to be
relatively ``heavy'', whereby restricting to $n=0$
should be a fair approximation).

The numerical importance of 
the virtual corrections can be appreciated from 
\fig\ref{fig:example2to2}(right).
Due to the presence of logarithmic and double-logarithmic 
IR divergences~\cite{master}, the inclusion of only $2\leftrightarrow 2$
scatterings would overestimate the correct result by a factor $\sim 10^3$
for these parameters. Moreover, $1\leftrightarrow 3$ rates, 
which are often overlooked, play an equally important role as 
$2\leftrightarrow 2$ scatterings. When virtual corrections 
are included, the result becomes much smaller than that from 
$1\leftrightarrow 2$ processes, and can be safely omitted. 
This conclusion is not changed by HTL resummation, 
as illustrated in \fig\ref{fig:example_subtractions}(right). 

The situation is very different for the parameters in 
\fig\ref{fig:example2to2}(left), where the non-equilibrium 
particle is ultrarelativistic ($M=0.3T$). Even though there is still 
a substantial cancellation between $2\leftrightarrow 2$ and 
$3\leftrightarrow 1$ rates and virtual corrections, 
the remainder is now larger than 
that from the $1\leftrightarrow 2$ processes, 
provided that $k \;\gsim\; (\mbox{a few}) \times T$.
This conclusion is not 
changed by HTL resummation, whose influence is smaller
at similar momenta, as is illustrated in 
\fig\ref{fig:example_subtractions}(left). 
Furthermore 
the conclusion turns out to be strengthened if $m^{ }_{\cS}$ 
is reduced towards its physical high-temperature value,  
$m^{ }_{\cS}\simeq 0.4T$.

To summarize, 
if the non-equilibrium particle is ultrarelativistic ($M \ll T$),
it is essential for quantitative investigations to include  
$2\leftrightarrow 2$ and $1\leftrightarrow 3$ rates as well as
virtual corrections to $1\leftrightarrow 2$ ones. 
If the non-equilibrium particle is non-relativistic ($M \gg T$), 
it would be dangerous to incorporate
$2\leftrightarrow 2$ or $1\leftrightarrow 3$ rates, 
{\em without} a full account of the
virtual corrections to $1\leftrightarrow 2$ processes that may cancel 
most of the result.\footnote{% 
 The latter statement applies as such
 to a mass spectrum similar to that in our example, where the 
 non-equilibrium particle can be heavier than the plasma particles
 and experiences only inelastic reactions. 
 If it can participate in elastic scatterings, or if one of the plasma
 particles has a mass close to that of the non-equilibrium one, 
 $2\leftrightarrow 2$ reactions can dominate even 
 in the non-relativistic regime. 
 } 

A {\sc c} code 
implementing the numerical parts of our procedure, 
and a {\sc form} code implementing the algebraic ones,  
are attached as ancillary files
to this paper.  
Even if their details are specific to the example 
in \eq\nr{example_1to3_alt},
relevant for leptogenesis scenarios, 
the structures and main steps are quite general.
Therefore we hope that they 
can be applied to other problems as well,
for instance in the context of freeze-in dark matter production, 
where the need for determining 
$2\leftrightarrow 2$
and $1\leftrightarrow 3$ scattering rates
has been underlined recently~\cite{freezein}. 

%%%%%%%%%%%%%%%%%%%%%%%%%%% SECTION %%%%%%%%%%%%%%%%%%%%%%%%%%%%%%%%%%
%
\section*{Acknowledgements}

G.J.\ was supported by the U.S.\
Department of Energy under Grant No.\ DE-FG02-00ER41132, and 
M.L.\ by the Swiss National Science Foundation
(SNF) under grant 200020B-188712.

%%%%%%%%%%%%%%%%%%%%%%% APPENDIX %%%%%%%%%%%%%%%%%%%%%%%%%%%%%%%%%%%
%
\appendix
\renewcommand{\thesection}{\Alph{section}} % {Appendix~\Alph{section}}
\renewcommand{\thesubsection}{\Alph{section}.\arabic{subsection}}
\renewcommand{\theequation}{\Alph{section}.\arabic{equation}}
%
%%%%%%%%%%%%%%%%%%%%%%%%%% SECTION %%%%%%%%%%%%%%%%%%%%%%%%%%%%%%%%%%%%%%%%%%

\section{Proof of thermal crossing relations}
\la{app:proof}

%%%%%%%%%%%%%%%%%%%%%%%%%%% SUBSECTION %%%%%%%%%%%%%%%%%%%%%%%%%%%%%%%%
%
\subsection{$2\leftrightarrow 2$ and $1\leftrightarrow 3$ real corrections}

In order to handle any $2\leftrightarrow2$ or $1\leftrightarrow 3$
reaction, we start by introducing the concept of a ``master'' sum-integral.
The master sum-integral originates by viewing the interaction rate as 
an imaginary part (``cut'') of a retarded correlator, 
in analogy with the optical theorem.
The retarded correlator can in turn
be represented as an analytic continuation of an imaginary-time (Euclidean)
correlator. This representation, even if sounding formal,
is quite helpful, 
as the imaginary-time expression automatically
encodes many independent reactions as well as the 
crossing symmetries between them. In particular, 
each such master structure is IR finite by itself,
containing no poles in accordance with the 
KLN theorem~\cite{kln1,kln2}. 

As can readily be verified pictorially, 
$2\leftrightarrow2$ and $1\leftrightarrow 3$ processes correspond to 
cuts of 2-loop diagrams. Any 2-loop contribution can, in turn, 
be represented as a linear combination of master sum-integrals. 
Inspired by ref.~\cite{asz}, we define 
a 2-loop master sum-integral as 
\ba
 && \hspace*{-1.5cm}
 I^{j_1\cdots j_6 }_{i_1\cdots i_6}(a^{ }_1,\ldots,a^{ }_6) 
 \nn 
 & \equiv & 
   \Tint{P,Q} \!\! 
   \frac{j^{ }_1 H_{i\tilde P}
        +j^{ }_2 H_{i\tilde Q}
        +j^{ }_3 H_{i(\tilde P-\tilde Q)} 
        +j^{ }_4 H_{-i(\tilde K+\tilde P)}
        +j^{ }_5 H_{-i(\tilde K+\tilde Q)}
        +j^{ }_6 H_{-i\tilde K} }
   { 
     \Delta^{i_1}_{P;a^{ }_1}
     \,\Delta^{i_2}_{Q;a^{ }_2}
     \,\Delta^{i_3}_{P-Q;a^{ }_3}
     \,\Delta^{i_4}_{-K-P;a^{ }_4}
     \,\Delta^{i_5}_{-K-Q;a^{ }_5}
     \,\Delta^{i_6}_{-K;a^{ }_6}
   }
 \;. \la{master_def} 
\ea
Here $\Tinti{P} \equiv T \sum_{p_n} \int_\vc{p}$ is a Matsubara 
sum-integral, with $p^{ }_n$ referring to a bosonic or 
fermionic Matsubara frequency, and $P \equiv (p^{ }_n,\vc{p})$.
The imaginary-time external four-momentum is denoted by $K$,
the corresponding Minkowskian four-momentum by $\K$.
The indices $i^{ }_x,j^{ }_x$ are integers, whereas 
the $a^{ }_x$ label particle species
($a^{ }_x \in \{ h,W,Z,Q,\nu^{ }_a,e^{ }_a,u,d,...\}$). 
Masses and chemical potentials appear through  
\be
 \Delta^{ }_{P;a_x} 
 \; \equiv \; (p^{ }_n + i \mu^{ }_{a_x})^2 + p^2 + m_{a_x}^2
 \;, \quad
 \Delta^{ }_{P;-a_x} 
 \; \equiv \; (p^{ }_n - i \mu^{ }_{a_x})^2 + p^2 + m_{a_x}^2
 \;, \la{prop}
\ee
where $\mu^{ }_{a_x}$ is the chemical potential, 
$m^{ }_{a_x}$ is the mass, 
and $p \equiv |\vc{p}|$. 
This notation implies that 
$
 \Delta^{ }_{-P;a_x} 
  = 
 \Delta^{ }_{P;-a_x}
$, 
where $-a^{ }_x$ labels an antiparticle. 
Momenta denoted by 
$\tilde P \equiv (p^{ }_n + i \mu^{ }_{a_x},\vc{p})$ 
include a shift by the chemical potential. 
The $H$ factors in the numerator denote
helicity projections, for instance for spin-$\frac{1}{2}$ particles
\be
 H^{ }_{i\tilde P} = \mathcal{E} \cdot i\tilde P
 \;, \la{ex_H}
\ee
where $\mathcal{E}$ is some external four-momentum, 
e.g.\ $\K$ or the medium four-velocity $\U$. 
For spin-1 fields the projectors are 
quadratic in momenta, for instance 
$H^{ }_{i\tilde P} = p^2 - (\vc{p}\cdot\vc{k})^2/k^2$ 
for the sum over transverse polarization of on-shell photons.
As the helicity projection is a linear operation, we have 
assumed a linear dependence on $H$ in \eq\nr{master_def}; 
this is a simplification, even if we believe that the result
holds more generally.  

Once the Matsubara sums are carried out, \eq\nr{master_def} contains 
spatial momentum
integrals, weighted by Bose-Einstein and Fermi-Dirac distribution functions. 
We now analytically continue $\tilde k^{ }_n$
to a Minkowskian frequency, and take the cut. 
The complete cut involves virtual
corrections to $1\leftrightarrow 2$ scatterings, as well as 
real processes, namely $2\leftrightarrow 2$ and $1\leftrightarrow 3$ 
scatterings. We first focus on the real processes, 
deferring virtual corrections to appendix~A.2.

Consider, for instance, a cut illustrated as 
\ba
 \nn[5mm] 
 \Cut
 \hspace*{1.2cm} \;. \nonumber 
\ea
The corresponding expression reads
\ba
 & & \hspace*{-1.5cm} 
 \im I^{0 j^{ }_2 j^{ }_3 j^{ }_4 0 j^{ }_6}
      _{i^{ }_1\! 1\, 1\, 1\, i^{ }_5 i^{ }_6}
        ( a^{ }_1,a^{ }_2,a^{ }_3,a^{ }_4,a^{ }_5,a^{ }_6)
 \bigr|^{ }_{\tilde k_n \to -i [\omega + i0^+]}
 \nn[2mm] 
 & \supset & 
 \bigl[ 
  \, \scat{1\to3}(a^{ }_2,a^{ }_3,a^{ }_4)
 \nn 
 & & \; 
 +\, \scat{2\to2}(-a^{ }_2;a^{ }_3,a^{ }_4)
 +\, \scat{2\to2}(-a^{ }_3;a^{ }_4,a^{ }_2)
 +\, \scat{2\to2}(-a^{ }_4;a^{ }_2,a^{ }_3)
 \nn 
 & & \; 
 +\, \scat{3\to1}(-a^{ }_2,-a^{ }_3;a^{ }_4)
 +\, \scat{3\to1}(-a^{ }_3,-a^{ }_4;a^{ }_2)
 +\, \scat{3\to1}(-a^{ }_4,-a^{ }_2;a^{ }_3)
 \bigr]
 \nn[2mm] 
 & & \; \times 
  \, \frac{  j^{ }_2 H^{ }_{ -\P^{ }_2 }
           + j^{ }_3 H^{ }_{ -\P^{ }_3 }
           + j^{ }_4 H^{ }_{ -\P^{ }_4 }
           + j^{ }_6 H^{ }_{ -\K }
          }{
         ({ \Sq{\s{23}}{m_{a_1}} })^{i_1}
         ({ \Sq{\s{34}}{m_{a_5}} })^{i_5}
         ({ \Sq{\MM}{m_{a_6}} })^{i_6}
          } 
 \;, \la{cut}
\ea
where we have denoted $\P^{ }_i \equiv \P^{ }_{a_i}$
and $s^{ }_{ij} \equiv (\P^{ }_i + \P^{ }_j)^2$.

Eq.~\nr{cut} shows that 
the cases $\scat{2\to2}$ and $\scat{3\to1}$ can be 
obtained from $\scat{1\to3}$ by pulling one or
two legs to the initial state, and inverting 
the signs of the corresponding momenta and chemical potentials,  
thereby proving \eq\nr{master} for a special example. 
The notation in \eq\nr{n_sigma} automatically takes care of 
minus signs associated with fermionic legs. 

Finally, consider the other diagonal cut, going through the lines 
$a^{ }_1,a^{ }_3,a^{ }_5$. By substituting 
$P \leftrightarrow Q$ in \eq\nr{master_def}, the result can be 
directly obtained from \eq\nr{cut}, just by inverting 
$a^{ }_3\to -a^{ }_3$ and setting $\P^{ }_3 \to -\P^{ }_3$
in the numerator. This confirms \eq\nr{structure_master}.

%%%%%%%%%%%%%%%%%%%%%%%%%%% SUBSECTION %%%%%%%%%%%%%%%%%%%%%%%%%%%%%%%%
%
\subsection{$1\leftrightarrow 2$ virtual corrections}

% Motivated by \eqs\nr{1to2_dir_final} and \nr{1to2_indir_final}
In analogy with the 2-loop master in \eq\nr{master_def}, we define
a 1-loop master sum-integral as  
\be
 J^{j_1 j_2 j_3}_{i_1 i_2 i_3}(a^{ }_1,a^{ }_2,a^{ }_3) 
 \; \equiv \; 
 \Tint{P} 
    \frac{j^{ }_1 H_{i\tilde P}
                    + j^{ }_2 H_{-i(\tilde K+\tilde P)}
                    + j^{ }_3 H_{-i\tilde K} }
   { 
     \Delta^{i_1}_{P;a^{ }_1}
     \,\Delta^{i_2}_{-K-P;a^{ }_2}
     \,\Delta^{i_3}_{-K;a^{ }_3}
   }
 \;, \la{master1}
\ee
where the propagator structures are defined according to \eq\nr{prop}. 

As originally demonstrated with scalar field theory~\cite{art}, 
the imaginary part of \eq\nr{master1} can be expressed in terms
of phase space integrals, in particular  
\ba
%% & & \hspace*{-1.5cm}
 \im J^{  j^{ }_1 j^{ }_2 j^{ }_3}_{1\, 1\, i^{ }_3}
        ( a^{ }_1,a^{ }_2,a^{ }_3 ) 
        \bigr|^{ }_{\tilde k_n \to -i [\omega + i0^+]}
 \!\! & = &  
  \scat{1\leftrightarrow2}(a^{ }_1,a^{ }_2)
  \, \frac{  j^{ }_1 H^{ }_{ - \P^{ }_1 }
           + j^{ }_2 H^{ }_{ - \P^{ }_2 }
           + j^{ }_3 H^{ }_{ -\K }
          }{
         ({ \Sq{\MM}{m_{a_3}} })^{i_3}
          } 
 \;. \hspace*{6mm} \la{ex}
\ea
Here we have set
$i^{ }_1 = i^{ }_2 = 1$; results for higher powers can be obtained 
by taking derivatives with respect to the masses $m^2_{a_1}$ and
$m^2_{a_2}$. This motivates \eq\nr{scat1lr2}. 

In order to verify \eq\nr{virtual_bubble}, 
we may start from \eq\nr{cut} but
choose, for instance, $i^{ }_5 = 0$, 
eliminating one of the poles. Simplifying other 
index choices as well, the part corresponding to 
$2\leftrightarrow 2$ and $1\leftrightarrow 3$ processes
then amounts to 
\ba
 & & \hspace*{-1.5cm} 
 \im I^{0 j^{ }_2 j^{ }_3 j^{ }_4 0 j^{ }_6}
      _{1\, 1\, 1\, 1\, 0\, 0}
        ( a^{ }_1,a^{ }_2,a^{ }_3,a^{ }_4,a^{ }_5,a^{ }_6)
 \bigr|^{ }_{\tilde k_n \to -i [\omega + i0^+]}
 \nn[2mm] 
 & \supset &  
  \scat{1\to3}(a^{ }_2,a^{ }_3,a^{ }_4)
  \, \frac{  j^{ }_2 H^{ }_{ -\P^{ }_2 }
           + j^{ }_3 H^{ }_{ -\P^{ }_3 }
           + j^{ }_4 H^{ }_{ -\P^{ }_4 }
           + j^{ }_6 H^{ }_{ -\K }
          }{
         ({ \Sq{\s{23}}{m_{a_1}} }) % ^{i_1}
          } 
 \;, \la{cut2}
\ea
with the other channels following by crossings according to \eq\nr{cut}.
We note that this channel has a pole of the type 
in \eq\nr{residue1}, with a residue $-1$. 
It is then sufficient to work out
the corresponding virtual contributions, i.e.\ the ones in which
the lines $a^{ }_1$ and $a^{ }_4$ are cut. This yields 
precisely the structure in \eq\nr{virtual_bubble}, 
with the same overall $-1$. 

Finally, for \eq\nr{virtual_triangle}, we need to consider two 
possibilities, cutting the lines $a^{ }_1,a^{ }_4$ on one
hand, and $a^{ }_2,a^{ }_5$ on the other. According to 
\eq\nr{master_def}, the results can be related to each other
through the substitution $P \leftrightarrow Q$. 
In any case, \eq\nr{virtual_triangle} can be confirmed.

%%%%%%%%%%%%%%%%%%%%%%%%%%% SECTION %%%%%%%%%%%%%%%%%%%%%%%%%%%%%%%%%%
%
\section{Further thermal averages}
\la{app:further}

In this appendix we supplement the procedure described in 
\se\ref{ss:phasespace_real} for $t$-channel $2\leftrightarrow 2$ scatterings, 
by working out thermal averages for the other channels
appearing in \eq\nr{master}. 

%%%%%%%%%%%%%%%%%%%%%%%%%%% SUBSECTION %%%%%%%%%%%%%%%%%%%%%%%%%%%%%%%%
%
\subsection{$1\to3$ reactions}
\la{app:1to3}

For $1\to 3$ reactions, all particles 
(apart from the ``external'' one, carrying the momentum~$\K$)
appear on one side and can thus be interchanged as
far as momentum labellings are concerned. Let us use this freedom 
to choose $s^{ }_{12} = (\P^{ }_1 + \P^{ }_2)^2$ 
as a potentially IR sensitive Mandelstam variable. If another
Mandelstam variable appears, it can be chosen as $s^{ }_{23}$
through further permutations. 
For the example in \eq\nr{example_1to3_alt}, 
with the momenta for $\scat{1\to3}(a^{ }_1,a^{ }_2,a^{} _3)$
labelled as $\P^{ }_i \equiv \P^{ }_{a_i}$, 
this gives 
\ba
 \frac{ \Gamma^\rmi{Born}_{1\to 3} }{  2 (g_1^2 + 3 g_2^2) }
 & \to & 
 \scat{1\to3}(\ala,\aQ,\aS) \, 
%  \nn[2mm] & \times & 
                \biggl\{ 
% \nn & & +
                      - \frac{\E\cdot\P^{ }_1}{\s{12} - m_\bla^2 }
                      + \frac{
                        (\MM - m_\cS^2 )\,
                        \E\cdot(
                        \P^{ }_2
                            + 
                         2 \P^{ }_1 )
                       }{ 
                         ({ \s{12} - m_\bla^2})
                         ({ \s{23} - m_{\tilde{\cS}}^2 }) }
                       \biggr\} 
%%%%%%%%%%%%%%%%
 \nn[2mm] 
 & + & 
 \scat{1\to3}(\aS,\aQ,\ala) \, 
%  \nn[2mm] & \times & 
                \biggl\{ 
                     - \frac{\E\cdot(\P^{ }_3 + \K)}
                       {\s{12} - m_{\tilde{\cS}}^2} 
                     - \frac{ 
                         2 m_\cS^2 \, \E\cdot \P^{ }_3    
                             }{ ({ \s{12} - m_{\tilde{\cS}}^2 })^2 }
                       \biggr\} 
 \;. \la{example_1to3p}  
\ea
 
The goal now is to have $\s{12}$ as an integration variable. 
To this end we introduce a four-momentum $\mathcal{Q}$ such 
that $\s{12} = \mathcal{Q}^2$, and write 
the phase space integration measure from \eq\nr{dO1to3} as   
\ba
 & & \hspace*{-1.0cm}
 \int \! {\rm d}\Omega^{ }_{1\to 3}
 \; = \; 
 \int \! \frac{{\rm d}^3\vc{p}_1^{ }
             \,{\rm d}^3\vc{p}_2^{ } 
             \,{\rm d}^3\vc{p}_3^{ }
             \, {\rm d}^4\mathcal{Q} }
         {8 (2\pi)^9 \epsilon^{ }_{1(\vc{p}_1)}
          \,\epsilon^{ }_{2(\vc{p}_2)}
          \,\epsilon^{ }_{3(\vc{p}_3)}}
  \, (2\pi)^4\, 
 \delta^{(4)}_{ }
   \bigl(\P^{ }_1 + \P^{ }_2 - \mathcal{Q}\bigr)\,
 \delta^{(4)}_{ } 
   \bigl(\mathcal{Q}  + \P^{ }_3 - \K \bigr)
 \nn 
 & = & 
 \frac{1}{8(2\pi)^5}
 \int\! \frac{{\rm d}^3\vc{p}_2^{ }
              \, {\rm d} q^{ }_0
              \, {\rm d}^3\vc{q} }
             {\epsilon^{ }_{1(\vc{q-p}_2)}
              \, \epsilon^{ }_{2(\vc{p}_2)} 
              \, \epsilon^{ }_{3(\vc{q-k})}
             }
 \, 
 \delta\bigl(   \epsilon^{ }_{1(\vc{q-p}_2)}
              + \epsilon^{ }_{2(\vc{p}_2)} 
              - q^{ }_0\bigr) 
 \delta\bigl( q^{ }_0  +  \epsilon^{ }_{3(\vc{q-k})} - \omega \bigr)
 \;, \hspace*{5mm} \la{dPhi13} 
\ea
where we have  
integrated over $\vc{p}^{ }_1$ and $\vc{p}^{ }_3$.
The Dirac-$\delta$'s fix two angles as 
\be
 \vc{q}\cdot \vc{p}_2^{ }  = 
  q^{ }_{0\,} \epsilon^{ }_{2(p^{ }_2)} 
 + \frac{m_1^2 - m_2^2 - \s{12} }{2} 
 \;, \quad
 \vc{q}\cdot\vc{k} = 
 q^{ }_{0\,} \omega 
 +  \frac{m_3^2 - M^2 - \s{12} }{2}
 \;. \la{angles_13}
\ee
The other Mandelstam variables can be expressed as 
\be
 \s{13} % = (\P^{ }_1 + \P^{ }_3)^2 
   = m_2^2 + M^2 + 2(\vc{k}\cdot\vc{p}^{ }_2 
                   - \omega \epsilon^{ }_{2(p^{ }_2)})
 \;, \quad
 %% \s{23} = m_1^2 + m_2^2 + m_3^2 + M^2 - \s{12} - \s{13}
 \s{23} = m_1^2 + m_3^2 - \s{12} - 2(\vc{k}\cdot\vc{p}^{ }_2 
                   - \omega \epsilon^{ }_{2(p^{ }_2)})
 \;. 
\ee
As discussed below \eq\nr{mandelstam_2to2_t}, the dependence on
$ \vc{k}\cdot\vc{p}^{ }_2 $ should be partial fractioned.

The azimuthal average of the angle 
between $\vc{k}$ and $\vc{p}^{ }_2$ can be worked out 
like in \eqs\nr{azimuthal_t_res}--\nr{azimuthal_t_res2}, 
with the exchange $\vc{p}^{ }_1 \leftrightarrow \vc{p}^{ }_2$.
Resolving the energy-conservation constraints in \eq\nr{dPhi13}, 
% 
% \ba
%  \int \! {\rm d}\Omega^{ }_{1\to 3}
%  & = & 
%  \frac{1}{(4\pi)^3k}
%  \int_{m^{ }_1 + m^{ }_2}^{\omega - m^{ }_3}\!{\rm d}q^{ }_0 
%  \int^{ k + \sqrt{(q^{ }_0 - \omega)^2 - m_3^2} }
%      _{|k - \sqrt{(q^{ }_0 - \omega)^2 - m_3^2}|} {\rm d}q  
%  \, 
%  \nn 
%  & \times  &   
%    \theta\bigl( \s{12} - (m^{ }_1 + m^{ }_2)^2\bigr)
%  \int^{[{q^{ }_0(\s{12}+m_2^2-m_1^2)+q\kallen({\s{12}},m_1^2,m_2^2)}]
%                                                 /({2\s{12}})} 
%      _{[{q^{ }_0(\s{12}+m_2^2-m_1^2)-q\kallen({\s{12}},m_1^2,m_2^2)}]
%                                                 /({2\s{12}})}
%  \! {\rm d} \epsilon^{ }_{2}
%  \;, \la{dPhi13_prefinal}
%  \hspace*{4mm} 
% \ea
% 
and replacing subsequently $q$ through $s^{ }_{12} = q_0^2 - q^2$,
the integration measure for the $1\to3$ channel becomes  
\ba
 && \hspace*{-1.5cm} 
 \int \! {\rm d}\Omega^{ }_{1\leftrightarrow 3}
 \; = \;  
 \frac{ \theta(M - m^{ }_1 - m^{ }_2 - m^{ }_3 ) }{(4\pi)^3k}
   \int_{(m^{ }_1 + m^{ }_2)^2}^{(M - m^{ }_3)^2} 
    \!\!\!  {\rm d}s^{ }_{12} \, 
   \int_{q_0^-}^{q_0^+} \! \frac{ {\rm d}q^{ }_0 }{2q} 
   \int_{\epsilon_2^-}^{\epsilon_2^+}
                                          \! {\rm d}\epsilon^{ }_2
 \;, \la{dPhi13_final} \hspace*{5mm} 
\ea
where $q = \sqrt{q_0^2 - s^{ }_{12}}$ and, 
making use of $\kallen$ from \eq\nr{kallen},
\ba
 q_0^{\pm} & \equiv & 
 \frac{ \omega (s^{ }_{12}+M_{ }^2-m_3^2)
                         \pm k\kallen(s^{ }_{12},M_{ }^2,m_3^2)}{2M^2}
 \;, \\
 \epsilon_2^{\pm} & \equiv & 
 \frac{ q^{ }_0(s^{ }_{12}+m_2^2-m_1^2)
                         \pm q\kallen(s^{ }_{12},m_1^2,m_2^2)}{2s^{ }_{12}}
 \;. 
\ea

The phase space distribution associated with 
$1\to 3$ scatterings, cf.\ \eq\nr{calN1to3}, 
is conveniently factorized as in \eq\nr{N_1to3_gen}, 
which after the insertion of  
$q^{ }_0 = \epsilon^{ }_1 + \epsilon^{ }_2 = \omega - \epsilon^{ }_3 $
from \eq\nr{dPhi13}, as well as the employment of 
$n^{ }_{\sigma}(-\epsilon) = -1 - n^{ }_{\sigma}(\epsilon)$, 
yields
\ba
 \mathcal{N}^{ }_{\sigma_1\sigma_2\sigma_3}
 \!\! & = & \!\!
 \bigl[
         n^{ }_{\sigma_1\sigma_2}(q^{ }_0 - \mu^{ }_1 - \mu^{ }_2 ) 
       - n^{ }_{\sigma_3}( q^{ }_0 - \omega + \mu^{ }_3 )
 \bigr]
 \bigl[
            n^{ }_{\sigma_2}(\epsilon^{ }_2 - \mu^{ }_2 )
          - n^{ }_{\sigma_1}(\epsilon^{ }_2 - q^{ }_0  + \mu^{ }_1 )
 \bigr]
 \;. \nn
\ea
Integrals over $\epsilon^{ }_2$ are 
exponentially convergent at large $\epsilon^{ }_2$; 
those over $q^{ }_0$ are localized close to 
the lower bound $q_0^-$.

In the massless limit, the integration range in \eq\nr{dPhi13_final}
collapses to a point. 
Therefore thermal IR divergences of the type discussed in 
\se\ref{sss:regge} are absent in $1\to 3$ decays. 

%%%%%%%%%%%%%%%%%%%%%%%%%%% SUBSECTION %%%%%%%%%%%%%%%%%%%%%%%%%%%%%%%%
%
\subsection{$s$-channel $2\leftrightarrow 2$ reactions}
\la{app:2to2_s}

The part of $2\leftrightarrow 2$ reactions that cannot be put in the
form treated in \se\ref{ss:phasespace_real} are $s$-channels reactions. 
With the labelling for 
$\scat{2\to2}(-a^{ }_1;b^{ }_1,b^{ }_2)$ chosen as 
$\K^{ }_1 \equiv - \P^{ }_{a_1^{ }}$, 
$\P^{ }_i \equiv \P^{ }_{b_i^{ }}$, 
$\nu^{ }_1 \equiv - \mu^{ }_{a_1^{ }}$, 
$\mu^{ }_i \equiv \mu^{ }_{b_i^{ }}$, 
so that the sign flips necessary for initial-state
momenta and chemical potentials are 
already explicit when using $\K^{ }_1$ and $\nu^{ }_1$,
the $s$-channel part originating from \eq\nr{example_1to3_alt} becomes
\ba
 \frac{ \Gamma^\rmi{Born}_{2\leftrightarrow 2(s)} }{ 2 (g_1^2 + 3 g_2^2) }
 & \to & 
 \scat{2\to2}(-\aS;\aQ,\ala) \, 
%  \nn[2mm] & \times & 
                \biggl\{ 
% \nn & & +
                      - \frac{\E\cdot\P^{ }_2}{s - m_\bla^2 }
                      + \frac{
                        (\MM - m_\cS^2 )\,
                        \E\cdot( \P^{ }_1 + 2 \P^{ }_2  )
                       }{ 
                        (s - m_\bla^2 )\, ({ t - m_{\tilde{\cS}}^2 }) }
                       \biggr\} 
%%%%%%%%%%%%%%%%
 \nn[2mm] 
 & + & 
 \scat{2\to2}(-\ala;\aQ,\aS) \, 
%  \nn[2mm] & \times & 
                \biggl\{ 
                       \frac{\E\cdot(\K^{ }_1 - \K)}
                            {s - m_{\tilde{\cS}}^2}
                      + \frac{ 
                         2 m_\cS^2 \, \E\cdot\K^{ }_1   
                             }{ ({ s - m_{\tilde{\cS}}^2 })^2 }
                       \biggr\} 
 \;. \la{example_2to2_s}  
\ea

The goal now is to have $s$ as an integration variable. 
For this aim we introduce a four-momentum $\mathcal{Q}$ such 
that $s = \mathcal{Q}^2$,
and define $s$-channel parametrization of 
the integration measure from \eq\nr{dO2to2} as  
\ba
 & & \hspace*{-1.2cm}
 \int \! {\rm d}\Omega^{s}_{2\leftrightarrow 2}
 \; \equiv \; 
 \int \! \frac{{\rm d}^3\vc{p}_1^{ }
             \,{\rm d}^3\vc{p}_2^{ } 
             \,{\rm d}^3\vc{k}_1^{ }
             \, {\rm d}^4\mathcal{Q} }
         {8 (2\pi)^9 \epsilon^{ }_{1(\vc{p}_1)}
          \,\epsilon^{ }_{2(\vc{p}_2)}
          \,E^{ }_{1(\vc{k}_1)}}
  \, (2\pi)^4\, 
 \delta^{(4)}_{ }
   \bigl(\P^{ }_1 + \P^{ }_2 - \mathcal{Q}\bigr)\,
 \delta^{(4)}_{ }\bigl(\mathcal{Q}  - \K^{ }_1 - \K \bigr)
 \nn 
 & = & \!\!
 \frac{1}{8(2\pi)^5}
 \int\! \frac{{\rm d}^3\vc{p}_2^{ }
              \, {\rm d} q^{ }_0
              \, {\rm d}^3\vc{q} }
             {\epsilon^{ }_{1(\vc{q-p}_2)}
              \, \epsilon^{ }_{2(\vc{p}_2)} 
              \, E^{ }_{1(\vc{q-k})}
             }
 \, 
 \delta\bigl(   \epsilon^{ }_{1(\vc{q-p}_2)}
              + \epsilon^{ }_{2(\vc{p}_2)} 
              - q^{ }_0\bigr) 
 \, 
 \delta\bigl( q^{ }_0  -  E^{ }_{1(\vc{q-k})} - \omega \bigr)
 \;, \hspace*{6mm} \nn \la{dPhi22_s}
\ea
where we have integrated over $\vc{p}^{ }_1$ and $\vc{k}^{ }_1$.
The Dirac-$\delta$'s fix two angles as 
\be
 \vc{q}\cdot \vc{p}_2^{ }  = 
  q^{ }_{0\,} \epsilon^{ }_{2(p^{ }_2)} 
 + \frac{m_1^2 - m_2^2 - s }{2} 
 \;, \quad
 \vc{q}\cdot\vc{k} = 
 q^{ }_{0\,} \omega 
 +  \frac{M_1^2 - M^2 - s }{2}
 \;. \la{angles_s}
\ee
The other Mandelstam variables can be expressed as 
\be
 t % = (\P^{ }_2 - \K^{ })^2 
   = m_2^2 + M^2 + 2(\vc{k}\cdot\vc{p}^{ }_2 
                   - \omega \epsilon^{ }_{2(p^{ }_2)})
 \;, \quad
 %% u = m_1^2 + m_2^2 + M_1^2 + M^2 - s - t
 u = m_1^2 + M_1^2 - s - 2(\vc{k}\cdot\vc{p}^{ }_2 
                   - \omega \epsilon^{ }_{2(p^{ }_2)})
 \;. \la{s-chan-t}
\ee
The azimuthal average of the angle 
between $\vc{k}$ and $\vc{p}^{ }_2$ can be worked out 
like in \eqs\nr{azimuthal_t_res}--\nr{azimuthal_t_res2}, 
with the exchange $\vc{p}^{ }_1 \leftrightarrow \vc{p}^{ }_2$.
Resolving the energy-conservation constraints 
in \linebreak \eq\nr{dPhi22_s}, 
the integration measure becomes  
\be
 \int \! {\rm d}\Omega^{s}_{2\leftrightarrow 2}
 = 
 \frac{1}{(4\pi)^3k}
 \int_{\omega + M^{ }_1}^{\infty}\!{\rm d}q^{ }_0 
 \int^{ k + \sqrt{(q^{ }_0 - \omega)^2 - M_1^2} }
     _{|k - \sqrt{(q^{ }_0 - \omega)^2 - M_1^2}|} {\rm d}q  
 \, 
   \theta\bigl( s - (m^{ }_1 + m^{ }_2)^2\bigr)
 \int^{ \epsilon_2^+ } 
     _{ \epsilon_2^- }
 \! {\rm d} \epsilon^{ }_{2}
 \;, \la{dPhi22_s_prefinal}
 \hspace*{4mm} 
\ee
where $ \epsilon_2^\pm$ are from 
\eq\nr{2to2s_e2pm}.
Subsequently we replace 
$q$ through $s = q_0^2 - q^2$,
resulting in 
\ba
 && \hspace*{-1.5cm} 
 \int \! {\rm d}\Omega^{s}_{2\leftrightarrow 2}
 \; = \;  
 \frac{1}{(4\pi)^3k}
   \int_{\rmi{max$((M +M_1^{ })^2,(m^{ }_1 + m^{ }_2)^2)$}}^{\infty} 
    \!\!\!  {\rm d}s \, 
   \int_{q_0^-}^{q_0^+} \! 
   \frac{ {\rm d}q^{ }_0  }{2q} 
   \int_{\epsilon_2^-}^{\epsilon_2^+}
                                          \! {\rm d}\epsilon^{ }_2
 \;, \la{dPhi22_s_final} \hspace*{5mm} 
\ea
where $q = \sqrt{q_0^2 - s}$ and 
\ba
 q_0^{\pm} & \equiv & 
 \frac{ \omega (s+M_{ }^2-M_1^2)
                         \pm k\kallen(s,M_{ }^2,M_1^2)}{2M^2}
 \;, \\[2mm]
 \epsilon_2^{\pm} & \equiv & 
 \frac{ q^{ }_0(s+m_2^2-m_1^2)
                         \pm q\kallen(s,m_1^2,m_2^2)}{2s}
 \;. \la{2to2s_e2pm}
\ea

When using the $s$-channel parametrization, 
the phase space distribution associated with 
$2\leftrightarrow 2$ scatterings from 
\eq\nr{calN2to2} is conveniently factorized as
in \eq\nr{N_2to2_s_gen}, 
which after the insertion of 
$q^{ }_0 = \epsilon^{ }_1 + \epsilon^{ }_2 = \omega + E^{ }_1 $
from \eq\nr{dPhi22_s} yields
\ba
 \mathcal{N}^{ }_{\tau_1;\sigma_1\sigma_2}
 \!\! & = & \!\!
 \bigl[
        n^{ }_{\tau_1}(q^{ }_0 - \omega - \nu^{ }_1 )
      - n^{ }_{\sigma_1\sigma_2}(q^{ }_0 - \mu^{ }_1 - \mu^{ }_2 ) 
 \bigr]
 \bigl[  
           n^{ }_{\sigma_2}(\epsilon^{ }_2 - \mu^{ }_2 )
         - n^{ }_{\sigma_1}( \epsilon^{ }_2 - q^{ }_0 + \mu^{ }_1 )
 \bigr]
 \;. \nn \la{N_2to2_s}
\ea
The latter factor guarantees that integrals
over $\epsilon^{ }_2$ are exponentially convergent at large 
$\epsilon^{ }_2$; 
those over $q^{ }_0$ are localized close to 
the lower bound $q_0^-$.

In the massless limit, the integration domain in \eq\nr{dPhi22_s_prefinal} 
becomes 
\be
  \lim_{m^{ }_i,M\to 0}
 \int \! {\rm d}\Omega^{s}_{2\leftrightarrow 2}
  =  
 \frac{1}{(4\pi)^3k}
 \int_{k}^{\infty}
      \! {\rm d}q^{ }_0 
 \int^{ q^{ }_0 }
     _{| 2 k - q^{ }_0 |}
     \! {\rm d}q  
 \int^{(q^{ }_0 + q)/2}_{(q^{ }_0 - q)/2}
     \! {\rm d}p^{ }_2
 \;. \la{dPhi22_s_massless}
\ee
Therefore $q^{ }_0$ is never small, and
$q$ could be small only in the vicinity of $q^{ }_0 = 2 k$, 
however around that point $s \approx 4k^2$ is large. 
Therefore $s$-channel $2\leftrightarrow 2$ scatterings
do not lead to IR divergences from small $q^{ }_0,q$, 
of the type that were discussed in \se\ref{sss:regge}.  
 
In contrast, if we expand $n^{ }_{\tau_1}$ from 
\eq\nr{N_2to2_s} to first order in $\nu^{ }_1$, a second order
pole emerges. 
Thus a logarithmic divergence can originate from the domain around 
$q^{ }_0\approx \omega$, as outlined in \se\ref{sss:off},  
which is regularized by finite masses~\cite{cptheory}. 
Likewise the final-state distributions 
$n^{ }_{\sigma_1}$, $n^{ }_{\sigma_2}$ can become singular if the
particles are massless bosons and they carry finite chemical potentials.

%%%%%%%%%%%%%%%%%%%%%%%%%%% SUBSECTION %%%%%%%%%%%%%%%%%%%%%%%%%%%%%%%%
%
\subsection{$t$-channel $3\to 1$ reactions}
\la{app:3to1_t}

Part of the $3\to 1$ reactions from \eq\nr{master} can be called 
$t$-channel ones, with the Mandelstam variables defined 
according to \eq\nr{mstam_3}. 
For \eq\nr{example_1to3_alt}, 
with momenta and chemical potentials labelled for 
$\scat{1\to3}(-a^{ }_1,-a^{ }_2;b^{ }_1)$
as 
$
 \K^{ }_i \equiv - \P^{ }_{a_i^{ }}
$, 
$
 \P^{ }_1 \equiv \P^{ }_{b_1^{ }}
$, 
$
 \nu^{ }_i \equiv - \mu^{ }_{a_i^{ }} 
$,
$
 \mu^{ }_1 \equiv \mu^{ }_{b^{ }_1}
$,
so that the sign flips necessary for initial-state
momenta and chemical potentials 
are already included when using $\K^{ }_i$ and $\nu^{ }_i$,
the $t$-channel processes amount to 
\ba
 && \hspace*{-1cm}
 \frac{ \Gamma^\rmi{Born}_{3\to 1(t)} }{ 2 (g_1^2 + 3 g_2^2) }
 \; \to \; 
 \scat{3\to1}(-\ala,-\aS;\aQ) \, 
%  2 (g_1^2 + 3 g_2^2)
%  \nn[2mm] & \times & 
                \biggl\{ 
% \nn & & +
                      \frac{ \E\cdot\K^{ }_1 }{t - m_\bla^2 } 
                   + 
                      \frac{
                       (\MM - m_\cS^2 )\,
                        \E\cdot ( \P^{ }_1 - 2 \K^{ }_1 ) 
                       }{ 
                        (t - m_\bla^2 )\, ({ u - m_{\tilde{\cS}}^2 }) }
                       \biggr\} 
%%%%%%%%%%%%%%%%
 \nn[2mm] 
 & + & 
 \scat{3\to1}(-\aQ,-\aS;\ala) \, 
%  2 (g_1^2 + 3 g_2^2)
%  \nn[2mm] & \times & 
                \biggl\{ 
% \nn & & +
                     - \frac{ \E\cdot\P^{ }_1 }{t - m_\bla^2} 
                     +
                      \frac{
                       ( m_\cS^2  - \MM )\,
                        \E\cdot ( \K^{ }_1 -2 \P^{ }_1) 
                       }{ 
                        (t - m_\bla^2 )\, ({ s - m_{\tilde{\cS}}^2 }) }
                       \biggr\} 
%%%%%%%%%%%%%%
  \la{example_3to1_t}  \\[2mm] 
 & + & 
 \bigl[\, \scat{3\to1}(-\aS,-\ala;\aQ) + 
          \scat{3\to1}(-\aQ,-\ala;\aS) 
 \;\bigr]\, 
%  2 (g_1^2 + 3 g_2^2)
%  \nn[2mm] & \times & 
                \biggl\{ 
%  \nn & & +  
                       \frac{\E\cdot(\K^{ }_2 - \K)}{ t - m_{\tilde{\cS}}^2 }
                      + \frac{ 
                         2 m_\cS^2\, \E\cdot\K^{ }_2  }
                         { ({ t - m_{\tilde{\cS}}^2 })^2 }
                       \biggr\} 
 \;. \nonumber
\ea
Some of these channels are not allowed kinematically, 
but this is taken care of automatically once we work out the integration
measure, cf.\ \eq\nr{dPhi31_t_final} below. 

The goal now is to have $t$ as an integration variable. 
To achieve this we introduce a four-momentum $\mathcal{Q}$ such 
that $t = \mathcal{Q}^2$, and write 
the phase space integration measure for ${3}\to{1}$
reactions from \eq\nr{dO3to1} as   
\ba
 & & \hspace*{-1.0cm}
 \int \! {\rm d}\Omega^{t}_{3\to 1}
 \; = \; 
 \int \! \frac{{\rm d}^3\vc{k}_1^{ }
             \,{\rm d}^3\vc{k}_2^{ } 
             \,{\rm d}^3\vc{p}_1^{ }
             \, {\rm d}^4\mathcal{Q} }
         {8 (2\pi)^9 E^{ }_{1(\vc{k}_1)}
          \,E^{ }_{2(\vc{k}_2)}
          \,\epsilon^{ }_{1(\vc{p}_1)}}
  \, (2\pi)^4\, 
 \delta^{(4)}_{ }
    \bigl(\K^{ }_1   - \P^{ }_1 - \mathcal{Q}\bigr)\,
 \delta^{(4)}_{ }\bigl(\mathcal{Q} + \K + \K^{ }_2 \bigr)
 \hspace*{5mm} \nn 
 & = & 
 \frac{1}{8(2\pi)^5}
 \int\! \frac{{\rm d}^3\vc{k}_1^{ }
              \, {\rm d} q^{ }_0
              \, {\rm d}^3\vc{q} }
             {   E^{ }_{1(\vc{k}_1)}
              \, E^{ }_{2(\vc{q+k})} 
              \, \epsilon^{ }_{1(\vc{q-k}_1)}
             }
 \, 
 \delta\bigl(   E^{ }_{1(\vc{k}_1)}
              - \epsilon^{ }_{1(\vc{q-k}_1)}
              - q^{ }_0\bigr) 
 \delta\bigl(   q^{ }_0   
              + \omega
              + E^{ }_{2(\vc{q + k})} \bigr)
 \;. \hspace*{5mm} \nn \la{dPhi31_t} 
\ea
The Dirac-$\delta$'s fix two angles as 
\be
 \vc{q}\cdot \vc{k}_1^{ }  = 
  q^{ }_{0\,} E^{ }_{1(k^{ }_1)} 
 + \frac{m_1^2 - M_1^2 - t }{2} 
 \;, \quad
 \vc{q}\cdot\vc{k} = 
 q^{ }_{0\,} \omega 
 +  \frac{t + M^2 - M_2^2 }{2}
 \;. \la{angles_31_t}
\ee
The other Mandelstam variables can be expressed as 
\be
 u 
   = M^2 + M_1^2 - 2(\vc{k}\cdot\vc{k}^{ }_1 
                   - \omega E^{ }_{1(k^{ }_1)}
                    )
 \;, \quad
 %% s = m_1^2 + M_1^2 + M_2^2 + M^2 - t - u
 s = m_1^2 + M_2^2 - t 
                   + 2( \vc{k}\cdot\vc{k}^{ }_1 
                      - \omega E^{ }_{1(k^{ }_1)}
                    )
 \;. 
\ee
The azimuthal average of the angle 
between $\vc{k}$ and $\vc{k}^{ }_1$ can be worked out 
like in \eqs\nr{azimuthal_t_res}--\nr{azimuthal_t_res2}, 
with $(\vc{k},\vc{p}^{ }_1)\to (\vc{k},\vc{k}^{ }_1)$.
Resolving the energy-conservation constraints in \eq\nr{dPhi31_t}, 
% 
% \ba
%  && \hspace*{-1.2cm}
%  \int \! {\rm d}\Omega^{t}_{3\to 1}
%   =  
%  \frac{1}{(4\pi)^3k}
%  \int_{-\infty}^{-\omega - M^{ }_2}\!{\rm d}q^{ }_0 
%  \int^{ k + \sqrt{(q^{ }_0 + \omega)^2 - M_2^2} }
%      _{|k - \sqrt{(q^{ }_0 + \omega)^2 - M_2^2}|}
%      {\rm d}q  
%  \, \biggl\{ 
%  \theta(-t)\int_{[{q^{ }_0(t-m_1^2+M_1^2)-q\kallen(t,m_1^2,M_1^2)}]/({2t})}
%                ^{\infty} 
%  \!\!\!\! {\rm d}E^{ }_1
%  \nn 
%  & +  & 
%    \theta(t)\,
%    \theta\bigl((M^{ }_1 - m^{ }_1)q^{ }_0\bigr)\, 
%    \theta\bigl((m^{ }_1 - M^{ }_1)^2-t\bigr)
%  \int^{[{q^{ }_0(t-m_1^2+M_1^2)+q\kallen(t,m_1^2,M_1^2)}]/({2t})} 
%      _{[{q^{ }_0(t-m_1^2+M_1^2)-q\kallen(t,m_1^2,M_1^2)}]/({2t})}
%  \! {\rm d} E^{ }_{1}
%  \biggr\} 
%  \;, \la{dPhi31_t_prefinal}
%  \hspace*{4mm} 
% \ea
%  
and replacing subsequently 
$q$ through $t = q_0^2 - q^2$, we note that $q^{ }_0$ is always
negative and $t$ is necessarily positive. 
The integration measure becomes 
\ba
 && \hspace*{-1.5cm} 
 \int \! {\rm d}\Omega^{t}_{3\to 1}
 \; = \;  
 \frac{ \theta(m^{ }_1 - M - M^{ }_1 - M^{ }_2 ) }{(4\pi)^3k}
   \int_{(M + M^{ }_2)^2}^{(m^{ }_1 - M^{ }_1)^2} 
    \!\!\!  {\rm d}t \, 
   \int_{q_0^-}^{q_0^+} \! \frac{ {\rm d}q^{ }_0 }{2q} 
   \int_{E_1^-}^{E_1^+}
                                          \! {\rm d}E^{ }_1
 \;, \la{dPhi31_t_final} \hspace*{5mm} 
\ea
where $q = \sqrt{q_0^2 - t}$ and,  
with $\kallen$ from \eq\nr{kallen},
\ba
 q_0^{\pm} & \equiv & 
 \frac{ - \omega (t+M_{ }^2-M_2^2)
                         \pm k\kallen(t,M_{ }^2,M_2^2)}{2M^2}
 \;, \\[2mm]
 E_1^{\pm} & \equiv & 
 \frac{ q^{ }_0(t+M_1^2-m_1^2)
                         \pm q\kallen(t,m_1^2,M_1^2)}{2t}
 \;. 
\ea

The phase space distribution associated with 
$3\to 1$ scatterings, from \eq\nr{calN3to1},  
is conveniently factorized as in 
\eq\nr{N_3to1_t_gen}, 
which after the insertion of  
$q^{ }_0 = E^{ }_1 -  \epsilon^{ }_1  = - \omega - E^{ }_2 $
from \eq\nr{dPhi31_t} yields
\ba
 \mathcal{N}^{ }_{\tau_1\tau_2;\sigma_1}
 \!\! & = & \!\!  
 \bigl[
        n^{ }_{\tau_1\sigma_1}(q^{ }_0 + \mu^{ }_1 - \nu^{ }_1) 
     -  n^{ }_{\tau_2}(q^{ }_0 + \omega + \nu^{ }_2)
 \bigr]
 \bigl[
            n^{ }_{\tau_1}(E^{ }_1 - \nu^{ }_1)
         -  n^{ }_{\sigma_1}(E^{ }_1 - q^{ }_0 - \mu^{ }_1)
 \bigr]
 \;. \nn
\ea
Integrals over $E^{ }_1$ are  
exponentially convergent at large $E^{ }_1$; 
those over $q^{ }_0$ are localized close to the upper bound $q_0^+$.

In the massless limit, the integration domain of 
\eq\nr{dPhi31_t_final} shrinks to a point.  
Thus there are no thermal IR problems of the type discussed
in \se\ref{sss:regge}.

%%%%%%%%%%%%%%%%%%%%%%%%%%% SUBSECTION %%%%%%%%%%%%%%%%%%%%%%%%%%%%%%%%
%
\subsection{$s$-channel $3\to 1$ reactions}
\la{app:3to1_s}

The remaining processes from \eq\nr{master} are $s$-channel
$3\to 1$ reactions.  
For the example of \eq\nr{example_1to3_alt}, 
with momenta and chemical potentials labelled for 
$\scat{1\to3}(-a^{ }_1,-a^{ }_2;b^{ }_1)$
as 
$
 \K^{ }_i \equiv - \P^{ }_{a_i^{ }}
$, 
$
 \P^{ }_1 \equiv \P^{ }_{b_1^{ }}
$, 
$
 \nu^{ }_i \equiv - \mu^{ }_{a_i^{ }} 
$,
$
 \mu^{ }_1 \equiv \mu^{ }_{b^{ }_1}
$,
so that the sign flips necessary for initial-state
momenta and chemical potentials 
have already been included when using $\K^{ }_i$ and $\nu^{ }_i$, 
they amount to 
\ba
 \frac{ \Gamma^\rmi{Born}_{3\to 1(s)} }{ 2 (g_1^2 + 3 g_2^2) }
 & \to & 
 \scat{3\to1}(-\aQ,-\ala;\aS) \, 
%  \nn[2mm] & \times & 
                \biggl\{ 
% \nn & & +
                      \frac{ \E\cdot\K^{ }_2 }{s - m_\bla^2 }
                    + 
                      \frac{
                        (  m_\cS^2 - \MM )\,
                        \E\cdot(\K^{ }_1 + 2\K^{ }_2)
                       }{ 
                        ( s - m_\bla^2 )\, ({ t - m_{\tilde{\cS}}^2 }) }
                       \biggr\} 
%%%%%%%%%%%%%%
 \nn[2mm] 
 & + & 
 \scat{3\to1}(-\aQ,-\aS;\ala) \, 
%  \nn[2mm] & \times & 
                \biggl\{ 
%  \nn & &  
                     -  \frac{\E\cdot(\P^{ }_1 + \K)}
                             {s - m_{\tilde{\cS}}^2}     
                     - \frac{ 
                         2 m_\cS^2 \, \E\cdot\P^{ }_1   
                             }{ ({ s - m_{\tilde{\cS}}^2 })^2 }
                       \biggr\} 
 \;. \la{example_3to1_s}  
\ea

The goal now is to have $s$ as an integration variable. 
We introduce a four-momentum $\mathcal{Q}$ such 
that $s = \mathcal{Q}^2$,
and write  
the phase space integration measure for ${3}\to{1}$
reactions from \eq\nr{dO3to1} as   
\ba
 & & \hspace*{-1.0cm}
 \int \! {\rm d}\Omega^{s}_{3\to 1}
 \; = \; 
 \int \! \frac{{\rm d}^3\vc{k}_1^{ }
             \,{\rm d}^3\vc{k}_2^{ } 
             \,{\rm d}^3\vc{p}_1^{ }
             \, {\rm d}^4\mathcal{Q} }
         {8 (2\pi)^9 E^{ }_{1(\vc{k}_1)}
          \,E^{ }_{2(\vc{k}_2)}
          \,\epsilon^{ }_{1(\vc{p}_1)}}
  \, (2\pi)^4\, 
 \delta^{(4)}_{ }
    \bigl(\K^{ }_1 + \K^{ }_2 - \mathcal{Q}\bigr)\,
 \delta^{(4)}_{ }\bigl(\mathcal{Q}  - \P^{ }_1 + \K \bigr)
 \hspace*{5mm} \nn 
 & = & 
 \frac{1}{8(2\pi)^5}
 \int\! \frac{{\rm d}^3\vc{k}_2^{ }
              \, {\rm d} q^{ }_0
              \, {\rm d}^3\vc{q} }
             {   E^{ }_{1(\vc{q-k}_2)}
              \, E^{ }_{2(\vc{k}_2)} 
              \, \epsilon^{ }_{1(\vc{q+k})}
             }
 \, 
 \delta\bigl(   E^{ }_{1(\vc{q-k}_2)}
              + E^{ }_{2(\vc{k}_2)} 
              - q^{ }_0\bigr) 
 \delta\bigl(   q^{ }_0  
              - \epsilon^{ }_{1(\vc{q+k})}
              + \omega \bigr)
 \;. \hspace*{5mm} \nn \la{dPhi31_s}
\ea
The Dirac-$\delta$'s fix two angles as 
\be
 \vc{q}\cdot \vc{k}_2^{ }  = 
  q^{ }_{0\,} E^{ }_{2(k^{ }_2)} 
 + \frac{M_1^2 - M_2^2 - s }{2} 
 \;, \quad
 \vc{q}\cdot\vc{k} = 
 q^{ }_{0\,} \omega 
 +  \frac{s + M^2 - m_1^2 }{2}
 \;. \la{angles_31_s}
\ee
The other Mandelstam variables can be expressed as 
\be
 t 
   = M^2 + M_2^2 - 2( \vc{k}\cdot\vc{k}^{ }_2 
                   -  \omega E^{ }_{2(k^{ }_2)}
                    )
 \;, \quad
 %% u = m_1^2 + M_1^2 + M_2^2 + M^2 - s - t
 u = m_1^2 + M_1^2 - s
                 + 2( \vc{k}\cdot\vc{k}^{ }_2 
                   -  \omega E^{ }_{2(k^{ }_2)}
                    )
 \;. 
\ee
The azimuthal average of the angle 
between $\vc{k}$ and $\vc{k}^{ }_2$ can be worked out 
like in \eqs\nr{azimuthal_t_res}--\nr{azimuthal_t_res2}, 
with $(\vc{k},\vc{p}^{ }_1)\to (\vc{k},\vc{k}^{ }_2)$.
Resolving the energy-conservation constraints in \eq\nr{dPhi31_s}, 
% 
% \ba
%  \int \! {\rm d}\Omega^{s}_{3\to 1}
%  & = & 
%  \frac{1}{(4\pi)^3k}
%  \int_{m^{ }_1 - \omega}^{\infty}\!{\rm d}q^{ }_0 
%  \int^{ k + \sqrt{(q^{ }_0 + \omega)^2 - m_1^2} }
%      _{|k - \sqrt{(q^{ }_0 + \omega)^2 - m_1^2}|} {\rm d}q  
%  \, 
%  \nn 
%  & \times  &   
%    \theta\bigl( s - (M^{ }_1 + M^{ }_2)^2\bigr)
%  \int^{[{q^{ }_0(s+M_2^2-M_1^2)+q\kallen(s,M_1^2,M_2^2)}]/({2 s})} 
%      _{[{q^{ }_0(s+M_2^2-M_1^2)-q\kallen(s,M_1^2,M_2^2)}]/({2 s})}
%  \! {\rm d} E^{ }_{2}
%  \;, \la{dPhi31_s_prefinal}
%  \hspace*{4mm} 
% \ea
% 
and replacing subsequently $q$ through $s = q_0^2 - q^2$, 
the integration measure becomes  
\ba
 && \hspace*{-1.5cm} 
 \int \! {\rm d}\Omega^{s}_{3\to 1}
 \; = \;  
 \frac{ \theta(m^{ }_1 - M - M^{ }_1 - M^{ }_2 ) }{(4\pi)^3k}
   \int_{(M^{ }_1 + M^{ }_2)^2}^{(m^{ }_1 - M)^2} 
    \!\!\!  {\rm d}s \, 
   \int_{q_0^-}^{q_0^+} \! \frac{ {\rm d}q^{ }_0 }{2q} 
   \int_{E_2^-}^{E_2^+}
                                          \! {\rm d}E^{ }_2
 \;, \la{dPhi31_s_final} \hspace*{5mm} 
\ea
where $q = \sqrt{q_0^2 - s}$ and, 
with $\kallen$ from \eq\nr{kallen},
\ba
 q_0^{\pm} & \equiv & 
 \frac{ - \omega (s+M_{ }^2-m_1^2)
                         \pm k\kallen(s,M_{ }^2,m_1^2)}{2M^2}
 \;, \\[2mm]
 E_2^{\pm} & \equiv & 
 \frac{ q^{ }_0(s+M_2^2-M_1^2)
                         \pm q\kallen(s,M_1^2,M_2^2)}{2s}
 \;. 
\ea

The phase space distribution associated with 
$3\to 1$ scatterings, from \eq\nr{calN3to1},  
is conveniently factorized as 
in \eq\nr{N_3to1_s_gen}, 
which after the insertion of
$q^{ }_0 = E^{ }_1 + E^{ }_2 = \epsilon^{ }_1 - \omega $
from \eq\nr{dPhi31_s} yields
\ba
 \mathcal{N}^{ }_{\tau_1\tau_2;\sigma_1}
 \!\! & = & \!\!  
 \bigl[
        n^{ }_{\tau_1\tau_2}(q^{ }_0 - \nu^{ }_1 - \nu^{ }_2) 
     -  n^{ }_{\sigma_1}(q^{ }_0 + \omega - \mu^{ }_1)
 \bigr]
 \bigl[
            n^{ }_{\tau_2}(E^{ }_2 - \nu^{ }_2)
         -  n^{ }_{\tau_1}(E^{ }_2 - q^{ }_0 + \nu^{ }_1)
 \bigr]
 \;. \nn
\ea
Integrals over $E^{ }_2$ are exponentially convergent at large $E^{ }_2$; 
those over $q^{ }_0$ are localized close to the lower bound $q_0^-$.

In the massless limit, the integration domain of 
\eq\nr{dPhi31_s_final} shrinks to a point. 
Thus there are no thermal IR problems of the type discussed
in \se\ref{sss:regge}.

%%%%%%%%%%%%%%%%%%%%%%%%%%% SECTION %%%%%%%%%%%%%%%%%%%%%%%%%%%%%%%%%%
%
\section{Angular averages for virtual corrections}

We define here angular averages that appear in the virtual 
corrections discussed in \se\ref{ss:phasespace_virtual}. 

Let $(\theta,\varphi)$ be the spherical angles associated with 
a loop momentum, which in the following we denote by $\vc{p}^{ }_a$. The 
axis with respect to which $\theta$ is measured will be specified later on,
but the choice plays no role, as all angles are integrated over. 
The angular average is taken in the presence of two further vectors, 
denoted by $\vc{p}^{ }_d$ and $\vc{k}$, and sometimes it is also
convenient to employ $\vc{p}^{ }_e \equiv \vc{k} - \vc{p}^{ }_d$. 
We are concerned with two
types of averages, 
\ba
 \mathcal{G}^{ }_{\vc{p}_a;\vc{p}_d,\vc{k};z} 
  \, P^{ }_n(\vc{p}^{ }_a\cdot\vc{k})
 & \equiv & 
  \int_{-1}^{+1} \! \frac{ {\rm d}\cos\theta }{2} \,  
  \int_0^{2\pi} \! \frac{ {\rm d}\varphi }{2\pi} \, \mathbbm{P}\,
 \frac{P^{ }_n(\vc{p}^{ }_a\cdot\vc{k})}
      { z - \vc{p}^{ }_a \cdot \vc{p}^{ }_d } 
 \;, \la{def_G} \\ 
%%%%
 \mathcal{H}^{ }_{\vc{p}_a;\vc{p}_d,\vc{k};z^{ }_1,z^{ }_2} 
  \, Q^{ }_n(\vc{p}^{ }_a\cdot\vc{k})
 & \equiv & 
  \int_{-1}^{+1} \! \frac{ {\rm d}\cos\theta }{2} \,  
  \int_0^{2\pi} \! \frac{ {\rm d}\varphi }{2\pi} \, \mathbbm{P}\,
 \frac{Q^{ }_n(\vc{p}^{ }_a\cdot\vc{k})}
      { ( z^{ }_1 - \vc{p}^{ }_a \cdot \vc{p}^{ }_d )
        ( z^{ }_2 - \vc{p}^{ }_a \cdot \vc{k} )   } 
 \;, \hspace*{6mm} \la{def_H}
\ea
where $P^{ }_n,Q^{ }_n$ are polynomials of degree $n$, 
and $\mathbbm{P}$ denotes the principal value. 

Starting with the latter average, the first step is to write
\ba
 Q^{ }_n(\vc{p}^{ }_a\cdot\vc{k}) 
 & = &  
 Q^{ }_n(z^{ }_2) + 
 Q^{ }_n(\vc{p}^{ }_a\cdot\vc{k}) - Q^{ }_n(z^{ }_2)
 \nn 
 & \equiv & 
 Q^{ }_n(z^{ }_2) + 
 (z^{ }_2 - \vc{p}^{ }_a\cdot\vc{k})\,
 \widetilde Q^{ }_{n-1}(\vc{p}^{ }_a\cdot\vc{k})
 \;. \la{factorize} 
\ea
If we express the original polynomial as 
$
 Q^{ }_n( \vc{p}^{ }_a\cdot\vc{k} ) 
 = 
 \sum_{i=0}^n a^{ }_i \, ( \vc{p}^{ }_a\cdot\vc{k} )^i_{ }
$, 
then 
$
 \widetilde Q^{ }_{n-1}( \vc{p}^{ }_a\cdot\vc{k} ) 
 = 
 \sum_{j=0}^{n-1} b^{ }_j \, ( \vc{p}^{ }_a\cdot\vc{k} )^j_{ }
$, 
where the coefficients read
$
 b^{ }_j = - \sum_{i=j+1}^n a^{} _i\, z_2^{i-j-1}
$.
Thereby \eq\nr{def_H} becomes 
\be
 \mathcal{H}^{ }_{\vc{p}_a;\vc{p}_d,\vc{k};z^{ }_1,z^{ }_2} 
  \, Q^{ }_n(\vc{p}^{ }_a\cdot\vc{k})
 = 
  \mathcal{H}^{ }_{\vc{p}_a;\vc{p}_d,\vc{k};z^{ }_1,z^{ }_2} 
  \, Q^{ }_n(z^{ }_2)
 + 
 \mathcal{G}^{ }_{\vc{p}_a;\vc{p}_d,\vc{k};z^{ }_1} 
  \, \widetilde Q^{ }_{n-1}(\vc{p}^{ }_a\cdot\vc{k})
 \;. \la{splitup}
\ee

For the first term, we combine the denominators with a Feynman parameter, 
\be
 \frac{1}{  
        ( z^{ }_1 - \vc{p}^{ }_a \cdot \vc{p}^{ }_d )
        ( z^{ }_2 - \vc{p}^{ }_a \cdot \vc{k} )  }
 = 
 \int_0^1 \! {\rm d}s \, 
 \frac{1}{
 [ 
   s\, z^{ }_1 + (1-s)\, z^{ }_2 - 
  \vc{p}^{ }_a\cdot 
  ( \vc{k} - s\, \vc{p}^{ }_e )
 ]^2
 }
 \;,
\ee
where we made use of 
$
 \vc{p}^{ }_e = \vc{k} - \vc{p}^{ }_d
$.
The angle is now chosen as 
$
 \theta \,\equiv\,
 \theta_{\vc{p}_a, \vc{k} - s\, \vc{p}^{ }_e }
$.
There is no dependence on $\varphi$, so that both integrals are 
readily carried out, 
\ba
 && \hspace*{-2.5cm}
  \int_{-1}^{+1} \! \frac{ {\rm d}\cos\theta }{2} \,  
  \int_0^{2\pi} \! \frac{ {\rm d}\varphi }{2\pi} \,
 \frac{1}{
  [
    s\, z^{ }_1 + (1-s)\, z^{ }_2 - 
   \vc{p}^{ }_a\cdot 
   ( \vc{k} - s\, \vc{p}^{ }_e )
  ]^2 }
 \nn
 && \hspace*{0.5cm} = \,    
 \frac{1}{
 [ s\, z^{ }_1 + (1-s)\, z^{ }_2 ]^2 
 - p_a^2 \,
 | \vc{k} - s\, \vc{p}^{ }_e  |^2
 } 
 \;. \la{theta_varphi_av}
\ea 
The denominator is a second order polynomial in $s$, 
and the integral over $s$ yields 
\be
 \int_0^{1} \! 
% {\rm d}s \,
% \{ \mbox{\eq\nr{theta_varphi_av}} \}
 \frac{{\rm d}s}{
 [ s\, z^{ }_1 + (1-s)\, z^{ }_2 ]^2 
 - p_a^2 \,
 | \vc{k} - s\, \vc{p}^{ }_e  |^2
 } 
 = 
 \mathcal{A}^{ }_{
 (z^{ }_1 - z^{ }_2)^2 - p_a^2 p_e^2 \,,\,
 2 [z^{ }_2(z^{ }_1-z^{ }_2) + p_a^2 \vc{k}\cdot\vc{p}_e^{ } ] \,,\,
 z_2^2 - p_a^2 k^2 
 }
 \;, \la{s_int}
\ee
where the value of 
$\mathcal{A}^{ }_{\alpha,\beta,\gamma}$ is given by
($\alpha,\beta,\gamma\in\mathbbm{R}$)
\ba
 \mathcal{A}^{ }_{\alpha,\beta,\gamma} 
 & \equiv & 
 \int_0^1 \! {\rm d}x \, \mathbbm{P}\, 
 \frac{1}{\alpha x^2 + \beta x + \gamma}
 \\
%%%% 
 & = & 
  \left\{ 
 \begin{array}{ll}
   \frac{1}{\sqrt{\beta^2 - 4\alpha\gamma}}
   \ln\biggl|
         \,
         \frac{\beta + 2\gamma + \sqrt{\beta^2 - 4\alpha\gamma}}
              {\beta + 2\gamma - \sqrt{\beta^2 - 4\alpha\gamma}} 
         \,
      \biggr|
   & 
   \; ,\quad
   \beta^2 > 4 \alpha\gamma \\ 
%%%%%%%%%%%%%%%%%%%%
%   \frac{4\alpha y}{\beta(2\alpha y + \beta)}
   \frac{2}{\beta + 2\gamma}
   & 
   \; ,\quad
   \beta^2 = 4 \alpha\gamma \\ 
%%%%%%%%%%%%%%%%%%%%
%   \frac{1}{i \sqrt{4\alpha\gamma - \beta^2}}
%   \ln\biggl(
%         \frac{\beta + 2\gamma + i\sqrt{4\alpha\gamma - \beta^2}}
%              {\beta + 2\gamma - i\sqrt{4\alpha\gamma - \beta^2}} 
%      \biggr)
   \frac{2}{\sqrt{4\alpha\gamma - \beta^2}}
   \arctan
      \biggl(
         \frac{\sqrt{4\alpha\gamma - \beta^2}}
              {\beta + 2\gamma} 
      \biggr)
   & 
   \; ,\quad
   \beta^2 < 4\alpha\gamma 
 \end{array}
 \right.
 \;. \la{A1}
\ea

Let us then turn to the average $\mathcal{G}$ 
from \eq\nr{def_G}. 
The idea here is to choose the angle 
$\theta^{ }_{\vc{p}_a,\vc{p}_d}$ to play the role of $\theta$.
The other scalar product can be expressed as 
\be
  \vc{p}^{ }_a \cdot \vc{k} 
 = 
 p^{ }_a k\, 
 \bigl( 
 \cos\theta^{ }_{\vc{p}_a,\vc{p}_d}
 \cos\theta^{ }_{\vc{k},\vc{p}_d}
 + 
 \sin\theta^{ }_{\vc{p}_a,\vc{p}_d} 
 \sin\theta^{ }_{\vc{k},\vc{p}_d}\, \cos\varphi 
 \bigr) 
 \;, 
\ee
where $ \theta^{ }_{\vc{k},\vc{p}_d} $ is fixed by the outer integral
(cf.\ \eq\nr{kdotpd}). 
Inserting this as the argument of~$P^{ }_n$
yields a polynomial in $\cos\varphi$. 
The azimuthal averages can now be carried
out, 
\be
 \int_0^{2\pi} \! \frac{{\rm d}\varphi}{2\pi} \, 
 \cos^n\!\varphi 
 = 
 \frac{[1+(-1)^n]^2 \Gamma(\frac{1+n}{2})}
      {4 \Gamma(\frac{1}{2})\Gamma(\frac{2+n}{2})}
 \;. \la{cosn}
\ee
As only even powers contribute, the dependence on 
$
 \sin\theta^{ }_{\vc{p}_a,\vc{p}_d} 
$
is quadratic, and can be expressed in terms of 
$\cos^2\!\theta^{ }_{\vc{p}_a,\vc{p}_d}$. It is 
convenient to write 
$
 \cos \theta^{ }_{\vc{p}_a,\vc{p}_d} = 
 \vc{p}^{ }_a\cdot\vc{p}^{ }_d/(p^{ }_a p^{ }_d)
$.
Thereby
\be
 \langle P^{ }_n (\vc{p}^{ }_a\cdot\vc{k}) \rangle^{ }_{\varphi}
 = 
 \sum_{k=0}^n c^{ }_k (\vc{p}^{ }_a\cdot\vc{p}^{ }_d)^k
 \; \equiv \; 
 R^{ }_n (\vc{p}^{ }_a\cdot\vc{p}^{ }_d)
 \;, 
\ee
and \eq\nr{def_G} takes the form
\be
 \mathcal{G}^{ }_{\vc{p}_a;\vc{p}_d,\vc{k};z} 
  \, P^{ }_n(\vc{p}^{ }_a\cdot\vc{k})
 = 
  \int_{-1}^{+1} \! 
  \frac{ {\rm d}\cos\theta^{ }_{\vc{p}_a,\vc{p}_d} }{2} \,
 \mathbbm{P} \,   
 \frac{ 
 R^{ }_n(  \vc{p}^{ }_a\cdot\vc{p}^{ }_d  ) }
 {z -  \vc{p}^{ }_a\cdot\vc{p}^{ }_d  }
 \;. 
\ee

To carry out the remaining integral, we repeat the logic 
of \eqs\nr{factorize} and \nr{splitup}. 
We write 
\be
 R^{ }_n( \vc{p}^{ }_a\cdot\vc{p}^{ }_d )
 = 
 R^{ }_n(z^{ }) + 
 (z^{ } - \vc{p}^{ }_a\cdot\vc{p}^{ }_d )\,
 \widetilde{R}^{ }_{n-1} 
 ( \vc{p}^{ }_a\cdot\vc{p}^{ }_d )
 \;, 
\ee
where 
$
  \widetilde{R}^{ }_{n-1}( \vc{p}^{ }_a\cdot\vc{p}^{ }_d )
  = 
  \sum_{l=0}^{n-1} d^{ }_l \, ( \vc{p}^{ }_a\cdot\vc{p}^{ }_d )^l_{ }
$, 
with the coefficients 
$
 d^{ }_l = - \sum_{k=l+1}^n c^{ }_k\, z^{k-l-1}
$.
Then 
\be
 \mathcal{G}^{ }_{\vc{p}_a;\vc{p}_d,\vc{k};z} 
  \, P^{ }_n(\vc{p}^{ }_a\cdot\vc{k})
 = 
  \int_{-1}^{+1} \! 
  \frac{ {\rm d}\cos\theta^{ }_{\vc{p}_a,\vc{p}_d} }{2} \,  
 \mathbbm{P} \,   
 \biggl\{ 
 \frac{ 
 R^{ }_n(z) }
 {z - p^{ }_a p^{ }_d \cos \theta^{ }_{\vc{p}_a,\vc{p}_d} }
 + \sum_{l = 0}^{n-1} d^{ }_l \,
 p^{l}_a p^{l}_d  \cos^l\! \theta^{ }_{\vc{p}_a,\vc{p}_d} 
 \biggr\}
 \;. 
\ee
Both parts are easily integrated, yielding 
\ba
 \mathcal{G}^{ }_{\vc{p}_a;\vc{p}_d,\vc{k};z} 
  \, P^{ }_n(\vc{p}^{ }_a\cdot\vc{k})
 & = & 
 \frac{R^{ }_n(z)}{2 p^{ }_a p^{ }_d}
 \ln \biggl|
       \frac{z + p^{ }_a p^{ }_d}{ z - p^{ }_a p^{ }_d } 
     \biggr|
 + \sum_{l = 0}^{n-1} 
 \frac{[ 1+(-1)^l ] 
 d^{ }_l \,
 p^{l}_a p^{l}_d  }{2(l+1)}
 \;. 
\ea

%%%%%%%%%%%%%%%%%%%%%%%% BIBLIO %%%%%%%%%%%%%%%%%%%%%%%%%%%%%%%%%%%%%%%%%%
%

\small{
 
}

\end{document}